\documentclass{aa}  
\usepackage{mathrsfs} 
\usepackage{esvect} 
\usepackage{amsmath} 
\usepackage{color}
\usepackage[dvipsnames]{xcolor} 
\usepackage{hyperref} 
\usepackage{array} 
\usepackage{graphicx} 
\usepackage{multirow}
\usepackage{natbib}
\bibpunct{(}{)}{;}{a}{}{,} 

\usepackage{comment}
\usepackage{graphicx}
\usepackage{txfonts}
\newcommand{\contributor}[1]{\textcolor{blue}{#1}}

\begin{document} 
\title{Bayesian self-calibration and imaging in Very Long Baseline Interferometry}
\titlerunning{Bayesian self-calibration in VLBI}
\authorrunning{Kim et. al.}
    

\author{Jong-Seo Kim \inst{1}\fnmsep\thanks{jongkim@mpifr-bonn.mpg.de}
\and Aleksei S. Nikonov \inst{1}
\and Jakob Roth \inst{2,3,4}
\and Torsten A. En{\ss}lin \inst{2,3}
\and Michael Janssen \inst{1,5}
\and Philipp Arras \inst{2,4}
\and Hendrik Müller \inst{1,6}
\and Andrei P. Lobanov \inst{1}
}

\institute{    
	    Max-Planck-Institut f\"ur Radioastronomie, Auf dem H\"ugel 69, D-53121 Bonn, Germany
	    \and
        Max-Planck-Institut f\"ur Astrophysik, Karl-Schwarzschild-Str. 1, 85748 Garching, Germany
        \and
        Ludwig-Maximilians-Universit\"at, Geschwister-Scholl-Platz 1, 80539 Munich, Germany
        \and
		  Technische Universität M\"unchen (TUM), Boltzmannstr. 3, 85748 Garching, Germany
        \and
        Department of Astrophysics, Institute for Mathematics, Astrophysics and Particle Physics (IMAPP), Radboud University, P.O. Box 9010, 6500 GL Nijmegen, The Netherlands
        \and 
        Jansky Fellow of National Radio Astronomy Observatory, 1011 Lopezville Rd, Socorro, NM 87801, USA
         }
   \date{Received February 19, 2024; accepted July 17, 2024}

 
\abstract
{Self-calibration methods with the \texttt{CLEAN} algorithm have been widely employed in Very Long Baseline Interferometry (VLBI) data processing in order to correct antenna-based amplitude and phase corruptions present in the data. However, human interaction during the conventional \texttt{CLEAN} self-calibration process can impose a strong effective prior, which in turn may produce artifacts within the final image and hinder the reproducibility of final results.}
{In this work, we aim to demonstrate a combined self-calibration and imaging method for VLBI data in a Bayesian inference framework. The method corrects for amplitude and phase gains for each antenna and polarization mode by inferring the temporal correlation of the gain solutions.}
{We use Stokes I data of M87 taken with the Very Long Baseline Array (VLBA) at 43\,GHz, pre-calibrated using the rPICARD CASA-based pipeline. For antenna-based gain calibration and imaging, we use the Bayesian imaging software \texttt{resolve}. To estimate gain and image uncertainties, we use a Variational Inference method. }
{We obtain a high-resolution M87 Stokes I image at 43\,GHz in conjunction with antenna-based gain solutions using our Bayesian self-calibration and imaging method. The core with counter-jet structure is better resolved, and extended jet emission is better described compared to the \texttt{CLEAN} reconstruction. Furthermore, uncertainty estimation of the image and antenna-based gains allows us to quantify the reliability of the result.}
{Our Bayesian self-calibration and imaging method is able to reconstruct robust and reproducible Stokes I images and gain solutions with uncertainty estimation by taking into account the uncertainty information in the data.}


\maketitle

\section{Introduction} \label{Chap1}

Calibration and imaging are closely interconnected in radio interferometry which combines signals received at multiple antennas in order to form a virtual telescope with the aperture effectively determined by the largest projected distance between the participating antennas. The reconstruction of high-fidelity images from radio interferometric observations requires appropriate calibration of the data. In the standard calibration process, amplitude and phase corruptions are corrected, and data are flagged and averaged by a deterministic algorithm. After calibration, image reconstruction is required as radio interferometers measure incomplete Fourier components of the sky brightness distribution instead of observing an actual image. Image reconstruction in radio interferometry is an ill-posed problem. Therefore, a unique solution does not exist. Furthermore, the standard data reduction process in radio interferometry still leaves a large amount of uncertainty due to instrumental and atmospheric errors. A probabilistic approach can be beneficial to properly deal with the incompleteness and uncertainty of radio interferometric data.

In Very Long Baseline Interferometry (VLBI), the distance between ground-based telescopes reaches the size of the earth to achieve high resolution. The data obtained from this set of telescopes are highly sparse owing to limited number of antennas and observation time. Due to this sparsity, it is challenging to reconstruct high-fidelity images without additional assumptions about the source structure and instrument effects. This additional knowledge is called prior knowledge in Bayesian statistics. Encoding physically sensible prior knowledge results in robust and consistent image reconstructions from highly sparse VLBI data sets. In addition to the sparsity, the data reduction process in VLBI is complicated due to large gain uncertainties, and low SNR measurements \citep{Janssen_2022}. Particularly, in millimeter VLBI, tropospheric effects cause significant visibility amplitude and phase errors. As a result, interpolating calibrator solutions to the science target is insufficient to correct time-dependent data corruption. 
Self-calibration methods within the \texttt{CLEAN} framework are the standard procedure to correct additional antenna-based gain corruption after the initial calibration.  

The \texttt{CLEAN} algorithm \citep{Hoegbom_1974_CLEAN, Clark_1980_CLEAN} is the de-facto standard for image reconstruction in radio interferometry due to its simplicity and interactivity. However, \texttt{CLEAN} suffers from several shortcomings \citep{Arras_2021_CygA}. First, it cannot produce optimal results because the algorithm does not produce images consistent with the data. In \texttt{CLEAN}, the sky brightness distribution is assumed to be a collection of point sources, which is incorrect for diffuse emission. Note that multi-scale \texttt{CLEAN} \citep{Cornwell_2008} is able to perform better than conventional \texttt{CLEAN} methods for extended emission. However, the conventional multi-scale \texttt{CLEAN} still does not compare the model with the data directly, the consistency between model and data therefore is not guaranteed. Furthermore, convolving with the \texttt{CLEAN} beam hinders achieving the optimal resolution.
Second, it is difficult to modify the model. We cannot explicitly utilize existing knowledge, such as closure amplitudes and phases, the positivity of the flux, and polarization constraints in image reconstruction by \texttt{CLEAN}.
Third, \texttt{CLEAN} requires the supervision of an experienced scientist. Hence, the results can be biased. For example, \texttt{CLEAN} windows and weighting schemes are user-dependent. Therefore, a strong effective prior is imposed onto the final image. Lastly, \texttt{CLEAN} does not provide an estimate of the uncertainties within an image. Due to these limitations, using \texttt{CLEAN} for self-calibration can have significant drawbacks. Human biases can be introduced into the final image during self-calibration since the data are modified by the inconsistent model image and flagged without objective criteria. Moreover, the simplicity of the \texttt{CLEAN} restricts the implementation of a more sophisticated optimization algorithm in self-calibration. 

Forward modeling imaging algorithms can be utilized for more robust self-calibration in VLBI. The forward modeling approach interprets image reconstruction as an optimization problem. Forward modeling fits the model to the data directly, thus ensuring consistency between the image and the data. There exist traditional forward modeling approaches, such as the Maximum Entropy Method (MEM; \cite{Cornwell_1985_MEM, Narayan_1986_MEM}. However, \texttt{CLEAN}-based algorithms have historically been favored due to the simplicity of implementation and the limited necessary computational resources. Recent developments in optimization theory and computer performance enable forward modeling algorithms to outperform \texttt{CLEAN}. There are a variety of different forward modeling methods, such as Regularized Maximum Likelihood (RML) and Bayesian imaging.

The RML method \citep{Wiaux_2009, Akiyama_2017_SMILI, Chael_2018_closure, Mueller_2022_Doghit} is a forward modeling approach based on ridge regression. The image is reconstructed by minimizing an objective function containing data fidelity terms and regularizers. The algorithm can be faster than Bayesian approaches because one final image is generated instead of a collection of posterior sample images. It can be interpreted as the maximum a posteriori (MAP) estimation in Bayesian statistics. Recently, \cite{Dabbech_2021_calibration&imaging} performed direction-dependent gain calibration and imaging jointly in the RML framework.

Bayesian imaging \citep{Junklewitz_2015, Arras_2021_CygA, Broderick_2020_THEMIS, Tiede_2022_Comrade} in radio interferometry is a probabilistic approach that provides samples of potential images that are consistent with the data and prior assumptions within the noise statistics instead of one image. The samples can be interpreted as being drawn from the posterior probability distribution. In Bayesian statistics, the probability distributuion of the reconstructed image is called the posterior, which is obtained from the prior distribution and the likelihood by Bayes' theorem. 

Bayesian imaging has an intrinsic advantage: the prior knowledge can be encoded explicitly into the image reconstruction. Knowledge about the source and instrument, such as closure quantities, can be utilized directly to obtain robust images from sparse VLBI data sets \citep{Arras_2022_M87*}. Moreover, uncertainty estimation can be obtained as a by-product of image reconstruction. The reliability of the results can be quantified by the uncertainty estimation. \cite{Arras_2019_calibration} performed calibration and imaging jointly in the Bayesian framework with Very Large Array (VLA) data. Recently, \cite{Roth_2023_DDE_calibration} proposed Bayesian direction-dependent calibration and imaging with VLA data. 

Building on the works of \cite{Arras_2019_calibration}, we propose a novel Bayesian self-calibration and imaging method for VLBI data sets. For our method, we require pre-calibrated data that have not undergone manual flagging in order to reduce human biases. Instead of the iterative procedure in the \texttt{CLEAN} self-calibration approach, our method infers the Stokes I image and antenna-based gain terms simultaneously. The spatial correlation between pixels in the sky model is inferred by the non-parametric Gaussian kernel following \cite{Arras_2021_CygA}. Furthermore, time-dependent antenna-based amplitude and phase gain terms are inferred by considering the temporal correlation between points in time-dependent gain solutions. We estimate the uncertainty of reconstructed parameters by the Variational Inference method \citep{Biel_2016_VI_review, Knollmueller_2019}. Estimated uncertainty of the image and antenna-based gain terms can provide valuable information about the image and each antenna condition.

This paper is structured as follows. In section 2, we describe the measurement equation in radio interferometry and interpretation of image reconstruction problem in Bayesian framework. In section 3, we describe our self-calibration and imaging prior model in VLBI and inference algorithm. We validate the Bayesian self-calibration and imaging with real VLBA M87 data at 43\,GHz in section 4 and synthetic data in section 5. In section 6, we summarize our results.



\section{Image reconstruction in radio interferometry} \label{Chap2}

\subsection{Radio Interferometer Measurement Equation (RIME)}

The mapping from a given sky brightness distribution to the data measured by the radio interferometer is represented by the radio interferometer measurement equation (RIME) and will be defined in this subsection. The two-point correlation function of the electric field is equivalent to the Fourier components of the sky brightness distribution by the van Cittert-Zernike theorem \citep{Hamaker_1996, Smirnov_2011, Thompson_2017_book}. A radio interferometer can obtain this correlation function called visibility $V(u,v)$. In VLBI, due to the small field of view, the relationship between the visibility $V(u,v)$ and sky brightness distribution $I(x,y)$ can be approximated with a two-dimensional Fourier transform \citep{Thompson_2017_book}:
\begin{align} \label{eq:measurement_equation}
   \mathcal{V} (u,v) = \mathbb{F}\,I := \int_{-\infty}^{\infty}\int_{-\infty}^{\infty} I(x,y)  e^{-2\pi i (ux+vy)} dx \, dy,
\end{align}
where $\mathbb{F}$ is the Fourier transform, $(u,v)=:\vv{k}$ are the Fourier and $(x,y)$ the image domain coordinates.

In an observation, a visibility data point from antenna pair $i,j$ at a time $t$ is 
the visibility at the Fourier space location $\vv{k}_{ij}(t):= (u,v)_{ij}(t) = \vv{b}_{\perp, ij}(t)/\lambda$, with $\vv{b}_{\perp, ij}(t)$ representing the vector baseline 
between the antenna pair at time $t$ projected orthogonal to the line of sight, and $\lambda$ being the observing wavelength.

The measured visibility, corrupted by errors in direction-independent antenna gains ($g_{i}, \, g_{j}$), and thermal noise, $n_{ij}$ is   
\begin{align} 
    V_{ij}(t) = g_{i}(t) \, g^*_{j}(t) \, \mathcal{V}(\vv{k}_{ij}(t)) + n_{ij}(t),
\end{align}
where $^*$ denotes complex conjugation.

The measurement equation can be written in shorthand notation:
\begin{align} \label{eq:measurement_eq_short} 
    V_{ij}(t)  = R^{(g,t)} I + n_{ij}(t) := g_{i}(t) \, g^*_{j}(t) \, B(t)[\mathbb{F}\,I] + n_{ij}(t),
\end{align}
with $R^{(g,t)}$ describing the measurement operator that is composed of sampling of the measured baseline visibilities for each sampled time t, $B(t)[\mathcal{V}]_{ij}:=\mathcal{V}(\vv{k}_{ij}(t))$, and the application of the antenna gains $g$.

Note that a straightforward inverse Fourier transform cannot produce high-fidelity images because the data are measured at sparse locations in the (u,v) grid, corrupted by gains, and contain uncertainty from thermal noise. Due to incompleteness and uncertainty in the data, multiple possible image reconstructions can be obtained that would be consistent with the same data. To address this problem, image reconstruction in radio interferometry aims to infer for example the most plausible image $I(x,y)$ from incomplete, distorted, and noisy data $V_{ij}(t)$.

\subsection{Bayesian imaging}
Bayes' theorem provides a statistical framework to construct the posterior probability distribution for the sky $I(x,y)$ from the incomplete visibility data  $V_{ij}(t)$.
Instead of computing a single final image, Bayesian imaging provides the probability distribution over possible images, called the posterior distribution. The posterior distribution of the image $I$ given the data $V$ can be calculated via Bayes' theorem:
\begin{align} \label{eq:Bayes' thm}
\mathcal{P}(I|V) = \frac{\mathcal{P}(V|I)\, \mathcal{P}(I)} {\mathcal{P}(V)},
\end{align}
where $\mathcal{P}(V|I)$ is the likelihood, $\mathcal{P}(I)$ is the prior, and $\mathcal{P}(V) = \int \mathcal{D}I\, \mathcal{P}(V|I) \,\mathcal{P}(I)$ the evidence acting as a normalization constant. In the case of Bayesian image reconstruction, the evidence involves an integral over the space of all possible images, hence $\int\mathcal{D}I \ldots\,$ indicates a path integral.\\

Bayes' theorem follows directly from the product rule of probability theory. It can be rewritten to resemble the Boltzmann distribution in statistical mechanics by introducing the information Hamiltonian \citep{Ensslin_2019}:

\begin{align}
    \mathcal{P}(I|V) = \frac{e^{-\mathcal{H}(V,I)}}{{\mathscr{Z}(V)}},
\end{align}
where $\mathcal{H}(V,I) = - \ln(\mathcal{P}(V,I))$ is the information Hamiltonian, which is the negative log probability, and $\mathscr{Z}(V) = \int \mathcal{D}I\, e^{-\mathcal{H}(V,I)}$ is the partition function. \\

For imaging with a large number of pixels, the posterior distribution of the image $\mathcal{P}(I|V)$ will be high dimensional. Due to the high dimensionality, it is not feasible to visualize the full posterior probability distribution. Thus, we compute and analyze summary statistics as the posterior mean 
\begin{align}
m =  \langle I \rangle_{\mathcal{P}(I|V)} 
\end{align}
and standard deviation
\begin{align}
\sigma_I = \sqrt{ \langle (I-m)^2 \; \rangle_{\mathcal{P}(I|V)}}, 
\end{align}
where $\langle f(I) \rangle_{\mathcal{P}(I|V)}:=\int \mathcal{D}I\, f(I) \, \mathcal{P}(I|V)$ denotes the posterior average of $f(I)$. \\

In Bayesian imaging, we can design the prior model to encode the desired prior knowledge, such as the positivity of flux density and the polarization constraints. The use of this additional prior information about the source and the instrument stabilizes the reconstruction of the sky image. For instance, diffuse emission can be well described by a smoothness prior on the brightness of nearby pixels. The details of our prior assumption are discussed in section \ref{Chap3}.

Furthermore, the mean and standard deviation of any parameter in the model can be estimated. The standard deviation can be utilized to quantify the uncertainty of reconstructed parameters. In other words, we can propagate the uncertainty information in the data domain into other domains, such as the image or the antenna gain.

For this, the set of quantities of interest that are to be inferred need to be extended, for example to include the sky intensity $I$ and  all antenna gains $g$. The gains enter the measurement equation by determining the response, $R\equiv R^{(g)}$, and thereby the likelihood, which becomes $\mathcal{P}(V|g,I)$.

The joint posterior distribution for gains and sky is again obtained by Bayes' theorem,
\begin{align}
\mathcal{P}(g,I|V) = \frac{\mathcal{P}(V|g,I) \,\mathcal{P}(g,I)}{\mathcal{P}(V)}, 
\end{align}
where $\mathcal{P}(V) = \int \mathcal{D}I \int \mathcal{D}g \, \mathcal{P}(V,g,I)$ is the joint evidence and $\mathcal{P}(g,I)=\mathcal{P}(g)\,\mathcal{P}(I)$ the joint prior, here assumed to be decomposed into independent sky and gain priors, $\mathcal{P}(I)$ and $\mathcal{P}(g)$ respectively. 

By marginalization of the joint posterior over the sky degrees of freedom, $\int \mathcal{D}I \, \mathcal{P}(I,g|V)=:\mathcal{P}(g|V)$, a gain (only) posterior can be obtained. From this,
the posterior mean $\langle g_{i} \rangle_{\mathcal{P}(g|V)}$ and standard deviation $\sigma_{g_{i}}$ can be calculated for any antenna $i$.
As a result, each antenna gain corruption mean and uncertainty can be inferred from the data by obtaining the joint posterior distribution. Uncertainty estimation is a distinctive feature of Bayesian imaging.

\subsection{Likelihood distribution} \label{Ch2.3}
For compact notation, we define the signal vector $s = (g, I)$ containing the gain $g$ and the image $I$ that carries all their components. The visibilities $V$ are in our case the data $d$. The likelihood distribution $\mathcal{P}(d|s)$ largely determines the resulting Bayesian imaging algorithm since it contains the information on the measurement process that will be inverted by the algorithm. 

The probability distribution of the noise $n$ in the measurement equation (Eq. \ref{eq:measurement_eq_short}) can often be approximated to be Gaussian distribution. From a statistical point of view, the effective noise level is a sum of many noise contributions and it can be approximated as Gaussian distribution by the central limit theorem, if sufficient SNR is provided.

We therefore assume that the noise $n$ is drawn from a Gaussian distribution with covariance $N$,
\begin{align}
n \curvearrowleft \mathscr{G}(n,N) = \frac{1}{\sqrt{|2\pi N|}} e^{-\frac{1}{2} n^{\dag}N^{-1}n}
\end{align}
where $|2 \pi N|$ is the determinant of the noise covariance (multiplied by $2\pi$) and $\dag$ indicates the conjugate transpose of a vector or matrix.

Under this assumption, the likelihood $\mathcal{P}(d|s)$ is a multivariate Gaussian distribution:
\begin{equation}
\begin{aligned} 
    \mathcal{P}(d|s) &=  \mathscr{G}(V-R^{(g)}I, N).
\end{aligned}
\end{equation}

The likelihood Hamiltonian 
\begin{equation}
\begin{aligned}
    \mathcal{H}(d|s) &= - \ln(\mathcal{P}(d|s)) \\
    &= \frac{1}{2} (V-R^{(g)}I)^{\dag}N^{-1}(V-R^{(g)}I) + \frac{1}{2} \ln |2\pi N|
\end{aligned}
\end{equation}
is then of quadratic form in $I$, but of fourth order in $g$, as $R^{(g)}$ is already quadratic in $g$.
This renders the joint imaging and calibration problem in radio interferometry a challenging undertaking. 

Furthermore, we assume that the noise is not correlated with time and different baselines. Time and baseline correlation of the noise could be inferred as well, similar to the inference of the signal angular power spectrum. However, this would be another algorithmic advancement in image reconstruction since it is significantly increasing the complexity of the algorithm. Therefore, we use the approximation that the noise is uncorrelated with time and different baselines, which is current standard in radio imaging. In other words, the noise covariance $N_{ij}=\delta_{ij}\,\sigma_i^2$ is diagonal. Under the assumption, denoting here with $n$ any datum in the data vector (that was previously indexed by the two antenna identities and a time stamp, $(i,j,t)$), the likelihood Hamiltonian can be interpreted as the data fidelity term:
\begin{align} \label{eq:data_fidelity}
    \mathcal{H}(d|s) \;=\; \frac{1}{2} \sum_{n} \frac{|V_{n} - (R^{(g)} I)_{n}|^{2}} {\sigma_{n}^{2}} = \frac{1}{2} \; \chi^{2}.
\end{align}

\subsection{Posterior distribution}

Bayesian imaging aims to calculate the posterior distribution $\mathcal{P}(s|d)$ from the visibility data $d=V$ and the prior $\mathcal{P}(s)$. 
Obtaining the posterior distribution $\mathcal{P}(s|d)$ is equivalent to calculating the posterior information Hamiltonian $\mathcal{H}(s|d)$.
The information Hamiltonian contains the same information as the probability density it derives from, but it is numerically easier to handle since the multiplication of two probability distributions is converted to an addition. 
The joint data and signal information Hamiltonian is composed of a likelihood and prior information term, $\mathcal{H}(s)$ and $\mathcal{H}(d|s)$, respectively, which are just added,
\begin{align}
    \mathcal{H}(d,s) = \mathcal{H}(d|s) + \mathcal{H}(s).
\end{align}
The posterior information Hamiltonian 
 \begin{align} \label{eq:posterior Hamiltonian}
    \mathcal{H}(s|d) \equiv - \ln(\mathcal{P}(s|d)) = \mathcal{H}(d|s) + \mathcal{H}(s) - \mathcal{H}(d)
\end{align}
differs from this only by the subtraction of $\mathcal{H}(d)$, the evidence Hamiltonian. \\

There are numerous algorithms that explore the posterior, such as Markov chain Monte Carlo (MCMC) and Hamiltonian Monte Carlo (HMC) methods. 
Those are, however, computationally very expensive for the ultra-high dimensional inference problems we are facing in imaging. Another method to obtain a signal estimate is to calculate the maximum a posteriori (MAP) estimation. The posterior Hamiltonian is minimized to calculate the MAP estimator of a signal
\begin{align}
	s_\text{MAP} := \text{argmin}_s \mathcal{H}(s|d).
\end{align}
 
The evidence term is independent of the signal and therefore has no influence on the location of the minimum in signal space. Thus, equivalently the joint information Hamiltonian $\mathcal{H}(d,s)$ of the signal and data can be minimized, as it only differs from the posterior information $\mathcal{H}(s|d)$ by the signal independent evidence $\mathcal{H}(d)$. MAP estimation is computationally much faster than sampling the entire posterior density, enabling high-dimensional inference problem. However, MAP does not provide any uncertainty quantification on its own \citep{Knollmueller_2019}. Note that RML methods can be interpreted as MAP estimation in Bayesian framework. The interpretation of RML method in terms of Bayesian perspective can be found in Appendix \ref{app:RML}.

Bayesian inference ideally explores the structure of the posterior around its maximum the MAP method focuses on. In case this has a nontrivial, non-symmetric structure, the MAP estimate can be highly biased w.r.t.~the more optimal posterior mean, which is optimal from an expected squared error functional perspective. In general, in Bayesian inference, one wants to be able to calculate expectation values for any interesting function $f(s)$ of the signal, be it the sky intensity, $f(s)=I$, its uncertainty dispersion $f(s)=(I-\overline{I})^2$, power spectrum $f(s)=P_I(\vv{k})=|\mathbb{F}\,I|^2(\vv{k})$, or any function of the gains. 
Thus,
\begin{align}
\overline{f}:=\langle f(s) \rangle_{(s|d)}
\end{align}
should somehow be accessible. 
For non-Gaussian posteriors, but also for non-linear signal functions, this will in general differ from $f(s_\text{MAP})$, and the difference can be substantial \citep{Ensslin_2011}.

A good compromise between efficiency (like that given by MAP) and accuracy (as HMC can provide) is Variational Inference method. These fit a simpler probability distribution to the posterior, one from which a set $S=\{s_1, \ldots \}$ of posterior samples can be drawn. 
The estimate of any posterior average of a signal function can then be approximated by the sample average
\begin{align}
\overline{f} \approx \frac{1}{|S|}\sum_{s\in S}f(s),
\end{align}
where $|S|=\sum_{s\in S}\, 1$ denotes the size of the set. To estimate the posterior distribution of the sky and gains, Metric Gaussian Variational Inference method \cite[MGVI,][]{Knollmueller_2019} is utilized in this work. For further details on the used Variational Inference scheme, see section \ref{Ch3.5}.

\subsection{Self-calibration in VLBI}

Before imaging, correlated radio interferometric data needs to be calibrated to correct for data corruption. For calibration, the target source of an observation and calibrator sources, which are commonly bright and point-like, are observed alternately. 
By comparing expected properties of the calibrators and observed data, instrumental effects can be identified. The calibrator solution, correcting instrumental effects, is then applied to the data of the target source. However, residual errors still exist after the initial calibration due to the time dependence of amplitude and phase corruptions and the angular separation between the target source and calibrators.

With a sufficient signal-to-noise ratio (SNR), the target source itself can be used as a calibrator. Using the target source itself for calibration is called self-calibration \citep{Cornwell_1981_selfcal, Brogan_2018}. Following the initial calibration, the conventional self-calibration method, iteratively correcting the antenna-based gain corruptions, works as follows: First, a model image is reconstructed from the target data $V^{(m)}_{ij}$. The image can be reconstructed by \texttt{CLEAN} or forward modeling algorithms. Second, the residual gains $g$ are estimated by minimizing the cost function \citep{Cornwell_1981_selfcal, Taylor_NRAO_book_1999}:

\begin{align}
\label{eq:selfcal_cost_function} 
    S = \sum_{k} \sum_{i,j}^{i \neq j} \frac{|V_{ij}^{(m)}(\vv{k}_{ij}(t_k)) - g_{i}(t_k) g_{j}^{\ast}(t_k) M_{ij}(\vv{k}_{ij}(t_k))|^2}{\sigma_{ij}^2 (t_k)},
\end{align}
where $g_i(t_{k})$ is the complex gain of the antenna $i$ at the time $t_{k}$, $M_{ij}(\vv{k}_{ij}(t_k))$ is the Fourier component at $\vv{k}_{ij}(t_k)$ of the model image, and $\sigma_{ij}$ is the noise standard deviation of the visibility data $V_{ij}$. \\

The cost function contains the difference between the data $V_{ij}$ and the Fourier-transformed model image $M_{ij}$ with gains $g$. Gains $g$ are free parameters in the minimization. Simple minimization schemes, such as the least-squares method, are utilized to infer the gains $g$ with a solution interval, which is the correlation constraint. Then the estimated residual gain corruption $g$ is removed from the data. The $m$-th iteration of self-calibration is

\begin{align}
    V_{ij}^{(m)}(t_k) = \frac{V_{ij}^{(m-1)}(t_k)}{g_{i}(t_{k}) \,g_{j}^{\ast}(t_{k})}.
\end{align}

There are different ways to propagate the error in CLEAN self-calibration. As an example, in \texttt{DIFMAP} \citep{1997ASPC..125...77S} software, phase self-calibration does not change the errors and amplitude self-calibration corrects errors by treating gains as constant within the solution interval. As a result, error of the self-calibrated data is 
\begin{align} 
\sigma_{ij, self} = \frac{\sigma_{ij}}{|g_{i}| \,|g_{j}|},
\end{align}
where $\sigma_{ij}$ is the error of the data before self-calibration.

Note that the SNR of the data is conserved in self-calibration, if the gains are treated as constant in error propagation. If the reported noise estimates are correct, it is a reasonable assumption. In \texttt{DIFMAP}, there is also an option to fix the weights to prevent the amplitude corrections from being applied to the visibility errors.

In conclusion, self-calibration consists of three steps: model image reconstruction from the data, residual gain estimation from the minimization, and modification of the data by removing estimated residual gains. Those three steps are continued iteratively until some stopping criterion is met.

Self-calibration works as long as the effective number of degrees of freedom for the visibility data for $N$ stations ($N(N-1)/2$) is larger than the degrees of freedom of antenna gains (amplitude: $N$, phase: $N-1$). The system of equations to obtain antenna-based gains from the data is over-determined. As a result, with a sufficient SNR and number of antennas, we can estimate the gains from the visibility data. The closure amplitude and phase are invariant under the antenna-based gain calibration. Therefore, we can calibrate antenna gains by self-calibration while conserving relative intensity information and relative position about the source. 

While self-calibration is required to obtain a high-fidelity VLBI image, it does have shortcomings. Strong biases might be imposed during the self-calibration due to the use of inconsistent model image from \texttt{CLEAN} reconstruction and manual setting of the solution interval for gain estimation. Furthermore, self-calibration hinders the reproduction of the result due to the human interactions in imaging. Some fraction of the data points are often flagged during the self-calibration, either manually or on the basis of non-convergence of the gain solutions. The flagging is however prone to errors, as it is difficult to distinguish between a bad data point and a data point inconsistent with the model image. 

In this paper, self-calibration and imaging are combined into a joint inference problem. In other words, the inference of gains and image is performed simultaneously. Data flagging is performed solely by the calibration pipeline. It is a generalized version of conventional self-calibration, as it combines image reconstruction and gain inference by the minimization of the data fidelity term. Note that the posterior distribution of gains are explored in Bayesian self-calibration instead of computing a point estimate in conventional iterative self-calibration approaches. Originally, the joint antenna-based calibration and imaging approach was presented in \citet{Arras_2019_calibration} and extended to also include direction-dependent effects in \cite{Roth_2023_DDE_calibration}. The Bayesian self-calibration and imaging allow us to reconstruct less-biased reproducible images with the help of a sophisticated prior model and minimization scheme.

\section{The algorithm} \label{Chap3}
\subsection{\texttt{resolve}}
The self-calibration approach developed in this paper is realized using the package \texttt{resolve}\footnote{\url{https://gitlab.mpcdf.mpg.de/ift/resolve}}, which is an open-source Bayesian imaging software for radio interferometry. It is derived and formulated in the language of information field theory \citep{Ensslin_2019}. The first version of the algorithm was presented by \citet{Junklewitz_2015,Junklewitz_2016}.
\citet{Arras_2019_calibration} added imaging and antenna-based gain calibration with Very Large Array (VLA) data.
Dynamic imaging with closure quantities was implemented in \cite{Arras_2022_M87*}.
In \texttt{resolve}, imaging and calibration are treated as a Bayesian inference problem. Thus from the data, \texttt{resolve} estimates the posterior distribution for the sky brightness distribution and calibration solutions. To obtain the posterior distribution and to define prior models, \texttt{resolve} builds on the a Python library NIFTy\footnote{\url{https://gitlab.mpcdf.mpg.de/ift/nifty}} \citep{Arras_2019_NIFTy5}. In NIFTy, Variational Inference algorithms such as Metric Gaussian Variational Inference \citep[][MGVI]{Knollmueller_2019} and geometric Variational Inference \citep[][geoVI]{Frank_2021}, as well as Gaussian process priors, are implemented.

\subsection{rPICARD}

For the signal stabilization and flux density calibration, we use the fully automated end-to-end rPICARD pipeline \citep{Janssen_2019_rPICARD}.
Using the input files stored in \url{https://zenodo.org/uploads/10190800}, our data calibration can be reproduced exactly with rPICARD version 7.1.2, available as Singularity or Docker container under \url{https://hub.docker.com/r/mjanssen2308/casavlbi}.\footnote{The container corresponding to v7.1.2 is tagged as ec77b9874a6de7d071c8a1d8b3816702d4c6fd9f.}

\subsection{Sky brightness distribution prior model} \label{Ch3.3}

We expect the Stokes I sky brightness distribution $I(\vv{x})$ to be positive, spatially correlated, and to vary over several orders of magnitude. We encode these prior assumptions into our sky brightness prior model. More specifically, to encode the assumption of positivity and variations over several orders of magnitude, we model the sky as
\begin{align} \label{eq:sky_prior}
    I(\vv{x}) = \exp(\psi(\vv{x})) \,,
\end{align}
where $\psi(\vv{x})$ is the logarithmic sky brightness distribution. \\

To also encode the spatial correlation structure into our prior model we generate the log-sky $\psi$ from a Gaussian process
\begin{align}
    \psi \curvearrowleft \mathscr{G}(\psi,\Psi) \, ,
\end{align}
where $\Psi$ is the covariance matrix of the Gaussian process. \\

The covariance matrix $\Psi$ represents the spatial correlation structure between pixels. Since the correlation structure of the source is unknown, we want to infer the covariance matrix $\Psi$, also called the correlation kernel, from the data. However, estimating the full covariance matrix for high-dimensional image reconstructions is computationally demanding since storing the covariance matrix scales quadratically with the number of pixels. To overcome this issue, the prior log-sky $\psi$ is assumed to be statistically isotropic and homogeneous. According to the Wiener-Khinchin theorem \citep{Wiener_1949, Khinchin_1934}, the spatial covariance $S$ of a homogeneous and isotropic Gaussian process becomes diagonal in Fourier space, and is described by a power spectrum $P_{\Psi}( | \vv{k} |)$,
\begin{equation}
\Psi(\vv{k},\vv{k}') = \langle \psi(\vv{k})\psi(\vv{k}')^{\dagger}\rangle  = (2\pi)^{d_{k}} \delta(\vv{k}-\vv{k}')P_{\Psi}(|\vv{k}|),
\end{equation}
where $d_{k}$ is the dimension of the Fourier transform. \\

The power spectrum $P_{\Psi}(|\vv{k}|)$ scales linearly with the number of pixels. Thus, inference of the covariance matrix assuming isotropy and homogeneity is computationally feasible for high-dimensional image reconstructions. In our sky prior model, the power spectrum is falling with $|\vv{k}|$, typically showing a power law shape. The falling power spectrum encodes smoothness in the sky brightness distribution $I$. Small-scale structures in the image $I$ are suppressed since high-frequency modes have small amplitudes due to the falling power spectrum. The correlation kernel in the prior can be interpreted as a smoothness regularizer in the RML method and vice versa.

The log-normal Gaussian process prior is encoded in \texttt{resolve} in the form of a generative model \citep{Knollmueller_2018}. This means that independently distributed Gaussian random variables $\xi = (\xi_\Psi, \xi_k)$ are mapped to the correlated log-normal distribution:

\begin{equation} \label{eq:sky_generative_model}
    I(\vv{x}) = \text{exp}(\psi(\vv{x})) = \text{exp}(\mathbb{F}[\sqrt{P_{\Psi}(\xi_\Psi)}\xi_k]) = I(\xi),
\end{equation}
where $\mathbb{F}$ is the Fourier transform operator, all $\xi$ are standard normal distributed, $P_{\Psi}(\xi_{\Psi})$ is the spatial correlation power spectrum of log-sky $\psi$, and $I(\xi)$ is the standardized generative model. \\

In \texttt{resolve}, the power spectrum model $P_{\Psi}$ is modeled non-parametrically. In the image reconstruction, the posterior parameters $\xi_{\Psi}$ modeling the power spectrum are inferred simultaneously with the actual image. More details regarding the Gaussian process prior model in \texttt{resolve} can be found in the methods section of \citep{Arras_2022_M87*}. 

Note that we can mitigate strong biasing since the reconstruction of the correlation kernel is a part of the inference process instead of assuming a fixed correlation kernel or a specific sky prior model. As an example, in \texttt{CLEAN}, the sky brightness distribution is assumed to be a collection of point sources. However, it is not a valid assumption for diffuse emission, and it therefore might create imaging artifacts, such as discontinuous diffuse emission with blobs. In \texttt{resolve}, the correlation structure in the diffuse emission can be learned from the data. As a result, the diffuse emission can be well described by the sky prior model. Furthermore, the Gaussian process prior model with non-parametric correlation kernel can also be used for the inference of other parameters, such as amplitude and phase gain corruptions, in order to infer the temporal correlation structure and to encode smoothness in the prior.

\subsection{Antenna-based gain prior model} \label{Ch3.4}
In this paper, we assume that the residual data corruptions can be approximately represented as antenna-based direction-independent gain corruptions. The measurement equation (Eq. \ref{eq:measurement_eq_short}) can be generalized for polarimetric visibility data with sky brightness distribution matrix including right-hand circular polarization (RCP) and left-hand circular polarization (LCP) antenna-based gain corruptions for antenna pair $i,j$ (\cite{Hamaker_1996}, \cite{Smirnov_2011}):

\begin{align} \label{eq:measurement_equation_with_gains}
   \mathbf{V}_{ij} = G_i(t) \, \biggl( \int_{-\infty}^{\infty}\int_{-\infty}^{\infty} \mathbf{I}(x,y) \, e^{-2\pi i (u_{ij}x+v_{ij}y)} dx \,dy \biggr) \,  G_j^{\dag}(t) + \mathbf{N}_{ij},
\end{align}
where $\mathbf{V}_{ij}$ is the visibility matrix with four complex correlation functions by the right-hand circularly polarized signal R and the left-hand circularly polarized signal L:
\begin{equation}
    \mathbf{V}_{ij} = \begin{pmatrix} \,R_{i}R_{j}^{*} & R_{i}L_{j}^{*}\, \\ \,L_{i}R_{j}^{*} & L_{i}L_{j}^{*}\, \end{pmatrix},
\end{equation}

$\mathbf{I}(x,y)$ is the sky brightness distribution matrix of the four Stokes parameters (namely, $I$, $Q$, $U$, and $V$):
\begin{equation}
    \mathbf{I} =\begin{pmatrix} I+V & Q+iU \\ Q-iU & I-V \end{pmatrix},
\end{equation}

$\mathbf{N}_{ij}$ is the additive Gaussian noise matrix, and $G_i(t)$ is the antenna-based gain corruption matrix:
\begin{equation}
    G_i(t) =\begin{pmatrix} g^{R}_i (t) & 0 \\ 0 & g^{L}_i (t) \end{pmatrix}.
\end{equation}

We model the complex gain $g(t)$ via the Gaussian process prior model described in the previous section. For instance, the $i$th antenna RCP gain $g_{i}^{R}(t)$ can be represented by two Gaussian process priors $\lambda$ and $\phi$: 
\begin{equation} \label{eq:gain_prior}
    g^{R}_i (t) = \text{exp}(\lambda^{R}_i (t) + i \phi^{R}_i (t)),
\end{equation}
where $\lambda$ is the log amplitude gain, and $\phi$ is the phase gain. \\

The Gaussian process priors $\lambda$ and $\phi$ are generated from multivariate Gaussian distributions with covariance matrices $\Lambda$ and $\Phi$: 
\begin{equation}
    \lambda \curvearrowleft \mathscr{G}(\lambda,\Lambda) \, ,
    \phi \curvearrowleft \mathscr{G} (\phi,\Phi) \, .
\end{equation}

The temporal correlation kernels $\Lambda$ and $\Phi$ are inferred from the data in the same way as the inference of spatial correlation of the log-sky $\psi$ (see Section \ref{Ch3.3}). The gain prior $g$ is represented in the form of a standardized generative model
\begin{align}
    g(\xi) = \text{exp}\biggl(\;\mathbb{F}\Bigl[ \sqrt{P_{\lambda}(\xi_{\Lambda})\vphantom{P_{\phi}(\xi_{\Phi})}}\xi_{k^{'}} + i \sqrt{P_{\phi}(\xi_{\Phi})}\xi_{k^{''}} \Bigr]\biggr),
\end{align}
where $\xi = (\xi_{\Lambda}, \xi_{k^{'}}, \xi_{\Phi}, \xi_{k^{''}})$ are standard normal distributed random variables, $P_{\lambda}(\xi_{\Lambda})$ is the temporal correlation power spectrum for log amplitude gain $\lambda$, and $P_{\phi}(\xi_{\Phi})$ is the temporal correlation power spectrum for phase gain $\phi$. \\

As we discussed before, in \texttt{resolve}, power spectra are modeled in a non-parametric fashion. Therefore, the temporal correlation structure of the amplitude and phase gains is determined automatically from the data. In \texttt{CLEAN} self-calibration, the solution intervals of the amplitude and phase gain solutions characterizes the temporal correlation. However, a fixed solution interval is chosen by the user without objective criteria; it might induce biases and create imaging artifacts from the noise in the data \citep{MartiVidal_2008, Popkov_2021}. This issue can be mitigated in Bayesian self-calibration by inferring the temporal correlation kernels for amplitude and phase gains from the data.

In this paper, only the total intensity image is  reconstructed. Therefore, non-diagonal terms in the visibility, which are related to linear polarization can be ignored. The visibility matrix is approximated as

\begin{equation} \label{eq:visibility matrix}
    \mathbf{V}_{ij} \approx \begin{pmatrix} R_{i}R_{j}^{*} & 0 \\ 0 & L_{i}L_{j}^{*} \end{pmatrix}.
\end{equation}

Similarly, Stokes Q, U, and V can be ignored in the sky brightness distribution:

\begin{equation}
    \mathbf{I}(\vv{x}) \approx \begin{pmatrix} I(\vv{x}) & 0 \\ 0 & I(\vv{x}) \end{pmatrix}.
\end{equation}

Therefore, the visibility matrix model is 

\begin{equation}\label{eq:model_visibility_matrix}
    \tilde{\mathbf{V}}_{ij}(t) = \begin{pmatrix} g^{R}_i (t) & 0 \\ 0 & g^{L}_i (t) \end{pmatrix} \, B(t) \, \left[ \mathbb{F}  \begin{pmatrix} I(\vv{x}) & 0 \\ 0 & I(\vv{x}) \end{pmatrix} \right] \, \begin{pmatrix} g^{R}_j (t) & 0 \\ 0 & g^{L}_j (t) \end{pmatrix}^{^{\dag}},
\end{equation}
where $B(t)$ is the sampling operator (see Eq. \ref{eq:measurement_eq_short}).

The visibility matrix model $\tilde{\mathbf{V}}_{ij}(t)$ can be calculated from the standardized generative sky model $I(\xi)$ and the gain model $g(\xi)$. Note that we aim to fit the model $\tilde{\mathbf{V}}_{ij}(t)$ in Eq. \ref{eq:model_visibility_matrix} containing the RCP and LCP gains and Stokes I image to the visibility matrix data $\mathbf{V}_{ij}$ in Eq. \ref{eq:visibility matrix} directly in a probabilistic setup. As a result, we can perform self-calibration (gain inference) and imaging simultaneously. In the next section, we describe the variational inference algorithm we use to approximate the posterior distribution of the gain and sky parameters given the data.

\subsection{Inference scheme} \label{Ch3.5}

Bayes' theorem allows us to infer the conditional distribution of the model parameters $\xi = (\xi_\Psi, \xi_k, \xi_{\Lambda}, \xi_{k^{'}}, \xi_{\Phi}, \xi_{k^{''}})$, also called posterior distribution, from the observed data. From the posterior distribution of $\xi$, we can obtain the correlated posterior distributions with inferred correlation kernels for the sky emission $I = I(\xi)$ and the gains $G = G(\xi)$. In order to estimate the posterior distribution for the high-dimensional image reconstruction, the MGVI algorithm \citep{Knollmueller_2019} is used. In MGVI, the posterior distribution $\mathcal{P}(\xi|V)$ is approximated as a multivariate Gaussian distribution $\mathscr{G}(\xi-\bar{\xi}, \Xi)$ with the inverse Fisher information metric $\Xi$ as a covariance matrix. 

The posterior distribution is obtained by minimizing the Kullback-Leibler (KL) divergence between the approximate Gaussian distribution and the true posterior distribution: 

\begin{align}
    D_{KL}(\mathscr{G}(\xi-\bar{\xi}, \Xi) || \mathcal{P}(\xi|V)) = \int d\xi \; \mathscr{G}(\xi-\bar{\xi}, \Xi) \; \text{ln} \left( \frac{\mathscr{G}(\xi-\bar{\xi}, \Xi)}{\mathcal{P}(\xi|V)} \right).
\end{align}
The KL divergence measures the expected information gain from the posterior distribution to the approximated Gaussian posterior distribution. By minimizing the KL divergence, we can find the closest Gaussian approximation to the true posterior distribution in the Variational Inference sense. \\

The KL divergence in the MGVI algorithm can be represented by the information Hamiltonians:

\begin{align}
    D_{KL} = \langle \mathcal{H}(\xi | V) \rangle_{\mathscr{G}(\xi -\bar{\xi}, \Xi)} - \langle \mathcal{H}(\xi - \bar{\xi}, \Xi) \rangle_{\mathscr{G}(\xi -\bar{\xi}, \Xi)},
\end{align}
where $\mathcal{H}(\xi | V)$ is the posterior Hamiltonian and $\mathcal{H}(\xi - \bar{\xi}, \Xi)$ is the approximated Gaussian posterior Hamiltonian.

We can express the posterior Hamiltonian in terms of the likelihood and prior Hamiltonians (see Eq. \ref{eq:posterior Hamiltonian}):

\begin{align}
    D_{KL} \cong \langle \mathcal{H}(V|\xi) + \mathcal{H}(\xi) \rangle_{\mathscr{G}(\xi -\bar{\xi}, \Xi)}.
\end{align}

The evidence Hamiltonian $\mathcal{H}(V)$ can be ignored because it is independent of the hyperparameters for the prior model. Note that the KL divergence contains the likelihood Hamiltonian, which is equivalent to the data fidelity term (see Section \ref{Ch2.3}), ensuring the consistency between the final image and the data. 

The MGVI algorithm infers samples $\xi$ of the normal distributed approximate posterior distribution. The posterior mean and standard deviation of the sky $I$ and the gain $G$ can be calculated from the samples of the normal distributed posterior distribution and those sky and gain posterior are consistent with the data. Note that MGVI allows us to capture posterior correlations between parameters $\xi$, although multimodality cannot be described, and the uncertainty values tend to be underestimated since it provides a local approximation of the posterior with a Gaussian \citep{Frank2021_paper2}. 
In conclusion, high-dimensional image reconstruction can be performed by the MGVI algorithm by striking a balance between statistical integrity and computational efficiency. A detailed discussion is provided in \cite{Knollmueller_2019}.

\begin{figure*}[h]
    \includegraphics[width=17cm] {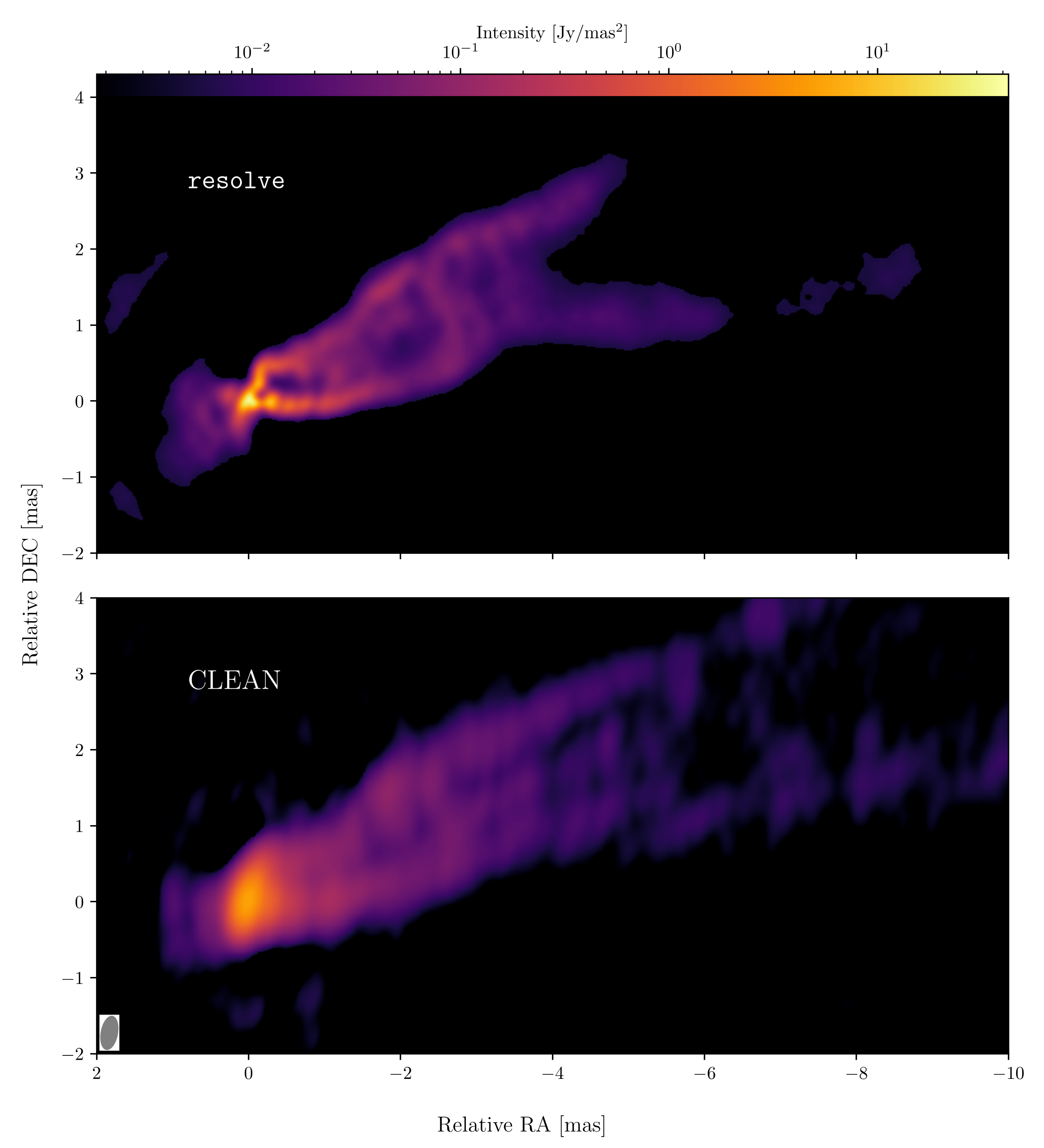}

    \caption{ M87: the posterior mean image by Bayesian self-calibration (top) and the self-calibrated \texttt{CLEAN} image (bottom) reconstructed from the same a-priori calibrated visibility data of VLBA observations at 43\,GHz. In the figure, intensities higher than 3$\sigma_{\textrm{rms}}$ (root mean square) of corresponding reconstruction were shown. The image obtained by \texttt{resolve} has $I_{\textrm{max}} = 35$\,Jy\,mas$^{-2}$ with the noise level of $\sigma_{\textrm{rms}} = 2$\,mJy\,mas$^{-2}$ which is calculated from the top left empty region in the posterior mean image. The \texttt{CLEAN} image restoring beam shown in the lower-left corner is 0.5$\times$0.2\,mas, Position angle (P.A.)~$= -11^{\circ}$. The \texttt{CLEAN} image has $I_{\textrm{max}} = 6$\,Jy\,mas$^{-2}$ with the noise level of $\sigma_{\textrm{rms}} = 0.6$\,mJy\,mas$^{-2}$.}
    \label{fig:M87_real_resolve_CLEAN}
\end{figure*}

\begin{figure}[h]
    \includegraphics[width=\columnwidth]
    {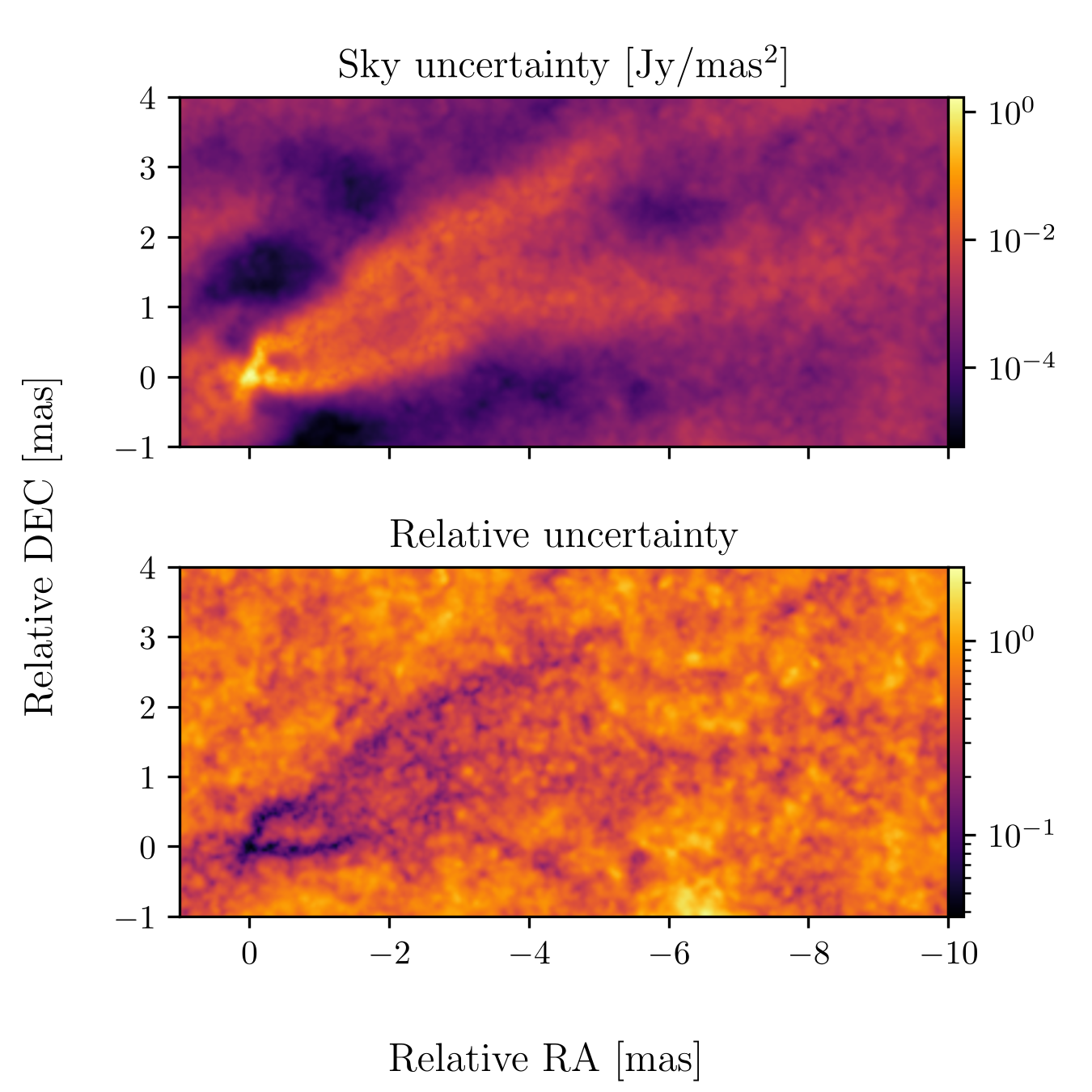}

    \caption{ M87: sky posterior pixel-wise standard deviation (top) and relative uncertainty, which is the sky posterior standard deviation normalized by the posterior mean (bottom) by \texttt{resolve} reconstruction from the top panel of \autoref{fig:M87_real_resolve_CLEAN}.}
    \label{fig:M87_sky_relative_uncertainty}
    
\end{figure}

\section{Image reconstruction: real data} \label{Chap_Image_real}
\subsection{M87 VLBA 7mm data}

To validate the Bayesian self-calibration and imaging method on the real data, we applied it to a Very Long Baseline Array (VLBA) observation of M87 in 2013 at 43\,GHz (7mm). The project code from the NRAO archive is BW098. A detailed description of the project can be found in \cite{Walker_2018}. The M87 VLBA data are chosen because it is a full track data set and the source has complex structures and a high-dynamic range. Therefore, we are able to test the performance of the Bayesian self-calibration algorithm in order to improve the image fidelity. We used correlated raw data from the NRAO archive and complete an a-priori and signal stabilization calibration using the rPICARD pipeline \citep{Janssen_2019_rPICARD}. All data and the results are archived in \url{https://zenodo.org/uploads/10190800}. 

\subsection{Reconstruction by \texttt{CLEAN}}
\label{chap:M87_CLEAN}

We used \texttt{DIFMAP} \citep{1997ASPC..125...77S} software to reconstruct a Stokes I image using hybrid mapping in combination with super-uniform, uniform and natural weighting with a pixel size of 0.03 milliarcsecond (mas) in \autoref{fig:M87_real_resolve_CLEAN}. Gain solutions from \texttt{CLEAN} self-calibration are shown in \autoref{fig:CLEAN_selfcal_solutions_amp} and \autoref{fig:CLEAN_selfcal_solutions_phase}. The resulting image corresponds to the result obtained by \cite{Walker_2018}. To facilitate a direct comparison with the results obtained by the \texttt{resolve} software, we convert the standard \texttt{CLEAN} image intensity unit, originally in Jy/beam, to Jy\,mas$^{-2}$. This conversion involved dividing the \texttt{CLEAN} output in Jy/beam by the beam area, which is calculated as $\frac{\pi}{4 \log{2}} \cdot$ BMAJ $\cdot$ BMIN, where BMAJ and BMIN represent the major and minor axes of the beam, respectively.
The noise level $\sigma_{rms} = 67$\,$\mu$Jy/beam was calculated 1 arcsecond from the phase center to minimize the influence of the bright VLBI core. Since \texttt{CLEAN} uses a conventional beam size, which does not always coincide with the effective resolution of the image, we display the \texttt{CLEAN} image convolved with 0.18 mas circular beam to show the resolved high signal-to-noise ratio (SNR) region. This ``over-resolved'' image is presented at the bottom of \autoref{fig:M87_real_resolve_CLEAN_appendix}. The image helps to compare \texttt{CLEAN} result of high flux density region with the \texttt{resolve} result.

\begin{figure*}
    \centering
    \includegraphics[width=1\linewidth]{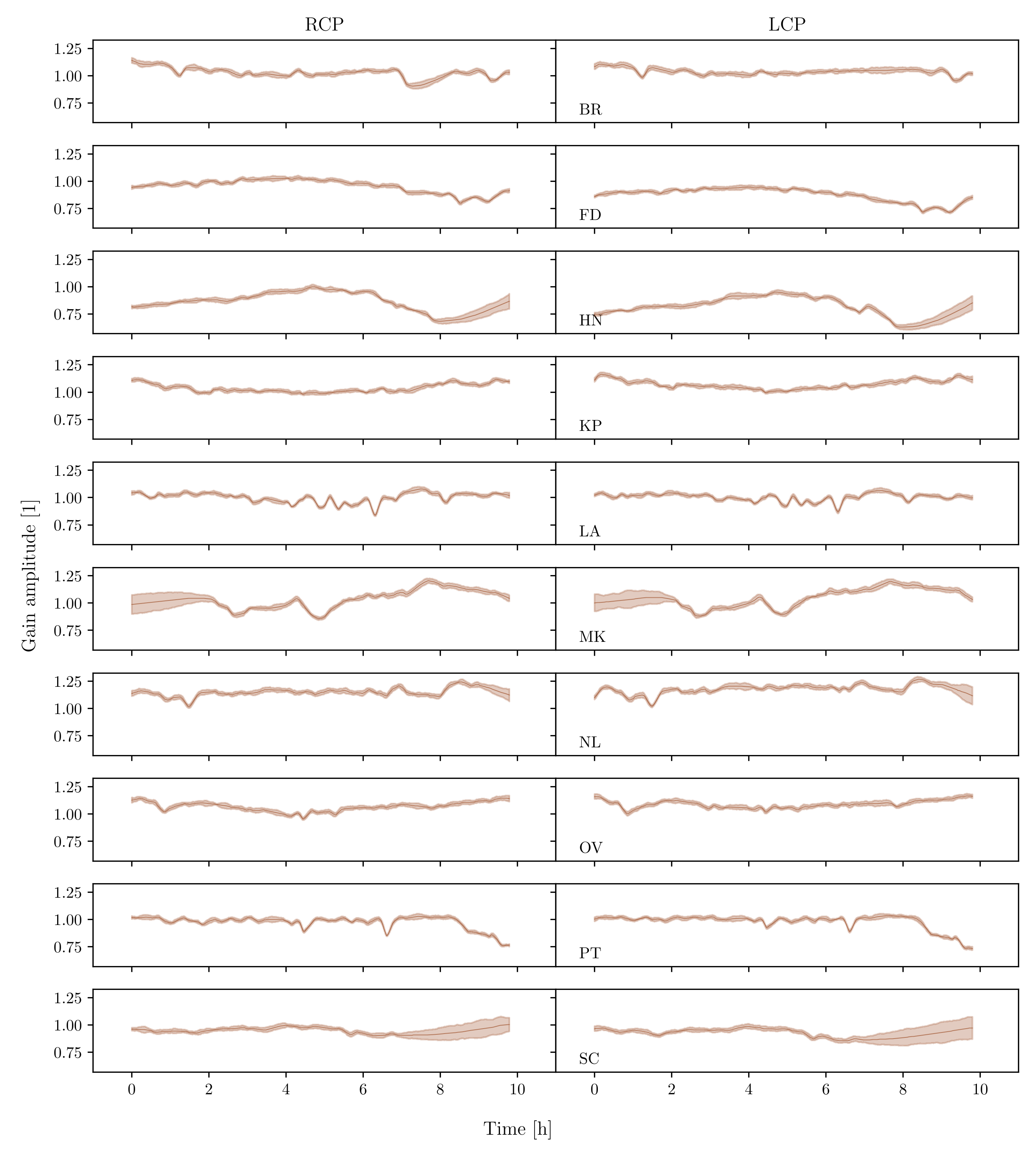}
    \caption{M87: Reconstructed posterior amplitude gains. The gain as a function of time is illustrated as a thin line with a semi-transparent standard deviation. The left and right columns of the figure show gains from the right (RCP) and left (LCP) circular polarizations correspondingly. Each row represents an individual antenna, whose abbreviated name is indicated in the bottom left corner of each LCP plot.  }
    \label{fig:amp_gains_realdata}
\end{figure*}

\begin{figure*}
    \centering
    \includegraphics[width=1\linewidth]{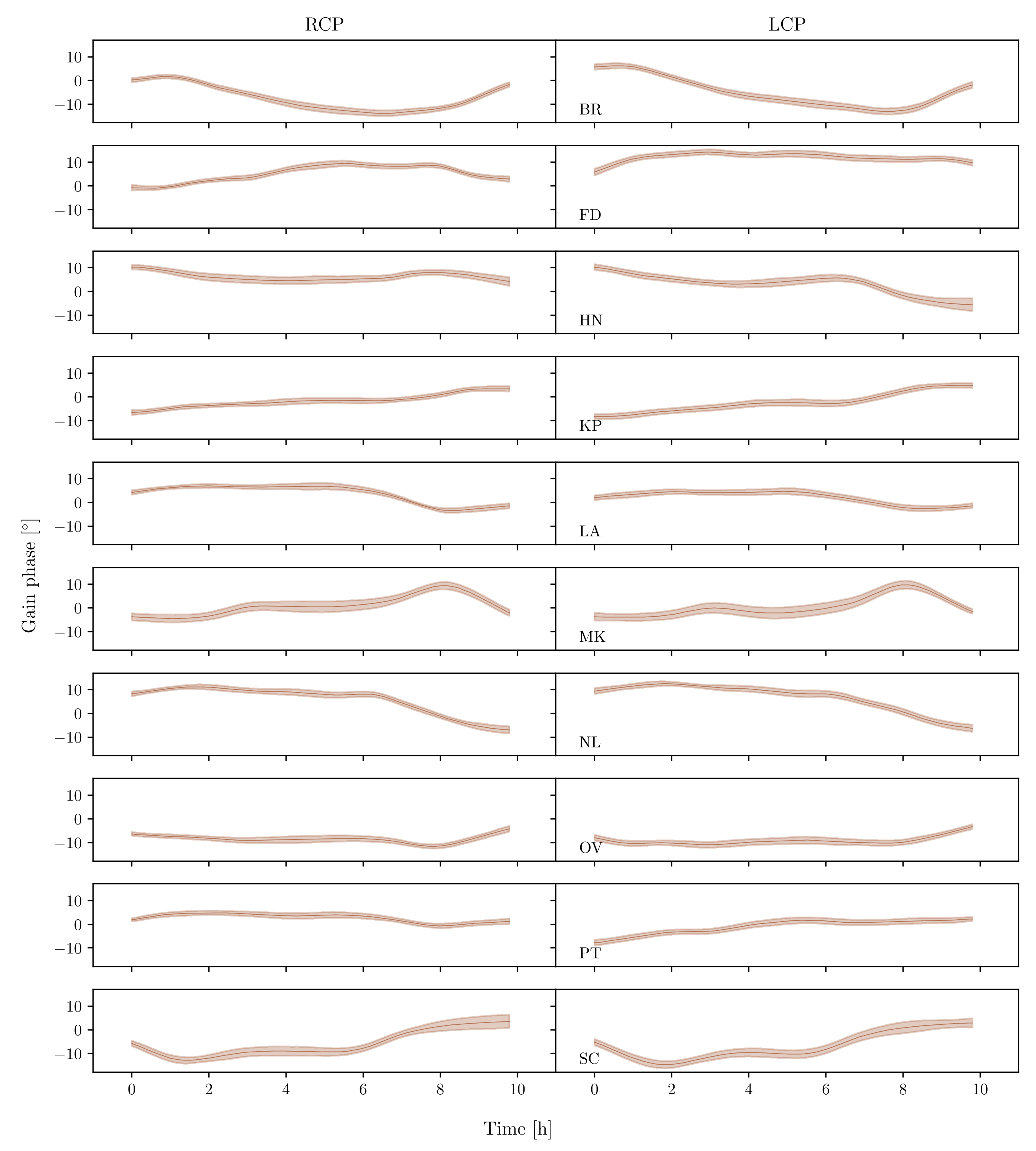}
    \caption{M87: Reconstructed posterior phase gains. The gain as a function of time is illustrated as a thin line with a semi-transparent standard deviation. The left and right columns of the figure show gains from the right (RCP) and left (LCP) circular polarizations correspondingly. Each row represents an individual antenna, whose abbreviated name is indicated in the bottom left corner of each LCP plot. }
    \label{fig:phase_gains_realdata}
\end{figure*}

\subsection{Reconstruction by \texttt{resolve}} \label{Section_resolve_real_data}

In \autoref{fig:M87_real_resolve_CLEAN}, the \texttt{resolve} posterior mean sky image is displayed. The \texttt{resolve} image is reconstructed with a spatial domain of $2048 \times 1024$ pixels and a field of view of $30 \; \text{mas} \times 15 \; \text{mas}$. The visibilities were time-averaged with a time interval of 10 seconds and frequency-averaged over 8 intermediate frequencies. For the \texttt{resolve} reconstruction, the number of data points is $\approx 1.5 \times 10^5$ (RR and LL components) and we used $20$ samples for the posterior estimation. As a result, the reduced $\chi^2$ value of the final result is $1.2$, which ensures the consistency between the reconstruction results and the data. The wall-clock time for \texttt{resolve} run is 2h 40min.

In addition to the posterior mean image, the uncertainty of the image can be estimated from the sky posterior samples in \texttt{resolve}. In \autoref{fig:M87_sky_relative_uncertainty}, the shown relative uncertainty is the pixel-wise posterior standard deviation normalized by the sky posterior mean. Note that low relative uncertainty values are estimated in the core and limb-brightened jet regions. 

During a conventional self-calibration for mm-VLBI data, low SNR data points and outliers are often flagged. However, it is often challenging to identify bad data points from data points inconsistent with the model image in self-calibration, and we therefore rely on expert's experience. Data flagging without objective criteria hinders the reproducibility of results and biased prior knowledge might be imposed in image reconstruction owing to the manual data flagging. For the \texttt{resolve} self-calibration and imaging, data were only flagged by the rPICARD pipeline during pre-calibration and not flagged manually during Bayesian self-calibration and imaging in order to minimize human interaction. In Bayesian imaging, high uncertainty in low SNR data points is taken into account naturally during image reconstruction. Therefore, high-fidelity images can be reconstructed without manual data flagging in Bayesian framework, if adequate information of the noise level is available. 3 percent of the visibility amplitude was added in the visibility noise as a systematic error budget, e.g., related to the uncalibrated polarimetric leakage \citep{EHT_2019_paper3_data}. 

The hyperparameter setup for log-sky $\psi$, log amplitude gain $\lambda$, and phase gain $\phi$ priors is described in Appendix \ref{Appendix_hyperparameter}. Four temporal correlation kernels (amplitude gain and the phase gain for RCP and LCP mode respectively) are inferred under the assumption that antennas from homogeneous array have similar amplitude and phase gain correlation structures per polarization mode. We chose a resolution of 7054 pixels for the temporal domain of $\lambda$ and $\phi$ and the time interval per each pixel is 10 seconds. Note that the time interval in gains is not directly related to the solution interval in \texttt{CLEAN} self-calibration method since the correlation structure is learned from the data automatically instead of using a fixed solution interval in \texttt{resolve}. We infer gain terms with two times of the observation time interval and crop only the first half since the Fast Fourier Transforms (FFT), which assumes periodicity, is utilized. The number of pixels for gain corrections is $7054 \times 10 \times 2 \times 2  \approx 2.8 \times 10^{5}$ ($2 \times$ observation interval $\times$ number of VLBI antennas $\times$ (phase, amplitude) $\times$ (RCP, LCP)). As a result, the degrees of freedom (DOF) for the inference are $2048 \times 1024 + 7054 \times 40 + \text{power spectrum DOF} \approx 2.9 \times 10^{6}$. 

The posterior mean and standard deviation of amplitude and phase gains per each antenna and polarization mode are shown in \autoref{fig:amp_gains_realdata} and \autoref{fig:phase_gains_realdata}. The uncertainty of the amplitude and phase gain solutions is estimated by the posterior samples. Therefore, we can quantify the reliability of the amplitude and phase gain solutions by Bayesian self-calibration method. In order to validate the result, self-calibration solutions from \texttt{CLEAN} algorithm and an RML method by \texttt{ehtim} software \citep{Chael_2018_closure} are shown in Appendix \ref{Appendix_CLEAN_selfcal} and Appendix \ref{Appendix_ehtim_selfcal}. Note that \texttt{resolve}, \texttt{CLEAN}, and \texttt{ehtim} self-calibration solutions are comparable qualitatively, although different imaging methods and minimization schemes are employed. In \autoref{fig:CLEAN_selfcal_solutions_amp} and \autoref{fig:CLEAN_selfcal_solutions_phase}, there is no gain solution when we do not have data (BR: 7-8h, HN: >8h, MK: <2h, NL: >9h, and SC: >7h). Gain solutions in those time intervals cannot be constrained by the data, standard deviations of the gain amplitude in \autoref{fig:amp_gains_realdata} are therefore increased. 

\begin{figure}[t]
    \includegraphics[width=
    9cm] {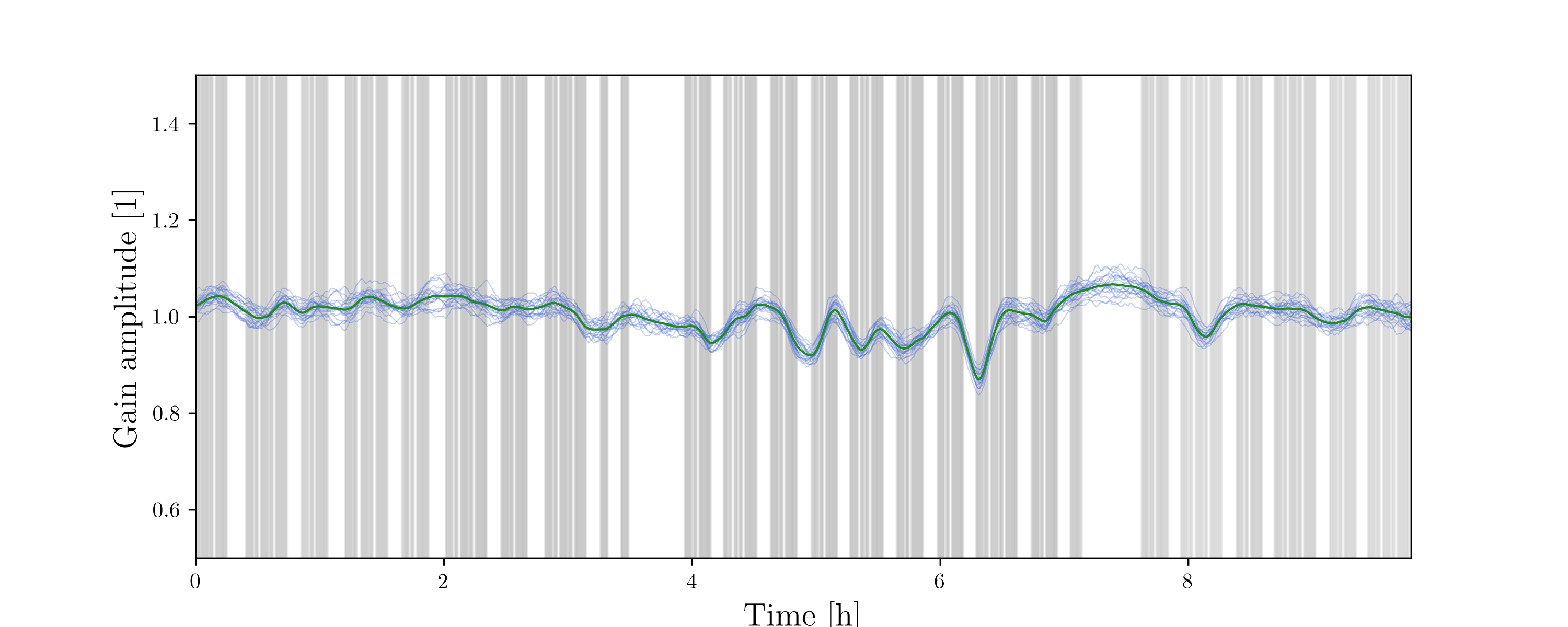}
    \includegraphics[width=
    9cm]{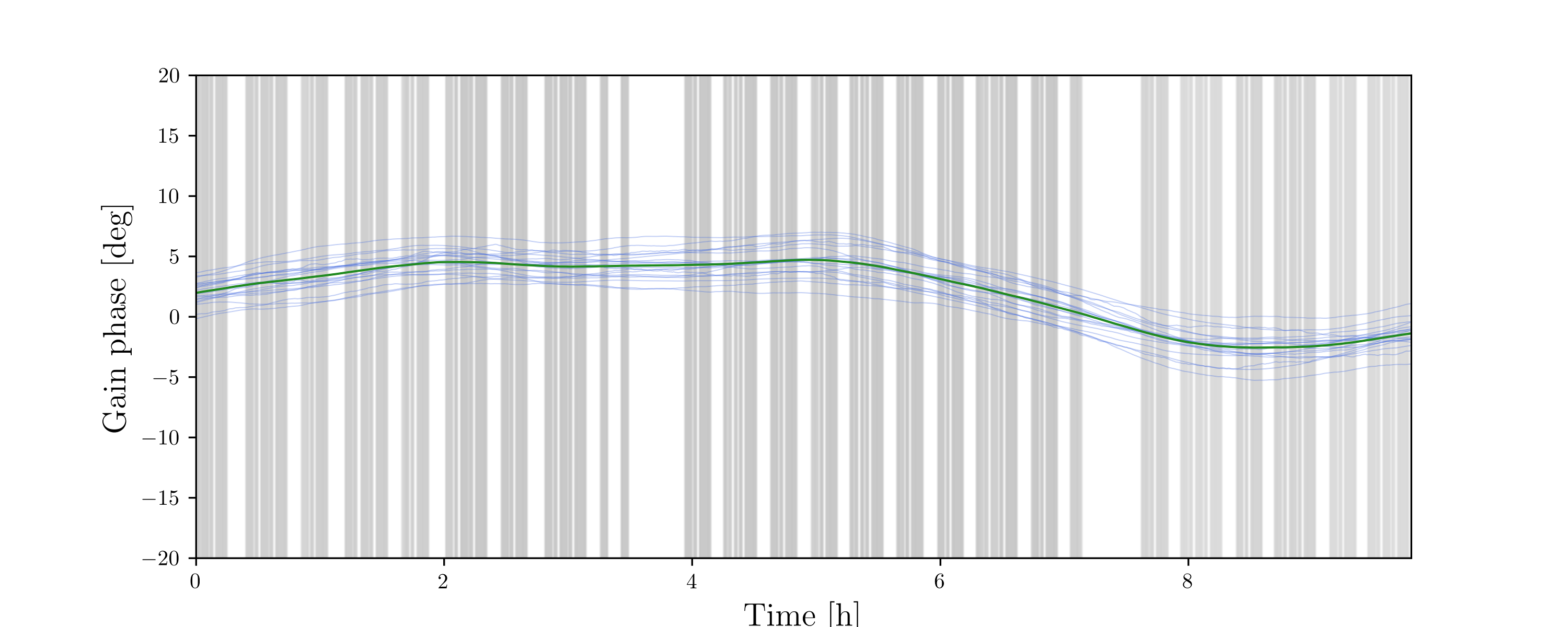}
    
    \caption{ M87: LCP amplitude (top) and phase (bottom) gain posterior mean and posterior samples for LA antenna by the Bayesian self-calibration. The thick green solid line represents the posterior mean value, while the thin blue solid lines depict individual samples. Grey vertical shades indicate areas with available data points.}
    
    \label{fig:real_amp_phase_gain_LA_LCP}
\end{figure}

\begin{figure}[t]
    \includegraphics[width=
    9cm] {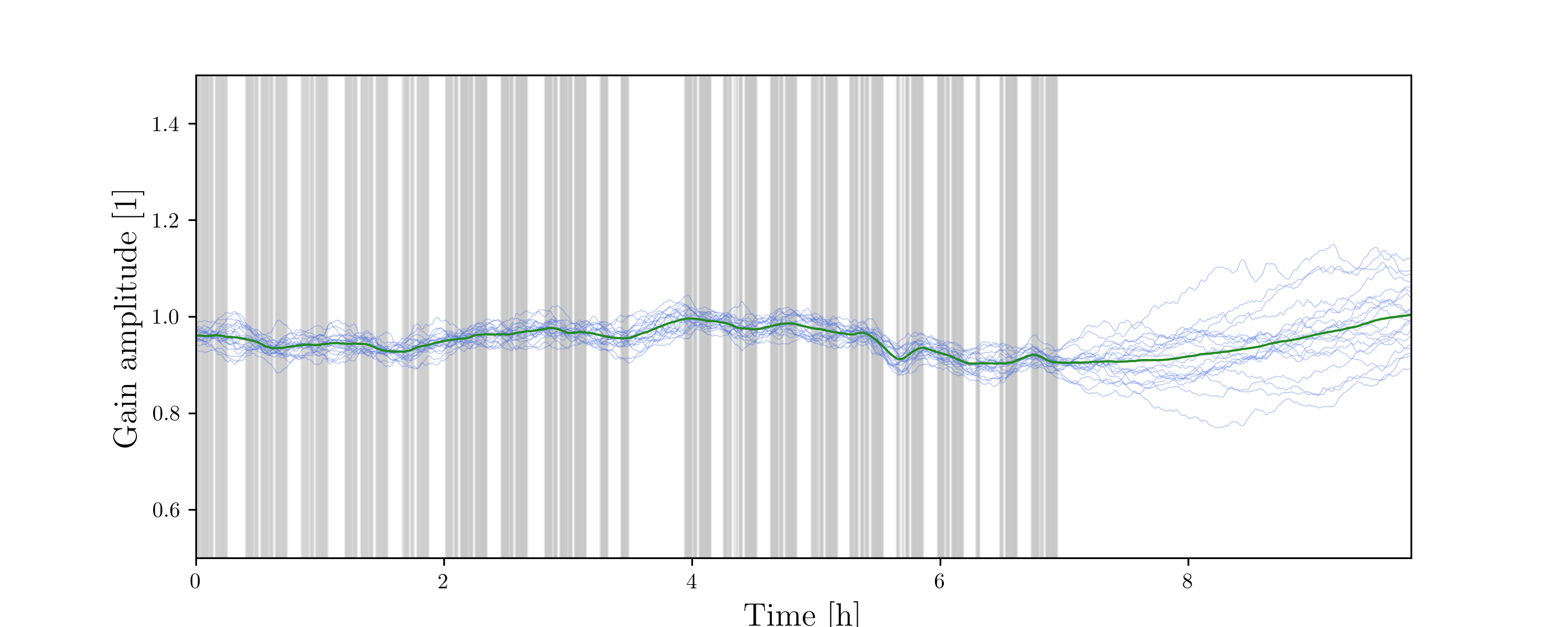}
    \includegraphics[width=
    9cm] {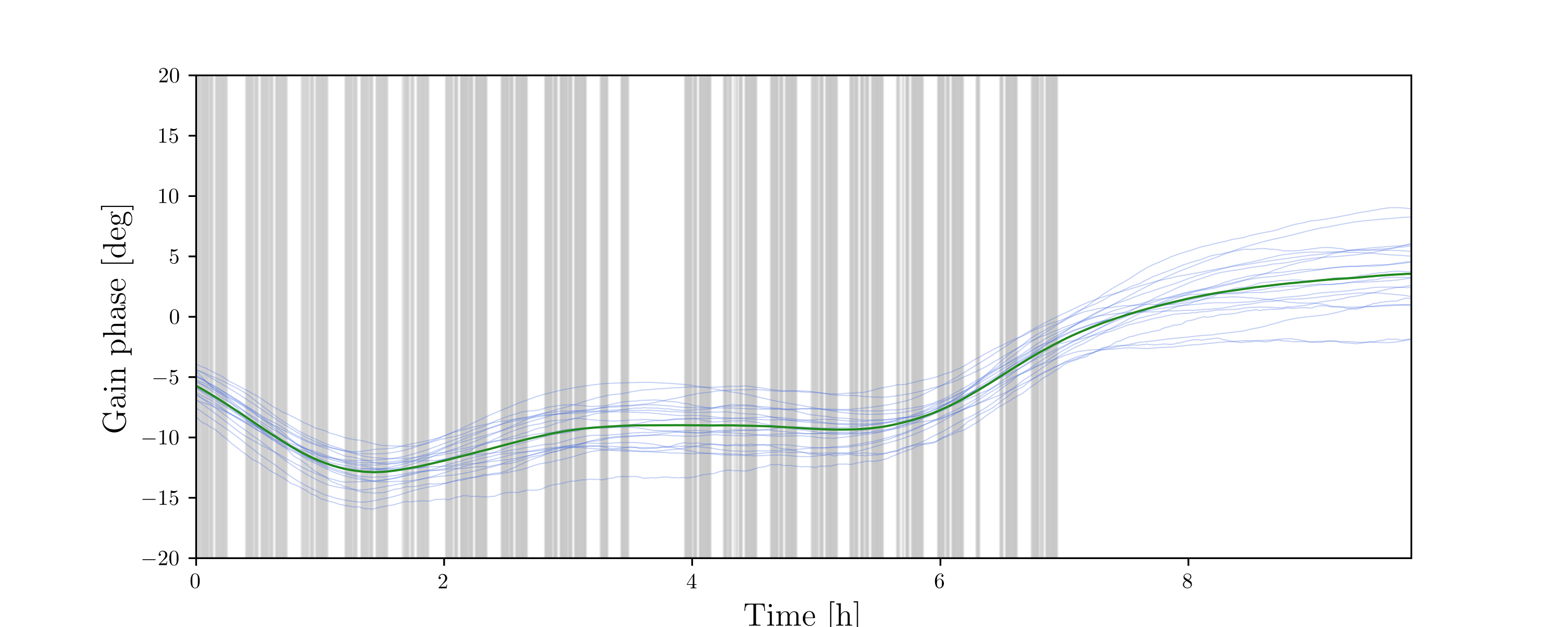}
    
    \caption{ M87: RCP amplitude (top) and phase (bottom) gain posterior mean and posterior samples for SC antenna by the Bayesian self-calibration. The thick green solid line represents the mean value, while the thin blue solid lines depict individual samples. Grey vertical shades indicate areas with available data points.}
    
    \label{fig:real_amp_phase_gain_SC_RCP}
\end{figure}

\autoref{fig:real_amp_phase_gain_LA_LCP} and \autoref{fig:real_amp_phase_gain_SC_RCP} show LCP and RCP gain solutions for Los Alamos (LA) and Saint Croix (SC) antenna, respectively. The amplitude and phase gain solutions are assumed to be smooth and to not vary much. The amplitude gains vary up to $20$ percent and the phase gains deviate up to $30$ degrees. 
Considering temporal correlation in amplitude gain and phase gain solutions for this data is reasonable because the amplitude and phase coherence time of the VLBI array at 43\,GHz is higher than 10 seconds.

\autoref{fig:M87_real_resolve_CLEAN} shows two image reconstructions by the \texttt{resolve} and \texttt{CLEAN}. In the \texttt{resolve} image, better resolution is achived in the core and limb-brightened regions and the counter jet is more clear than \texttt{CLEAN} image. In \autoref{fig:M87_real_resolve_CLEAN_appendix}, even the over-resolve \texttt{CLEAN} image cannot achieve the high resolution comparable with the \texttt{resolve} image. The extended jet emission looks more consistent in \texttt{resolve} image since the correlation structure between pixels are inferred by the Gaussian process prior with correlation kernel in \texttt{resolve}. Furthermore, the \texttt{resolve} image does not have negative flux because the positivity of the flux is enforced in the log-normal sky prior. 

For the residual gain inference by \texttt{resolve}, the correlation structure between the gain solutions is inferred by the data. In other words, we do not need to choose solution interval of the gains manually but rather amplitude and phase gain solutions can be estimated by the Gaussian process prior model and more sophisticated inference scheme than the conventional \texttt{CLEAN} self-calibration. The consistency between data and the image with gain solutions is ensured since we fit the model to the data directly. 

In order to obtain high-fidelity image, it is crucial to distinguish the uncertainty of gains and image from the VLBI data. From the perspective of statistical integrity, self-calibration and imaging should be performed simultaneously. Conventional iterative self-calibration estimates gain as a point and often flag outliers manually. This can impose a strong effective prior and thereby hinder proper accounting of the uncertainty information in the data. Furthermore, a variety of different images result from different ways \texttt{CLEAN} boxes are placed or different solution interval are chosen for amplitude and phase gains. Bayesian self-calibration and imaging can reduce such biases and provides reasonable uncertainty estimation of the gain solutions and image. This example with VLBA data set demonstrates that \texttt{resolve} is not only able to reconstruct images from real VLBI data but perform robust joint self-calibration and image reconstruction from sparse VLBI data set without iterative manual procedures.

\begin{figure}[t]
    \includegraphics[width=\linewidth] {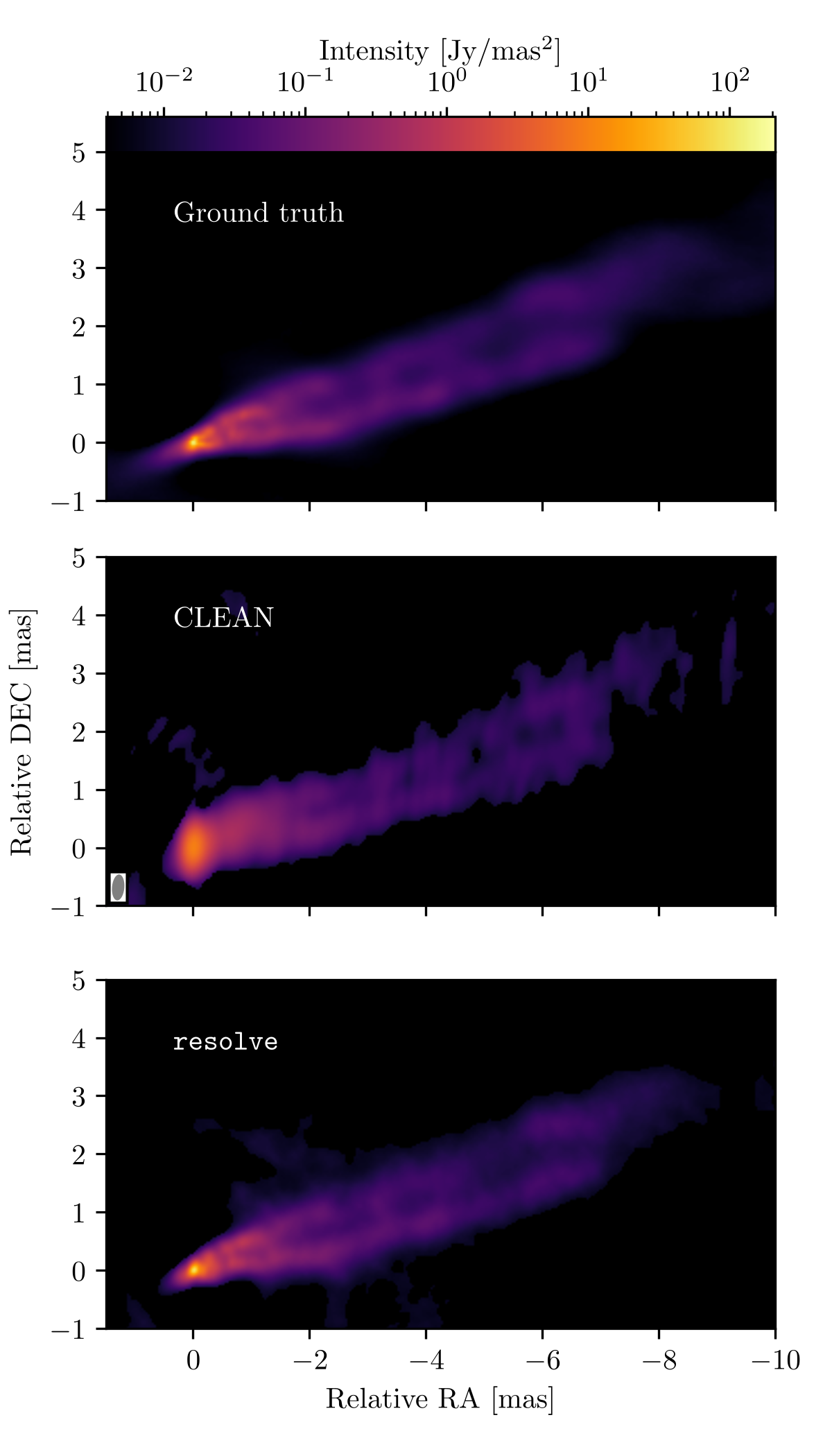}{\caption{ Synthetic data: ground truth (top) and reconstructed images obtained using \texttt{CLEAN} (middle) and \texttt{resolve} (bottom) self-calibration. The restoring \texttt{CLEAN} beam illustrated in the bottom left corner of the plot is $0.5\times0.2$\,mas, P.A.\,$= -5^{\circ}$. All images in the figure were masked at 3$\sigma_{\textrm{rms}}$ level of a corresponding image. The unified color bar on the top of the figure shows an intensity range of the ground truth (GT) image, where maximum intensity is $I_{\textrm{max}}^{\textrm{GT}} = 209$\,Jy\,mas$^{-2}$, the rms noise level is $\sigma_{\textrm{rms}}^{\textrm{GT}} = 1$\,mJy\,mas$^{-2}$. The noise level of the reconstructed images are $\sigma_{\textrm{rms}}^{\textrm{\texttt{CLEAN}}} = 3$\,mJy\,mas$^{-2}$, $\sigma_{\textrm{rms}}^{\texttt{resolve}} = 2$\,mJy\,mas$^{-2}$. Maximum intensity values are $I_{\textrm{max}}^{\textrm{\texttt{CLEAN}}} = 10$\,Jy\,mas$^{-2}$, $I_{\textrm{max}}^{\texttt{resolve}} = 111$\,Jy\,mas$^{-2}$ correspondingly.}
    \label{fig:synthetic_sky}
    }
\end{figure}

\begin{figure}[t]
    \includegraphics[width=
    9cm] {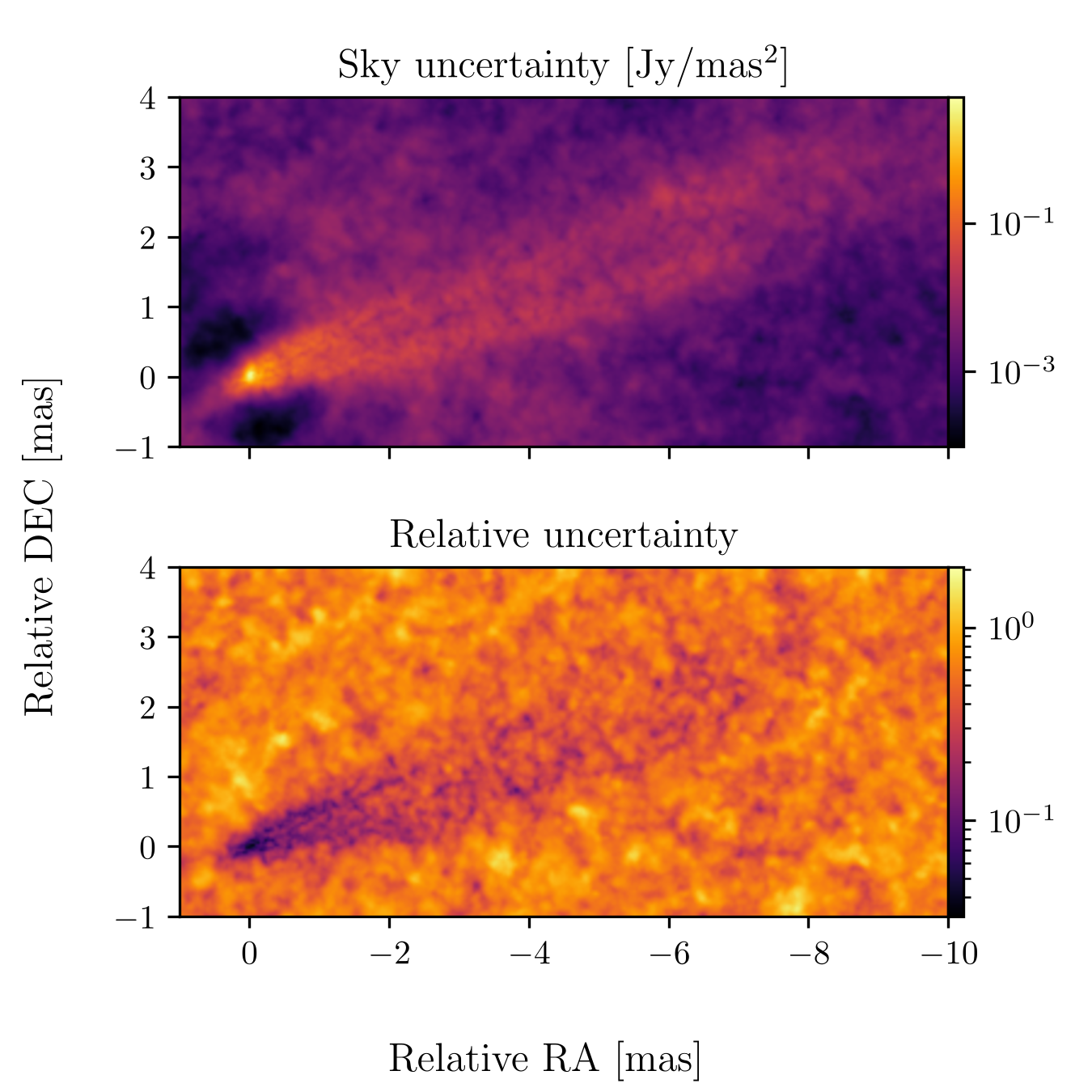}{\caption{ Synthetic data: sky posterior pixel-wise standard deviation (top) and relative uncertainty, which is the sky posterior standard deviation normalized by the posterior mean (bottom) by \texttt{resolve} reconstruction from the bottom panel of \autoref{fig:synthetic_sky}. \label{fig:synthetic_sky_relative_uncertainty}}}
\end{figure}

\begin{figure*}[h]
    \centering
    \includegraphics[width=1\linewidth]{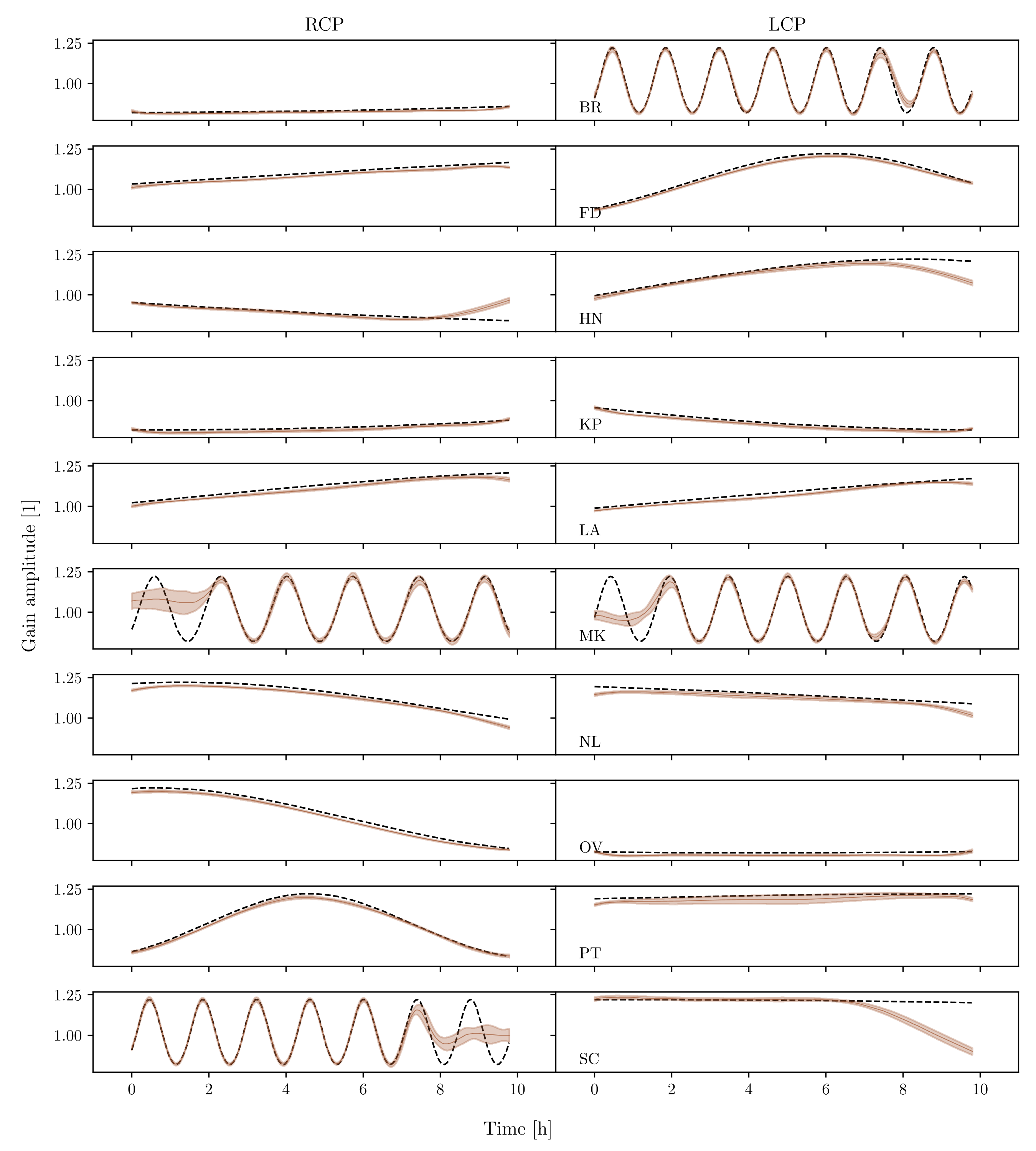}
    \caption{Ground truth and posterior amplitude gains from the synthetic data. The left and right columns of the figure show amplitude gains from the right (RCP) and left (LCP) circular polarizations correspondingly. Each row represents an individual antenna, whose abbreviated name is indicated in the bottom left corner of each LCP plot. The black dashed lines denote the ground truth amplitude gain corruptions, and posterior mean amplitude gains and their standard deviations by \texttt{resolve} are presented as a brown solid lines with shades. The visible discrepancies between the ground truth and reconstructed amplitude gains in the several baselines at specific time intervals (BR 7-8h, HN >8h, MK <2h, NL >9h, and SC >7h) are due to data gaps there (see Fig. \ref{fig:synthetic_amp_phase_gain_BR_RCP}- ]\ref{fig:synthetic_amp_phase_gain_SC_LCP}).}
    \label{fig:synthetic_gains_amp}
\end{figure*}

\begin{figure*}[h]
    \centering
    \includegraphics[width=1\linewidth]{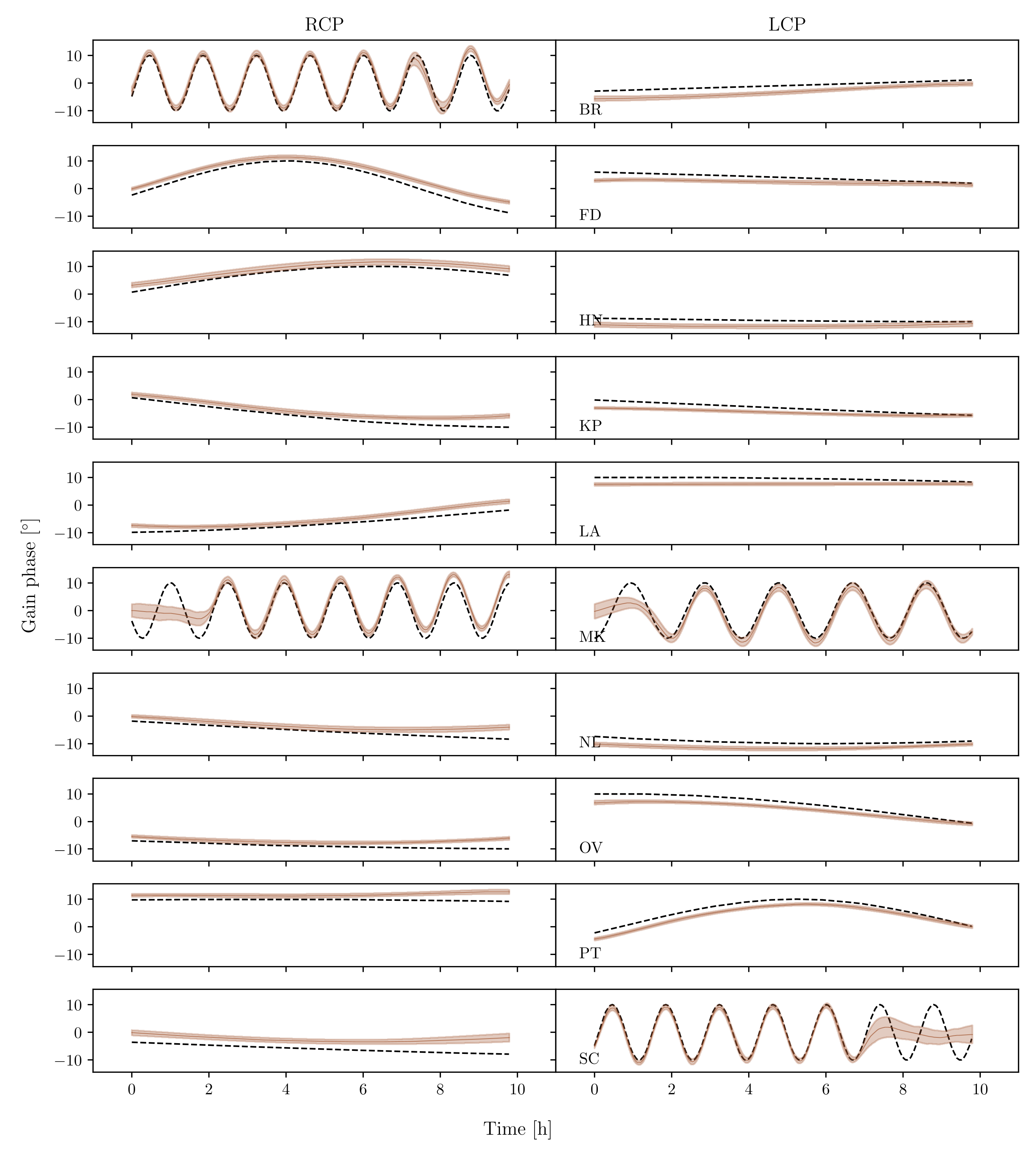}
    \caption{Ground truth and posterior phase gains from the synthetic data. The left and right columns of the figure show phase gains from the right (RCP) and left (LCP) circular polarizations correspondingly. Each row represents an individual antenna, whose abbreviated name is indicated in the bottom left corner of each LCP plot. The black dashed lines denote the ground truth phase gain corruptions, and posterior mean phase gains and their standard deviations by \texttt{resolve} are presented as a brown solid lines with shades. The visible discrepancies between the ground truth and reconstructed phase gains in the several baselines at specific time intervals (BR 7-8h, HN >8h, MK <2h, NL >9h, and SC >7h) are due to data gaps there (see Fig. \ref{fig:synthetic_amp_phase_gain_BR_RCP}- ]\ref{fig:synthetic_amp_phase_gain_SC_LCP}).}
    \label{fig:synthetic_gains_phase}
\end{figure*}

\section{Image reconstruction: synthetic data} 
\label{Chap_Image_synthetic}

\subsection{Synthetic data}
In order to validate the method, it is crucial to test the Bayesian self-calibration algorithm by applying it to synthetic visibility data with a known ground truth image. The synthetic data test is conducted in a semi-blind way. The metadata, including $uv$-coverage, frequency, and the error associated with each visibility point, was imported from the real observation data discussed in Section \ref{Chap_Image_real}. For the ground truth image, we chose the 15\,GHz intensity image obtained by \texttt{resolve} using full-track VLBA May 2009 observations used in \citet{nikonov2023properties}. The ground truth image shows a great variety of scales from small and bright filaments to extended faint structures. To align the ground truth image with the real 43\,GHz data, we scaled it down linearly to ensure that the jet lengths roughly correspond to each other. The resulting ground truth image is displayed in \autoref{fig:synthetic_sky}. The $uv$-data was created from this image using eht-imaging \citep{ehtim}.
To simulate the atmospheric, pointing and other antenna-based errors, we corrupted the data with periodic time-dependent complex antenna gains using CASA software. The periods of the gain functions for an individual antenna are defined between 1 and 12 hours to mimic inhomogeneous statistics, which is commonly found in data from inhomogeneous arrays. The degree of gain variation was chosen based on the real observations, where one can observe the change of amplitude gain approximately 20\% and for the phase around 10$^{\circ}$. The final gain corruption for each antenna and polarisation mode is presented as black dashed lines in \autoref{fig:synthetic_gains_amp} and \autoref{fig:synthetic_gains_phase}.

\subsection{Reconstruction by \texttt{CLEAN} and \texttt{resolve}} \label{Section_CLEAN_resolve_synthetic_data}

The \texttt{CLEAN} and \texttt{resolve} image with the synthetic data are depicted in \autoref{fig:synthetic_sky}. The iterative self-calibration and image reconstruction setup by \texttt{CLEAN} are similar to the one discussed in Section \ref{chap:M87_CLEAN}. For Bayesian self-calibration and imaging by \texttt{resolve}, the hyper parameters of the log-sky prior $\psi$, log-amplitude gain prior $\lambda$, and phase gain prior $\phi$ for the synthetic data are the same as the reconstruction for the real data (in Tables \ref{table:hyper_logsky} and \ref{table:hyper_gains}). However, for all the antenna and their polarization modes, individual temporal correlation kernels for the amplitude and phase gains were employed for the synthetic data in order to infer gain corruptions with different correlation structures. 

We choose a resolution of $2048 \times 1024$ pixels for the Stokes I image with a field of view $30$ mas $\times$ 15 mas. The reduced $\chi^2$ of the \texttt{resolve} reconstruction with the synthetic data is $0.6$. 

\autoref{fig:synthetic_sky} shows a comparison of images of ground truth, \texttt{CLEAN} reconstruction, and \texttt{resolve} reconstruction. The \texttt{CLEAN} algorithm tends to reconstruct blobby extended structure since \texttt{CLEAN} reconstructs a collection of delta components and the delta components are convolved with the \texttt{CLEAN} beam to visualize the image. Furthermore, the core region is not optimally resolved even with over-resolved beam in \autoref{fig:synthetic_sky_appendix}. Note that multi-scale \texttt{CLEAN} might be able to recover the extended jet structure with better-resolved core, but iterative user-dependent self-calibration steps are still required in order to obtain high fidelity images. In the \texttt{resolve} image, the core and the extended jet structure in the ground truth image are recovered better than the \texttt{CLEAN} reconstruction. 

\begin{figure}[t]
    \includegraphics[width=9cm] {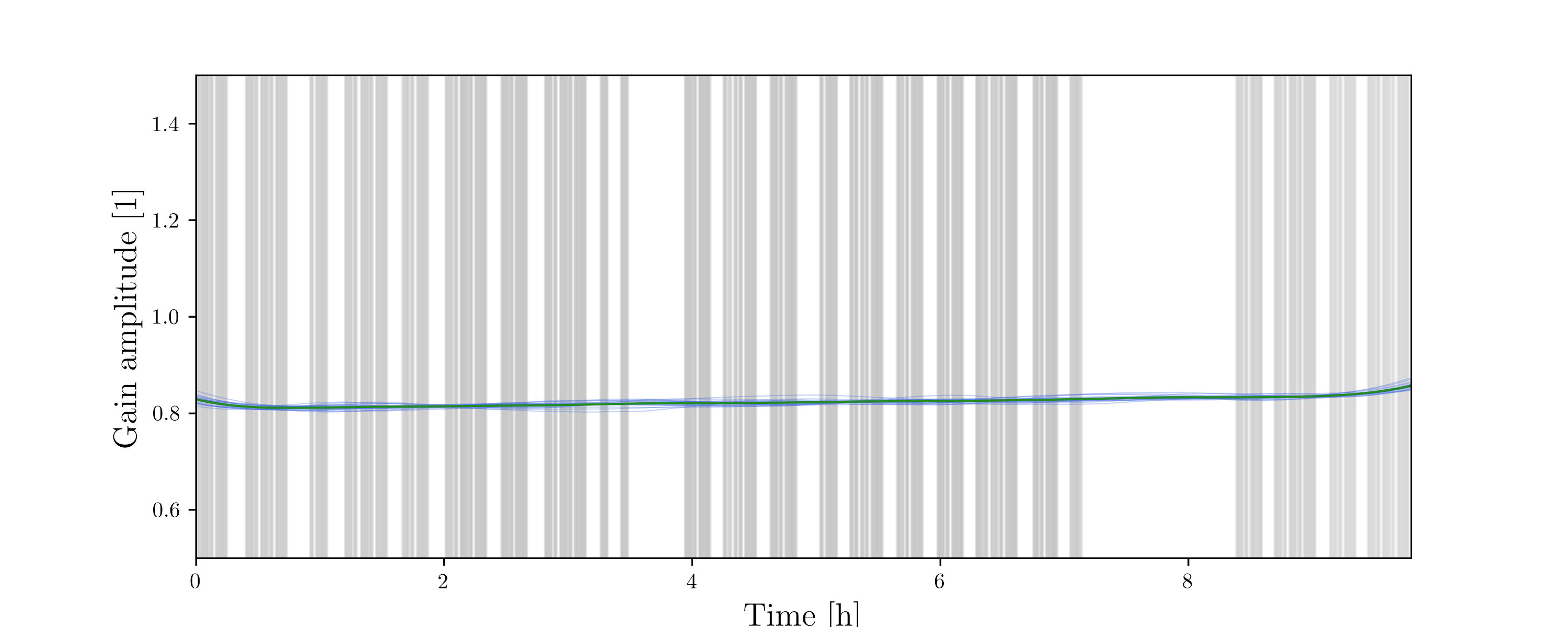}
    \includegraphics[width=9cm]{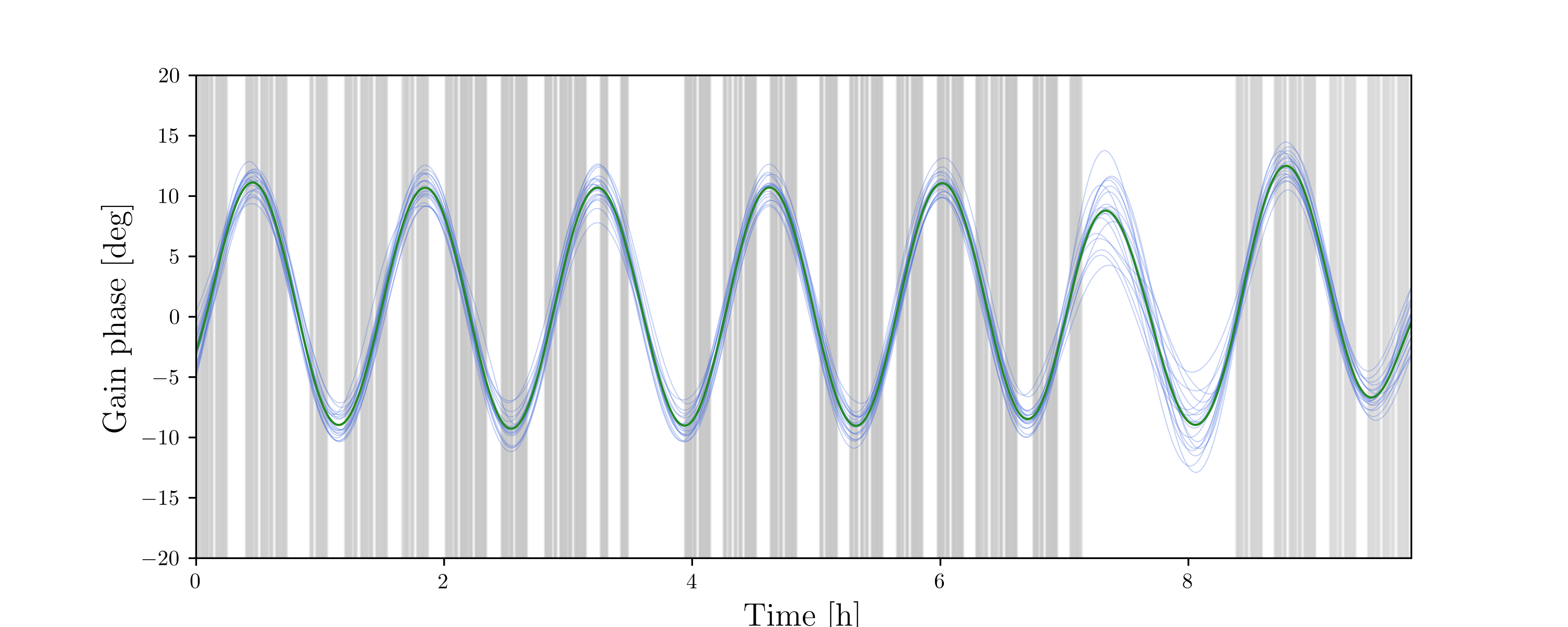}
    
    \caption{Synthetic data: RCP amplitude (top) and phase (bottom) gain posterior mean and posterior samples for BR antenna by the Bayesian self-calibration. The thick green solid line represents the posterior mean, while the thin blue solid lines depict individual posterior samples. Grey vertical shades indicate areas with available data scans.}
    \label{fig:synthetic_amp_phase_gain_BR_RCP}
\end{figure}

\begin{figure}[t]
    \includegraphics[width=9cm] {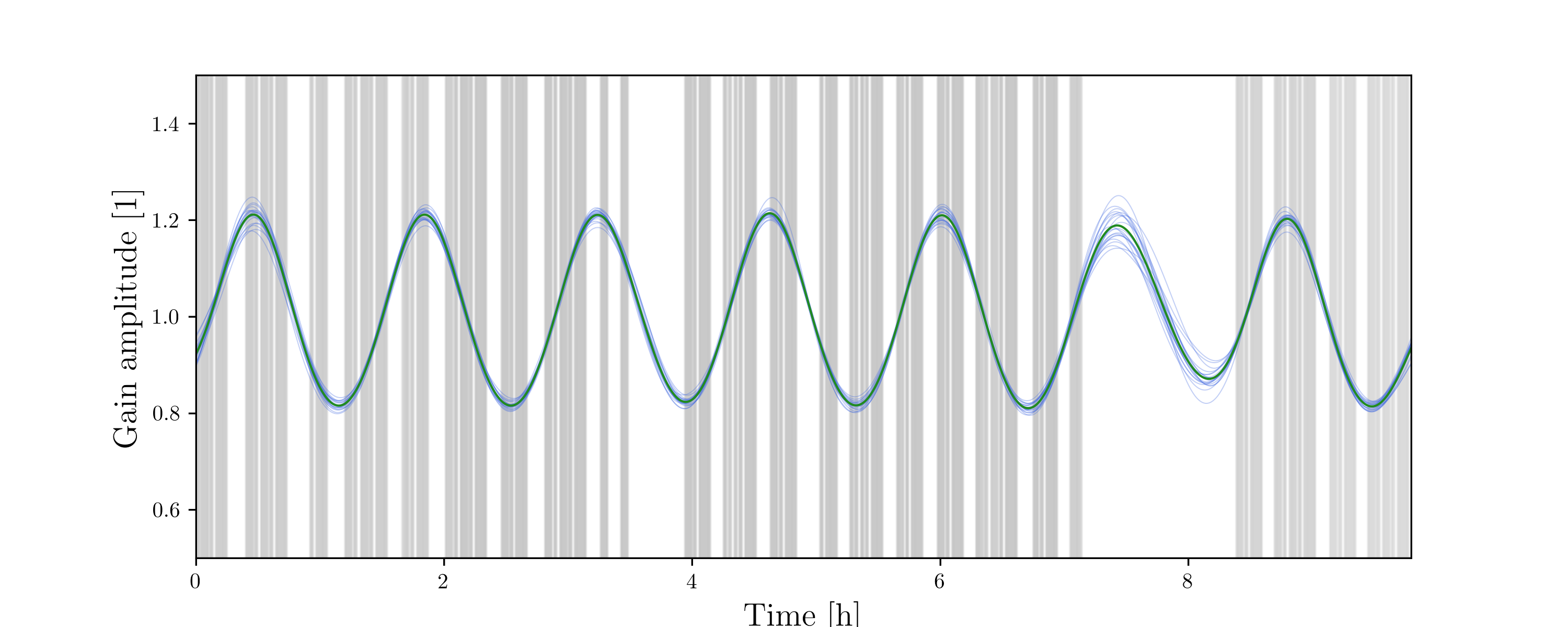}
    \includegraphics[width=9cm]{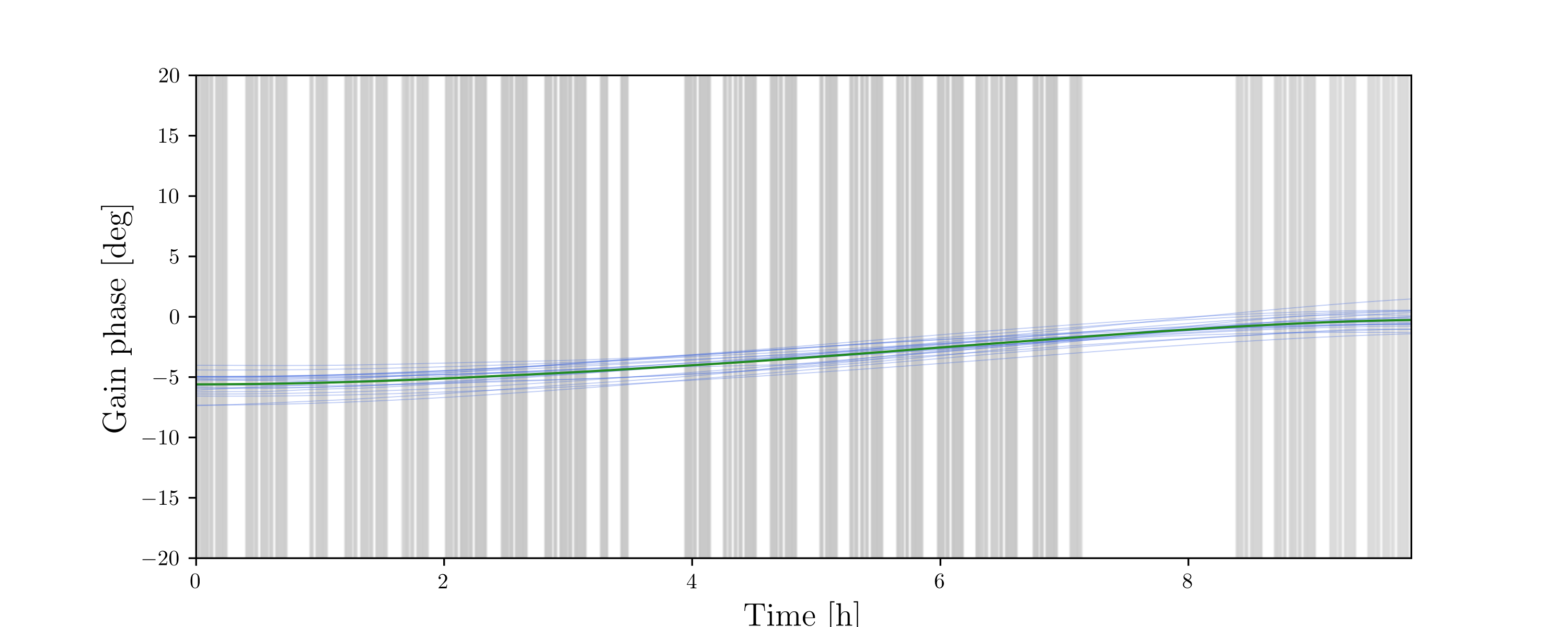}
    
    \caption{Synthetic data: Same as \autoref{fig:synthetic_amp_phase_gain_BR_RCP}, but with LCP amplitude (top) and phase (bottom) gain for BR antenna.}
    \label{fig:synthetic_amp_phase_gain_BR_LCP}
\end{figure}

\begin{figure}[t]
    \includegraphics[width=9cm] {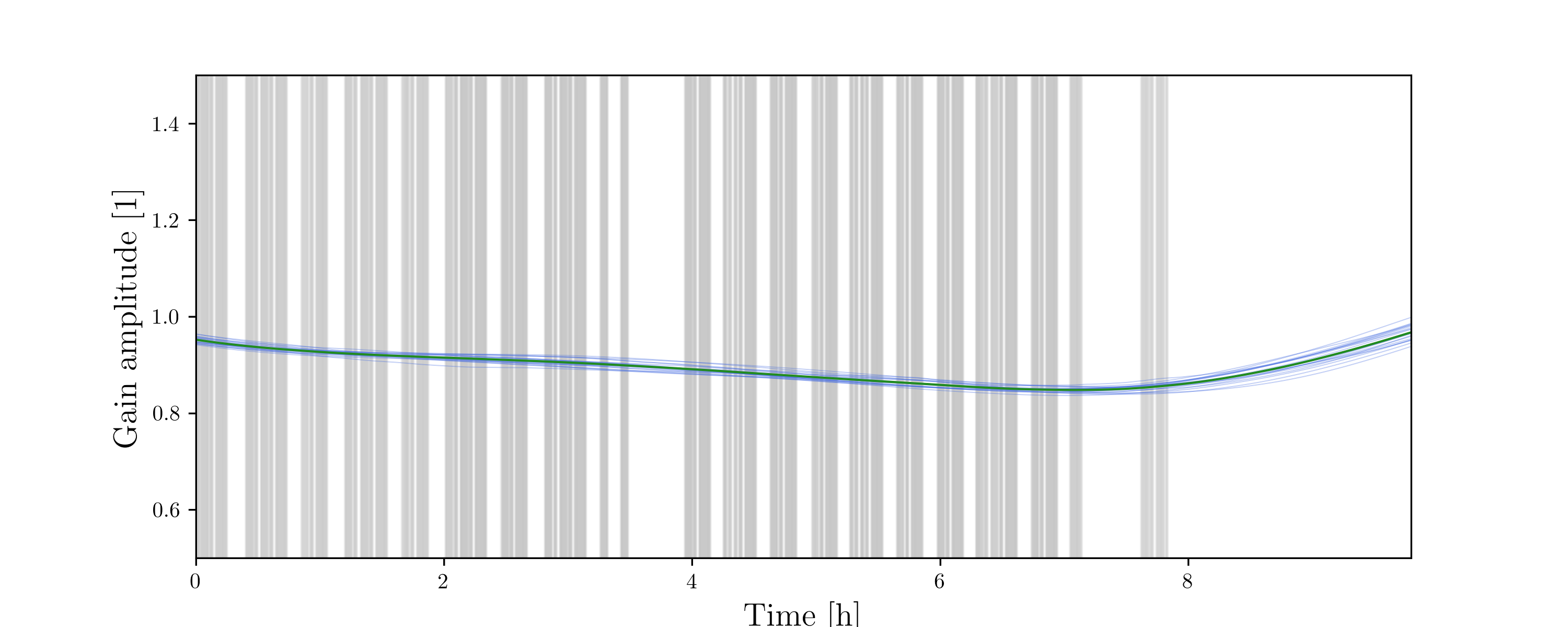}
    \includegraphics[width=9cm]{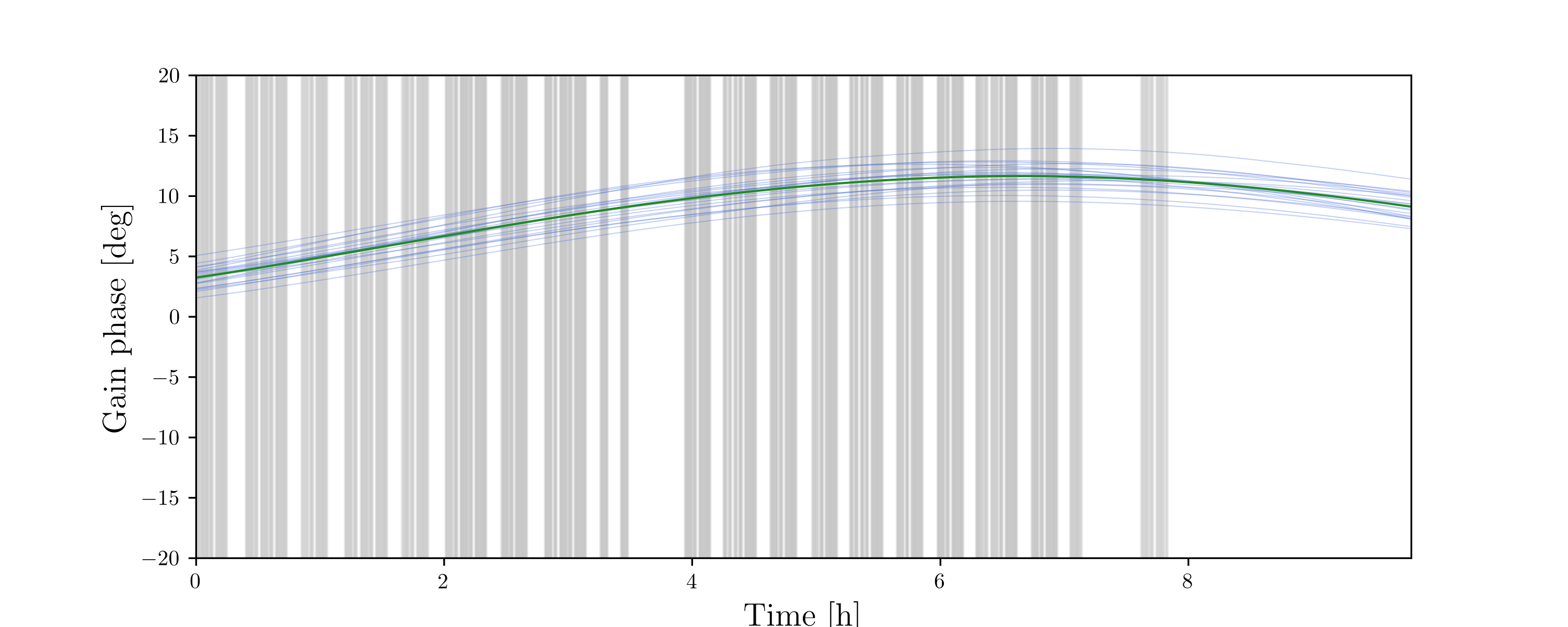}
    
    \caption{Synthetic data: Same as \autoref{fig:synthetic_amp_phase_gain_BR_RCP}, but with RCP amplitude (top) and phase (bottom) gain for HN antenna.}
    \label{fig:synthetic_amp_phase_gain_HN_RCP}
\end{figure}

\begin{figure}[t]
    \includegraphics[width=9cm] {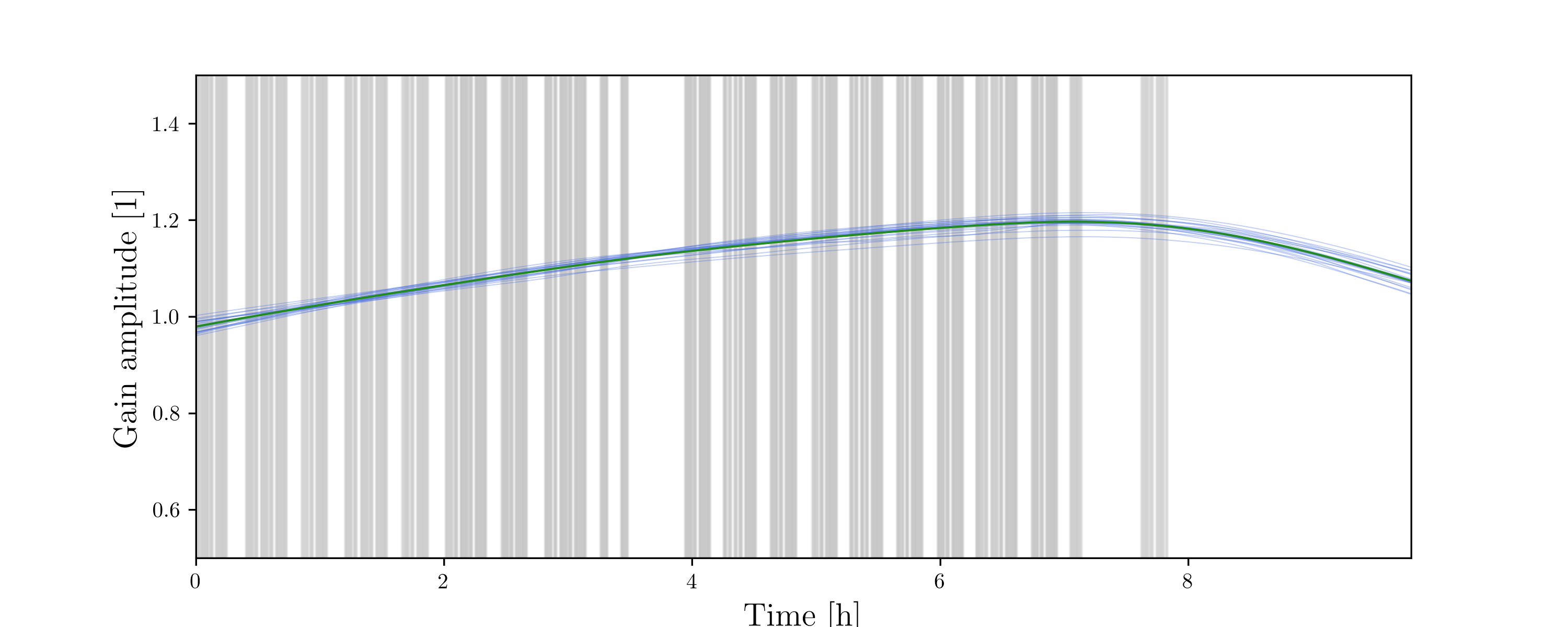}
    \includegraphics[width=9cm]{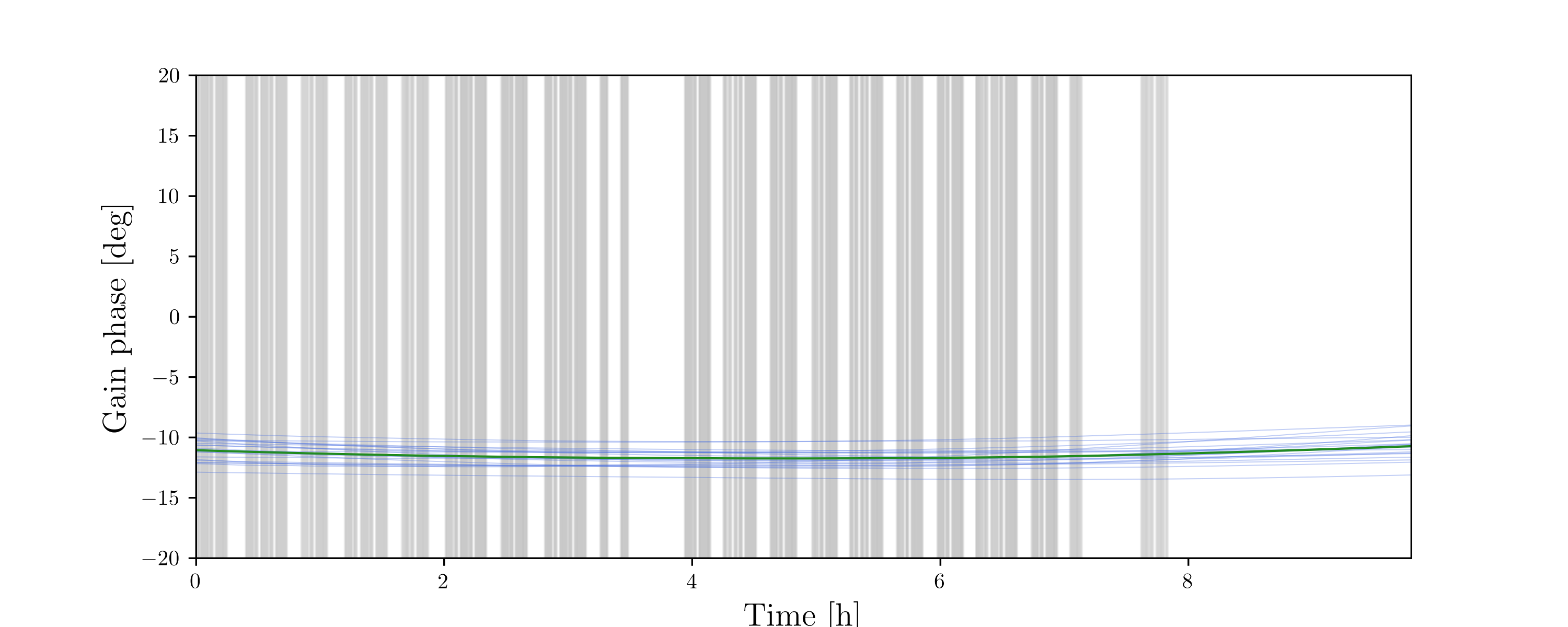}
    
    \caption{Synthetic data: Same as \autoref{fig:synthetic_amp_phase_gain_BR_RCP}, but with LCP amplitude (top) and phase (bottom) gain for HN antenna.}
    \label{fig:synthetic_amp_phase_gain_HN_LCP}
\end{figure}

\begin{figure}[t]
    \includegraphics[width=9cm] {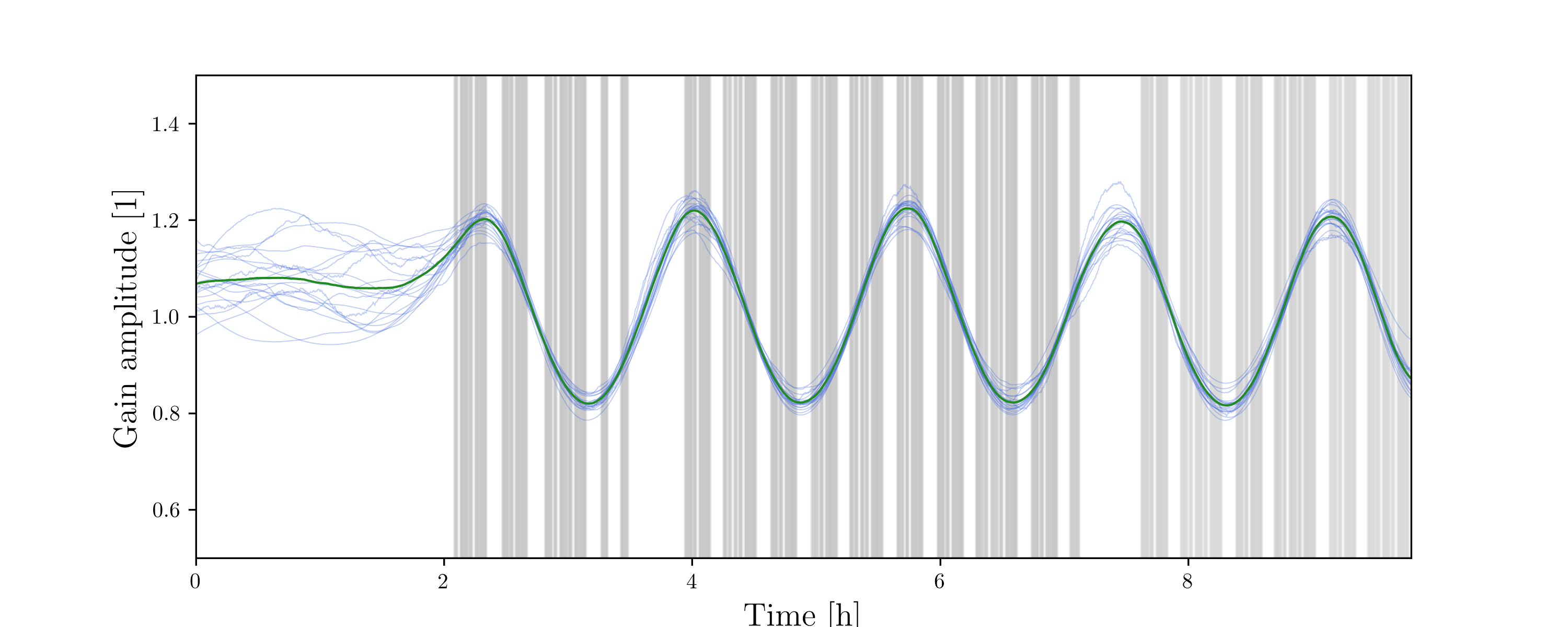}
    \includegraphics[width=9cm]{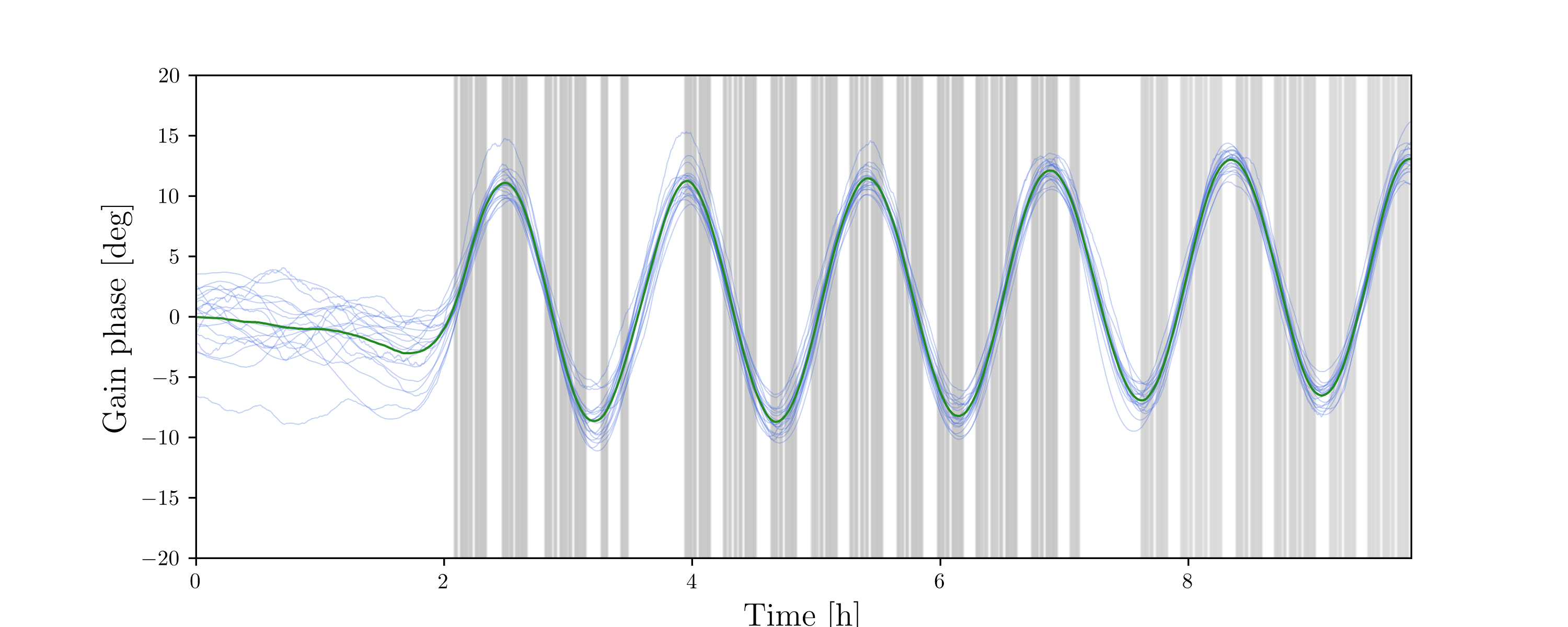}
    
    \caption{Synthetic data: Same as \autoref{fig:synthetic_amp_phase_gain_BR_RCP}, but with RCP amplitude (top) and phase (bottom) gain for MK antenna.}
    \label{fig:synthetic_amp_phase_gain_MK_RCP}
\end{figure}

\begin{figure}[t]
    \includegraphics[width=9cm] {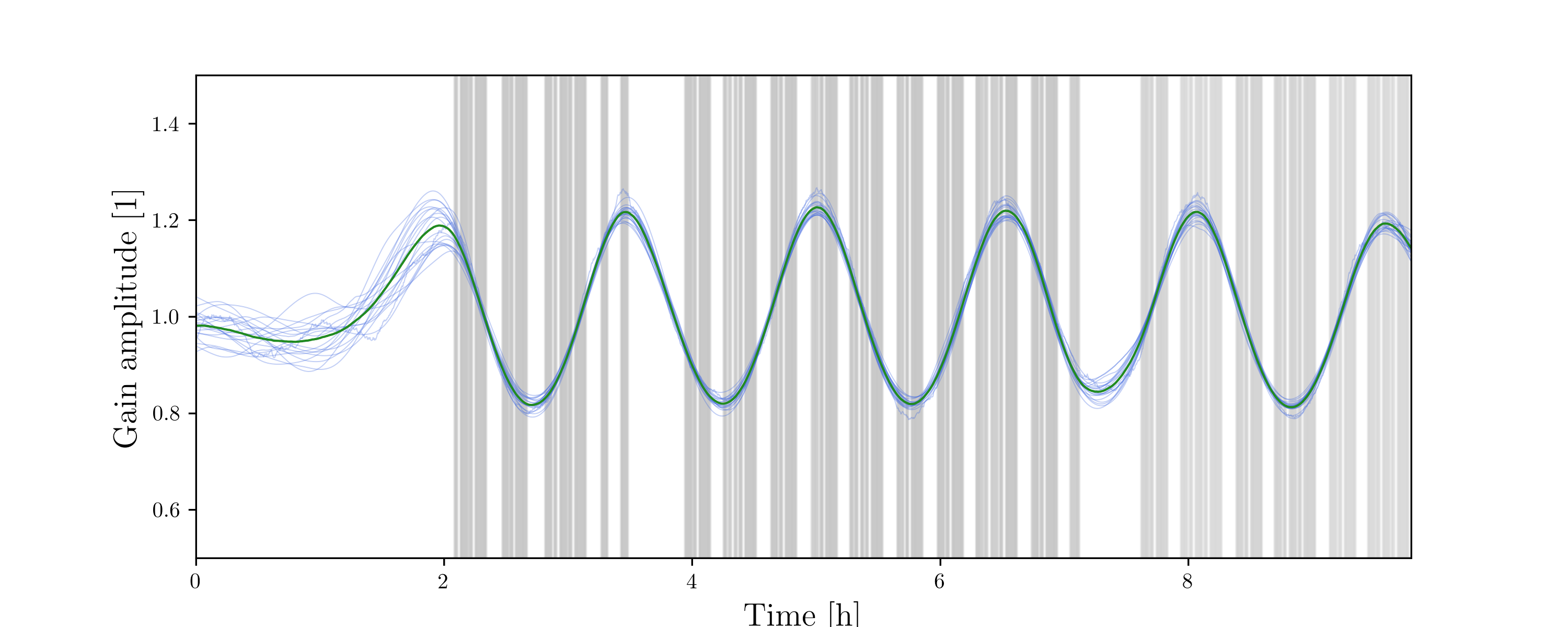}
    \includegraphics[width=9cm]{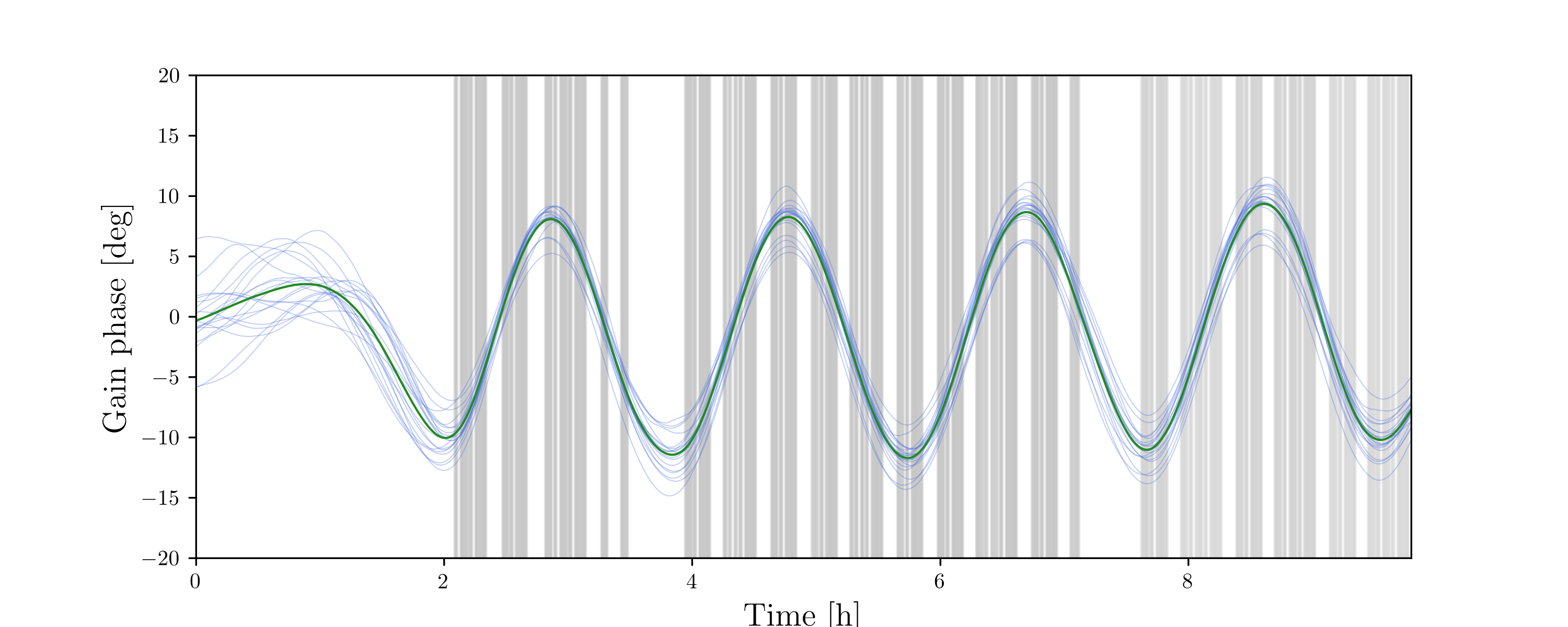}
    
    \caption{Synthetic data: Same as \autoref{fig:synthetic_amp_phase_gain_BR_RCP}, but with LCP amplitude (top) and phase (bottom) gain for MK antenna.}
    \label{fig:synthetic_amp_phase_gain_MK_LCP}
\end{figure}

\begin{figure}[t]
    \includegraphics[width=9cm] {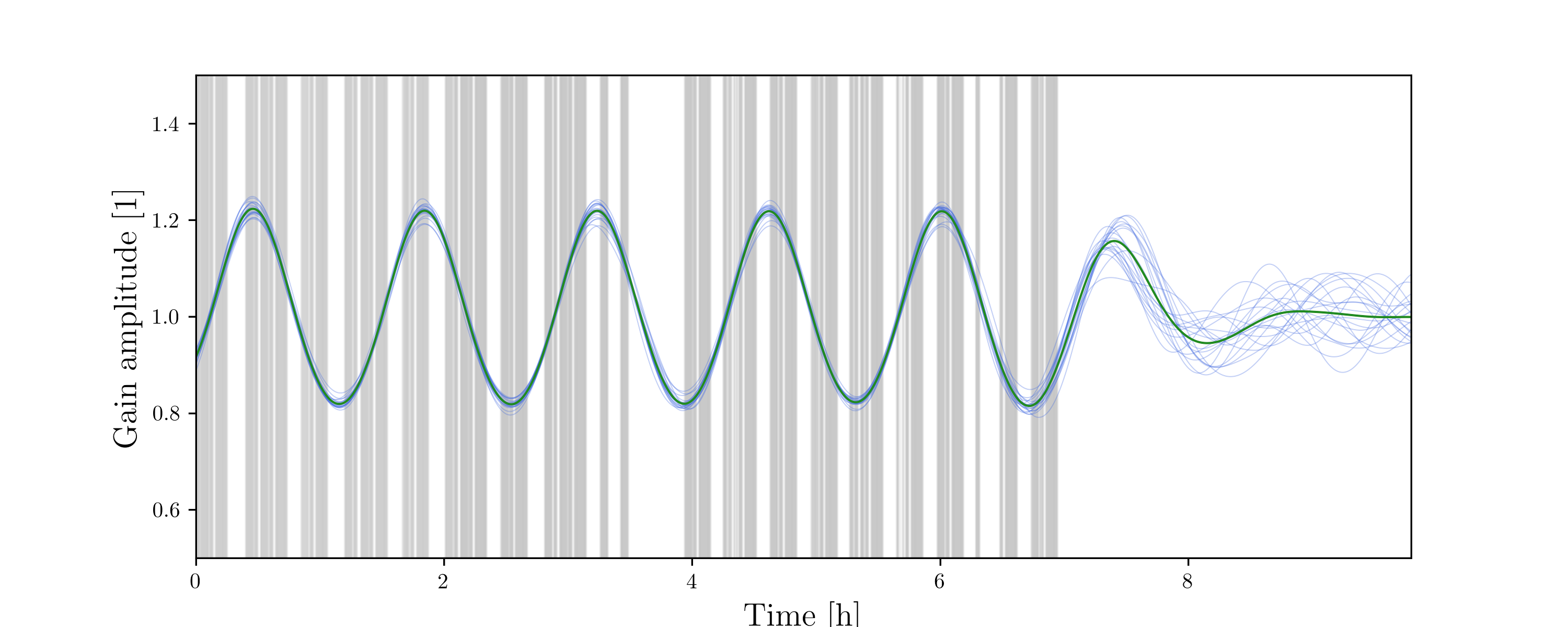}
    \includegraphics[width=9cm]{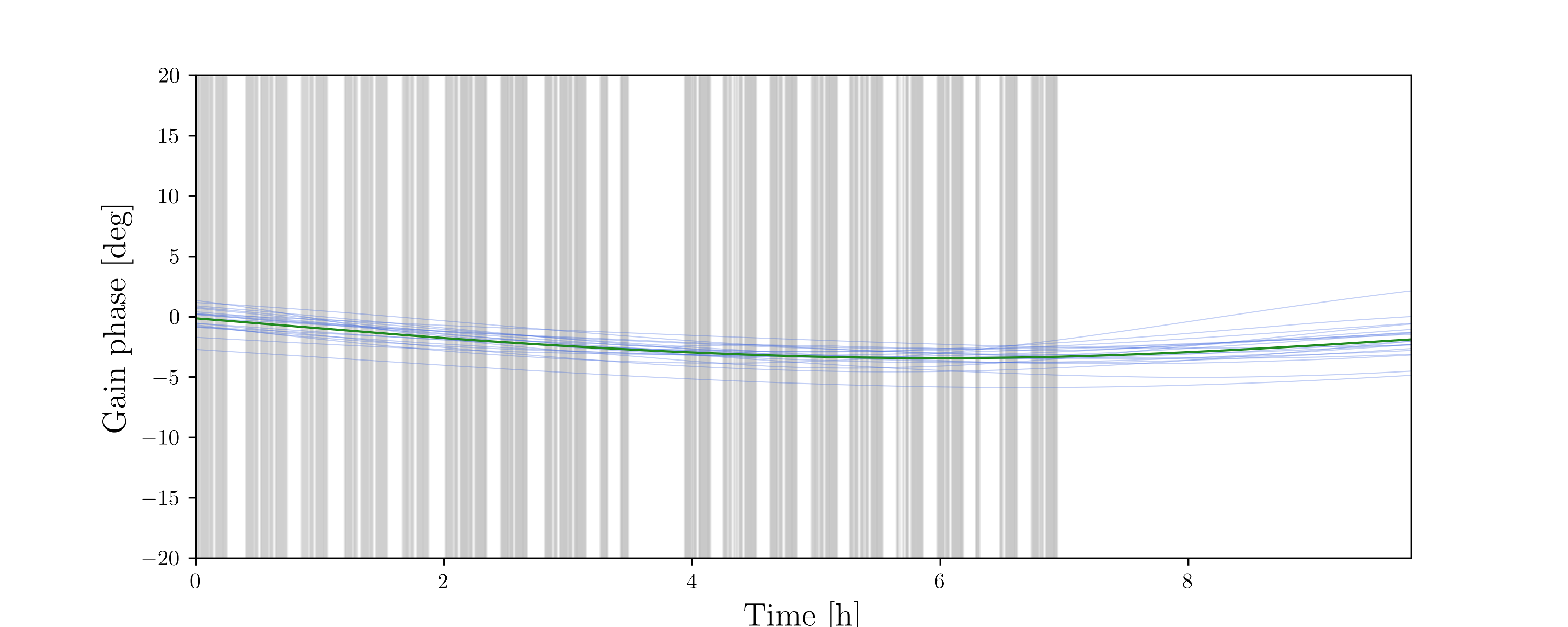}
    
    \caption{Synthetic data: Same as \autoref{fig:synthetic_amp_phase_gain_BR_RCP}, but with RCP amplitude (top) and phase (bottom) gain for SC antenna.}
    \label{fig:synthetic_amp_phase_gain_SC_RCP}
\end{figure}

\begin{figure}[t]
    \includegraphics[width=9cm] {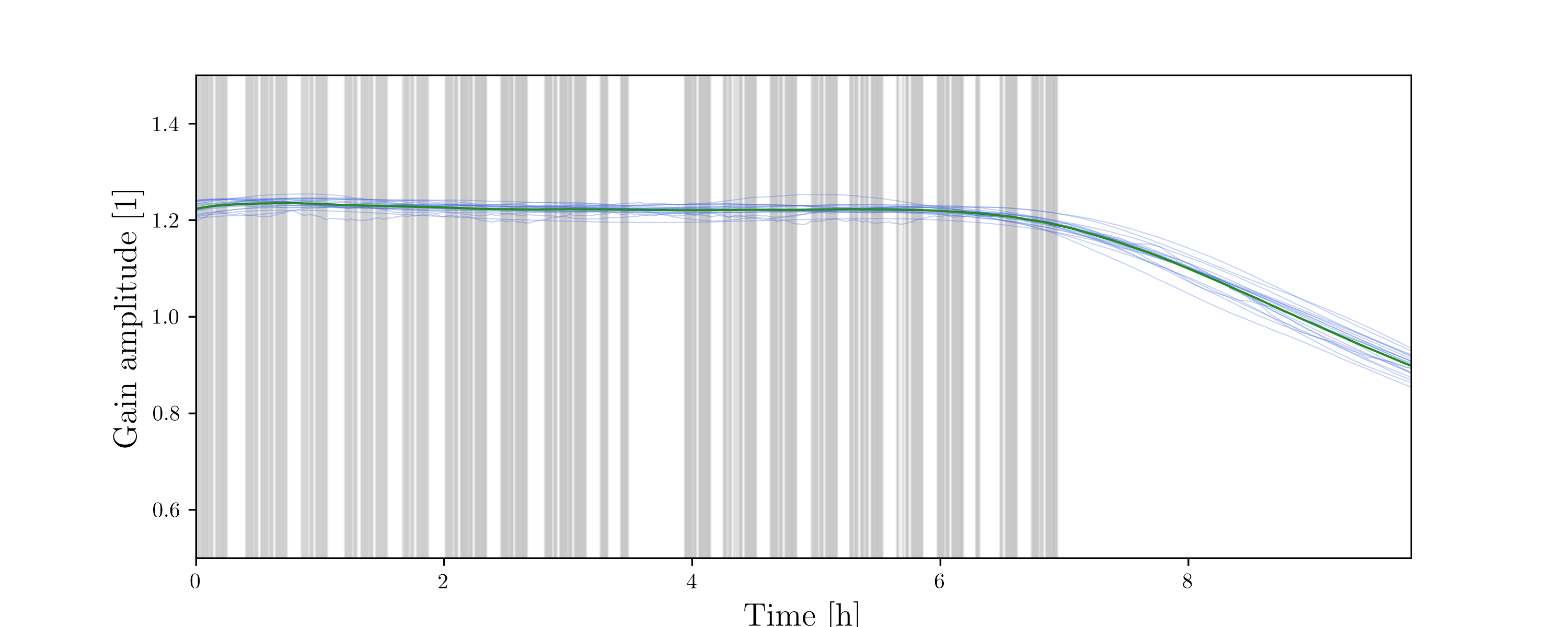}
    \includegraphics[width=9cm]{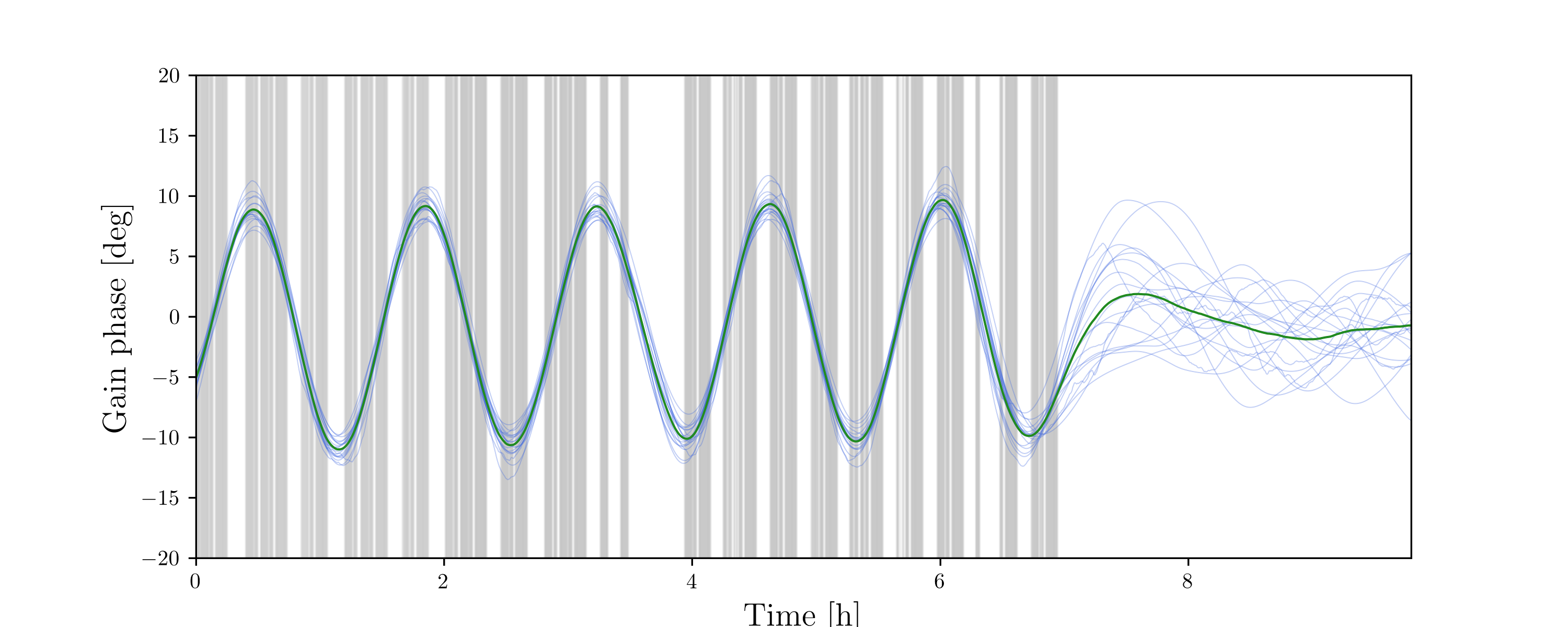}
    
    \caption{Synthetic data: Same as \autoref{fig:synthetic_amp_phase_gain_BR_RCP}, but with LCP amplitude (top) and phase (bottom) gain for SC antenna.}
    \label{fig:synthetic_amp_phase_gain_SC_LCP}
\end{figure}

The ground truth and reconstructed amplitude and phase gains are presented in \autoref{fig:synthetic_gains_amp} and \autoref{fig:synthetic_gains_phase}. In the amplitude and phase gain prior models, different correlation kernels per each antenna and polarization mode are employed in order to infer different temporal correlation structure. The reconstructed amplitude and phase gain solutions by the Bayesian self-calibration in the time coverage with data are reasonably consistent with the ground truth. The inconsistency between the gain solution and ground truth for BR (7-8h), HN (> 8h), MK (< 2h), NL (< 9h), and SC (> 7h) is due to the absence of these data (see \autoref{fig:synthetic_amp_phase_gain_BR_RCP} - \autoref{fig:synthetic_amp_phase_gain_SC_LCP}). As an example, \autoref{fig:synthetic_amp_phase_gain_SC_LCP} shows the posterior mean and posterior samples of LCP gain amplitude and phase for SC antenna. In the time coverage without data (> 7h), the uncertainties of amplitude and phase gain are higher than the gain solutions with data and the posterior mean amplitude and phase gains are inconsistent with the ground truth. However, this does not affect the fidelity of the reconstructed image because those gain solutions are not applied due to the absence of these data. 

In conclusion, this example illustrates that a high fidelity image with robust amplitude and phase gain solutions with different correlation structure can be reconstructed from the corrupted synthetic data set by the Bayesian self-calibration and imaging method. Bayesian self-calibration may be utilized for inhomogeneous arrays, such as global mm-VLBI array (GMVA) and EVN (European VLBI network) to reconstruct reliable image and gain solutions with uncertainty estimation in the future.

\section{Conclusion} \label{Chap7}

We have presented the Bayesian self-calibration and imaging method by \texttt{resolve} and applied it to real and synthetic VLBI data. VLBA M87 data at 43\,GHz were pre-calibrated by the rPICARD CASA-based pipeline and imaging with self-calibration was performed by Bayesian imaging software \texttt{resolve}. The data flagging was done by rPICARD pipeline without manual flagging and the image and gain solutions with uncertainty estimation were reconstructed jointly in the \texttt{resolve} framework. 

Self-calibration solutions by \texttt{resolve} from real data are consistent with the conventional \texttt{CLEAN} and an RML method \texttt{ehtim}, despite the fact that different imaging approaches and gain inference schemes are utilized. In Bayesian self-calibration, we do not have to choose the solution interval of gain solutions but rather the time correlation structure of the gain solutions is inferred from the data. The synthetic data test shows that the Bayesian self-calibration method is able to infer different correlation structure of gain solutions per antenna and polarization mode, which is common in inhomogeneous VLBI arrays. Furthermore, in \texttt{resolve}, the uncertainty estimation of the gain solutions and image is provided on top of the posterior mean gains and image, which is important for the scientific analysis. On the other hand, in \texttt{CLEAN} self-calibration, residual gains are estimated and removed iteratively by using the \texttt{CLEAN} reconstructed model images as prior. Therefore, inconsistencies in the CLEAN model image resulting from user choices in the manual setting of \texttt{CLEAN} windows, weighting schemes, data flagging, and solution intervals for gain solutions may introduce bias. In RML-based \texttt{ehtim} self-calibration, the model image for self-calibration is consistent with the data since the model fits the data directly. However, solution interval is still chosen manually and only a point estimate of the antenna gain temporal evolution is provided, lacking any uncertainty information on the provided gain solution.

As a result, our example with the VLBA M87 data shows that a robust image and antenna-based gain solutions with uncertainty estimation were obtained by taking into account the uncertainty information in the data. The VLBA M87 \texttt{resolve }image has better resolved core and counter jet, and consistent extended jet emission than the one provided by the conventional \texttt{CLEAN} reconstruction. 
This demonstrates the potential of the proposed method also for applications to other data sets. From our perspective, future work is needed on testing Bayesian self-calibration and imaging method on more challenging VLBI data, such as from sparse and inhomogeneous antenna array, and at high frequencies, and to incorporate polarization calibration in the \texttt{resolve} framework.

\begin{acknowledgements} 
      We thank the anonymous referee who suggested the validation with an RML-based method and constructive comments, Jack Livingston for feedback on drafts of the manuscript, Jakob Knollmüller for discussions on resolve software and feedback, Thomas Krichbaum and Martin Shepherd for discussions on \texttt{CLEAN} algorithm and \texttt{DIFMAP} software. J. K. and A. N. received financial support for this research from the International Max Planck Research School (IMPRS) for Astronomy and Astrophysics at the Universities of Bonn and Cologne. This work was supported by the M2FINDERS project funded by the European Research Council (ERC) under the European Union's Horizon 2020 Research and Innovation Programme (Grant Agreement No. 101018682). J.R. acknowledges financial support from the German Federal Ministry of Education and Research (BMBF) under grant 05A23WO1 (Verbundprojekt D-MeerKAT III). P.A. acknowledges financial support from the German Federal Ministry of Education and Research (BMBF) under grant 05A20W01 (Verbundprojekt D-MeerKAT). The National Radio Astronomy Observatory is a facility of the National Science Foundation operated under cooperative agreement by Associated Universities, Inc.
\end{acknowledgements}

%
%

\bibliographystyle{aa} 
\bibliography{reference.bib}
\appendix

\section{Weighting scheme in \texttt{resolve}}

The statistics of the posterior sky distribution determined by Bayesian inference has some analogy to robust weighting in the \texttt{CLEAN} algorithm. The posterior distribution will include Fourier scales with high signal-to-noise, while Fourier scales with low signal-to-noise will be damped with the signal-to-noise ratio. This has some analogy to the \texttt{CLEAN} algorithm with robust weighting where Fourier scales with high signal-to-noise are uniformly weighted and scales with low signal-to-noise are weighted naturally. Nevertheless, in contrast to the \texttt{CLEAN} algorithm, this behavior of the resulting sky reconstruction is intrinsic to Bayesian inference or other methods following some form of regularized maximum likelihood approach.

In the appendix A.4 of \cite{Junklewitz_2016}, this analogy between robust weighting in \texttt{CLEAN} and the posterior obtained from Bayesian inference was made more explicit. More specifically, \cite{Junklewitz_2016} derived the Bayesian Wiener Filter operation for estimating sky brightness, finding that it takes the same mathematical form of robust weighting when expressed in Fourier space. Thereby the robust parameter, which in the case of \texttt{CLEAN} needs to be chosen by the user is determined by the prior distribution for the sky brightness.

\section{Bayesian perspective on Regularized Maximum Likelihood}
\label{app:RML}

In this subsection, we aim to investigate the RML method from a Bayesian perspective. In the RML method, an objective function $J(d,s)$ is to be minimized w.r.t. to the unknown signal $s$. Therefore, the RML estimator is
\begin{align}
s_\text{RML} := \text{argmin}_{s} J(d,s).
\end{align}

The objective function consists of a data fidelity term $\chi^2(d,s)$ and a regularizing term $r(s,\beta)$ that should prevent irregular solutions  \citep{Chael_2018_closure}:
\begin{align}
	J(d,s) =  \chi^2(d,s) + r(s,\beta),
\end{align}
where $\beta$ denotes the set of parameters the regularizer might depend on, including a potential parameter determining the relative weight of the regularization w.r.t.~the data fidelity term. \\

The data fidelity term $\chi^2(d,s)$ is equivalent to the likelihood Hamiltonian $\mathcal{H}(d|s)$ in Bayesian inference up to irrelevant additive and multiplicative constants (see Eq. \ref{eq:data_fidelity}). In other words, the data fidelity term is the negavie log likelihood $\chi^2(d,s) = -\ln \mathcal{P}(d|s) + \text{const.}$. It is natural to regard the regularizing term $r(s,\beta)$ as the negative log prior $r(s,\beta) = -\ln \mathcal{P}(s|\beta) + \text{const.}$. In this case, the RML estimator would be identical to the Bayesian maximum a posteriori (MAP) estimator for the corresponding signal prior $\mathcal{P}(s|\beta)$, with fixed hyperparameters $\beta$.

Therefore, RML method can be regarded as Bayesian methods in which the regularizing term $r(s,\beta)$ specifies the prior assumption on the signal and that exploits the MAP approximation.   
From this perspective, the regularization terms can be translated into prior assumptions, via
\begin{align}\label{eq:reconstructedRMLprior}
	\mathcal{P}(s|\beta)=e^{-\mathcal{H}(s|\beta)}\equiv\frac{e^{-r(s,\beta)}}{\int \mathcal{D}s\,e^{-r(s,\beta)}}.
\end{align} 
Here, the denominator in the last expression ensures proper normalization and would be essential to a Bayesian determination of the parameters $\beta$ of the regularizer, which, however, is usually done in RML practice by trial and error and visual inspection of the results. Note that \cite{Mueller_2023_GA} employed the genetic algorithm in order to automate the parameter $\beta$ selection.

In RML methods, several regularizers are often combined in order to encode different prior knowledge about the source. Let us therefore revisit some of the commonly used regularizers in RML methods, such as the entropy, total variation (TV), and total squared variation (TSV) regularization and interpretation from the Bayesian perspective:

\textbf{Entropy regularization} of a discretized intensity field $s(x,y) = I(x,y)$ , with $x,y\in \mathbb{Z}$, is given by 
\begin{eqnarray}
	r_{\text{entropy}}(I,\beta) &=&  T\,\sum_{x,y} I_{x,y} \ln \frac{I_{x,y}}{M_{x,y}} \,,
\end{eqnarray}
with $\beta=(T,M)$ consisting of a "temperature" $T$ that determines the strength of the regularization and a reference image  $M$, with respect to which this relative entropy like expression is evaluated. Converting this to a prior according to Eq.~\ref{eq:reconstructedRMLprior} yields
\begin{eqnarray}
	\label{eq:RMLentropyprior}
	\mathcal{P}(I|\beta)&\propto & {e^{-T\,\sum_{x,y} I_{x,y} \ln \frac{I_{x,y}}{M_{x,y}}}}
	= \prod_{x,y} e^{-T\, I_{x,y} \ln \frac{I_{x,y}}{M_{x,y}}}
	\nonumber \\
	&=&
	 \prod_{x,y}  \left(\frac{I_{x,y}}{M_{x,y}} \right)^{-T\, I_{x,y}} =:   \prod_{x,y}\mathcal{P}(I_{x,y}|\beta).
\end{eqnarray} 
This means that all pixels are assumed to be uncorrelated, since the prior is a direct product of individual pixel priors
\begin{align}
	\mathcal{P}(I_{x,y}|\beta) \propto \left(\frac{I_{x,y}}{M_{x,y}} \right)^{-T\, I_{x,y}}.
\end{align}
These assign nearly constant probabilities for any flux value for which $ I_{x,y} \ll 1/T$ and a sharper than exponential cut off  of the prior probabilities for intensities beyond $I_{x,y} = 1/T$. 
High flux values are therefore strongly suppressed by the entropy regularizer as was already pointed out by \citep{Junklewitz_2016}.

Here, we would like to point out in addition to this that the reference image has a very mild effect on the results, as this super exponential cut off is fully determined by $T$.
Approximately, the entropy regularization therefore corresponds to assuming the intensity field to be white noise with values roughly uniformly distributed between $0$ and $1/T$.
For these reasons, the entropy RML is not expected to provide improved images for most radio astronomical observations of diffuse sources. 
The reputation of the entropy RML method to produce smooth images results probably from the usage of a large $T$ parameter, which strongly discourages extreme brightnesses and therefore encourages the neighboring pixel of a strong flux location to explain the flux of that.
A good feature of the entropy RML method is, however, that negative intensities are excluded  by it a priori.

The \textbf{total squared variation regularization}
\begin{equation}
	r_\text{TSV}(I,\beta) = \alpha \sum_{x,y} \left[\left(I_{x+1,y} - I_{x,y}\right)^{2} + \left(I_{x,y+1} - I_{x,y}\right)^2 \right]\,,
\end{equation}
can be regarded in the continuum limit as requesting intensity gradients to be minimal  
\begin{eqnarray}
	r_\text{TSV}(I,\beta) &\equiv& \alpha' \int dx\int dy \left[ \left(\frac{\partial I(x,y)}{\partial x}\right)^{2} + \left(\frac{\partial I(x,y)}{\partial y}\right)^{2} \right]\\
	&=& \alpha' \int dx\int dy \left|\nabla I(x,y) \right|^{2}\,,
\end{eqnarray}
where $\beta = \alpha$ in the discrete case and $\beta = \alpha'$ in the corresponding continuum case determine the strength of the regularization.

The total squared variation regularization corresponds  certainly to a more appropriate prior for diffuse emission, as it couples nearby locations and enforces some smoothness of the reconstruction.

This regularization actually becomes diagonal in Fourier space, with  
\begin{eqnarray}
	r_\text{TSV}(I,\beta) &=&  \alpha' \int \frac{dk^2}{(2\pi)^2} \left|\vec{k} I(\vec{k})\right|^{2} \,.
\end{eqnarray}
This turns out to be exactly (up to negligible additive terms) the log prior of a statistical homogeneous and isotropic Gaussian random field 
\begin{eqnarray}
	\mathcal{G}(I,P_I(k)) &\propto& \exp\left( - \frac{1}{2} \int \frac{dk^2}{(2\pi)^2} \frac{| I(\vec{k})|^2}{P_I(k)} \right), 
\end{eqnarray}
 with an intensity power spectrum 
\begin{eqnarray}
	P_I(k) &=&  \frac{ 1}{2 \alpha'} \,k^{-2} \,. 
\end{eqnarray}
In one dimension, a Gaussian process with such a power spectrum would be equivalent to a Wiener process, which is known to produce continuous, but rough structures.

Changing to the \textbf{total variation regularizer}, in the continuum representation written as
\begin{eqnarray}
	r_\text{TV}(I,\beta) &\equiv& \alpha' \int dx\int dy \left|\nabla I(x,y) \right|\,,
\end{eqnarray}
enhances the tendency to allow rough structures, but it favors roughness to be more localized. This is because the $L_1$ norm underlying the TV regularizer is more tolerant to few large intensity gradients and the $L_2$ norm underlying the TSV regularizer instead prefers many, smaller gradients.
Note that the TSV regularizer does not have a simple Fourier space representation. Both, TV and TSV, regularizers are utilized to consider the correlation between neighboring pixels. However, they do not enforce the positivity of the sky intensity and tend to generate rough structures.

\section{Hyperparameter setup for sky and gain priors} \label{Appendix_hyperparameter}

\begin{figure}[t]
    \includegraphics[width=9cm] {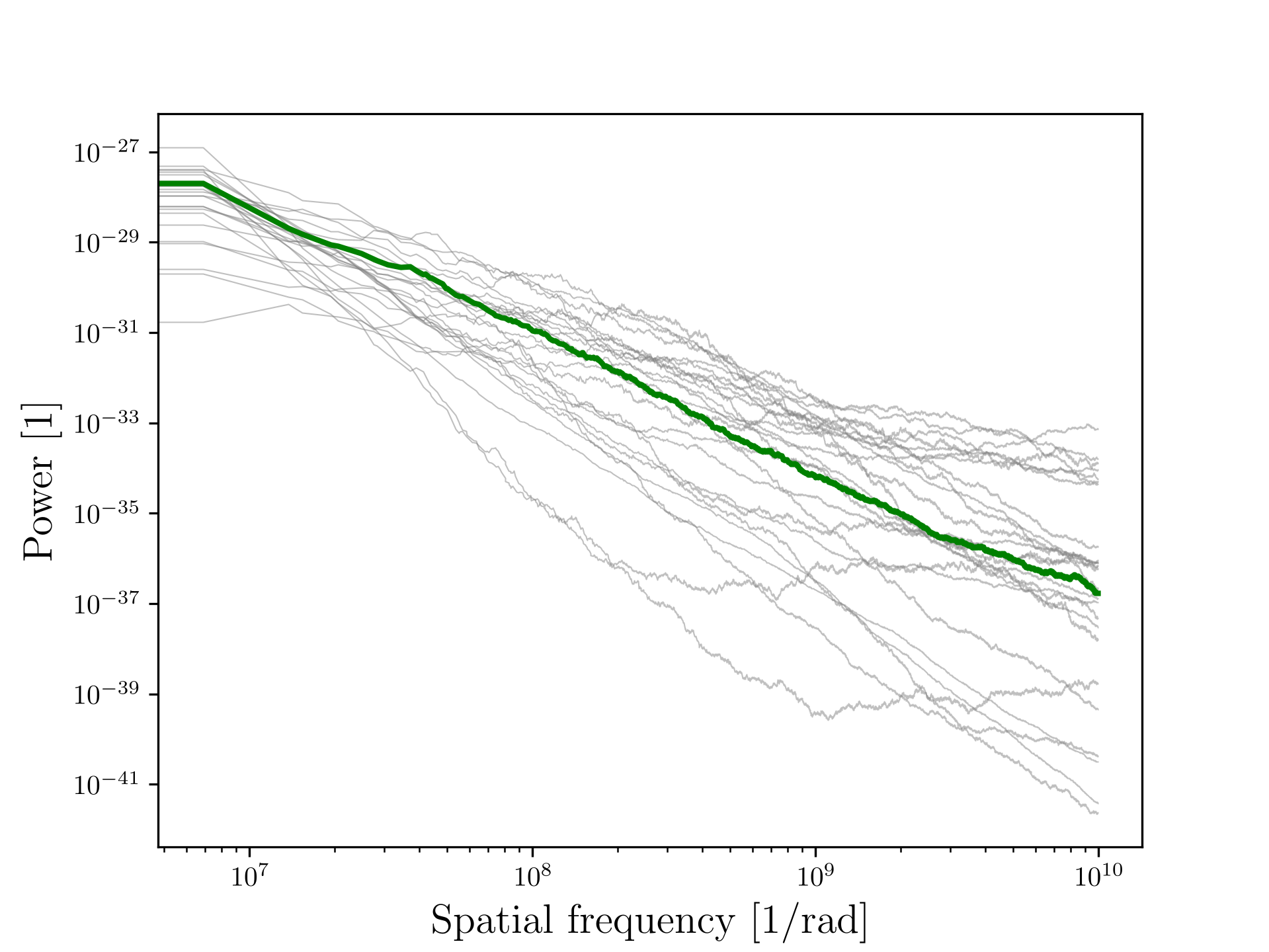}{\caption{ M87: posterior and prior power spectra of logarithmic sky brightness distribution $\psi$. The green line denotes posterior mean power spectrum; grey lines denote prior power spectrum samples. \label{fig:logsky_pspec}}}
\end{figure}

\begin{table}[h]
	\centering
	\begin{tabular}{lrr}
		\hline
		$\:$               & $\psi$ mean & $\psi$ std \\
		\hline
		Offset 	           & 35 & --- \\
		Zero mode variance & 1 & 0.1  \\
		Fluctuations       & 3 & 1  \\
		Flexibility        & 1.2 & 0.4  \\
		Asperity           & 0.4 & 0.4  \\
		Average slope      & -3 & 1  \\
		\hline
	\end{tabular}
	\caption{Hyper parameters for the log-sky prior $\psi$. All hyperparameters are unitless. Detailed description of each parameter can be found in Section 3.4 of \citet{Arras_2021_CygA}.}
	\label{table:hyper_logsky}
\end{table}

\begin{table}[h]
	\centering
	\begin{tabular}{lrrrr}
		\hline
		$\:$               & $\lambda$ mean & $\lambda$ std & $\phi$ mean  & $\phi$ std \\
		\hline
		Offset 	           & 0 & --- & 0 & --- \\
		Zero mode variance & 0.2 & 0.1 & 1e-3 & 1e-6 \\
		Fluctuations       & 0.2 & 0.1 & 0.2 & 0.1 \\
		Flexibility        & 0.5 & 0.2 & 0.5 & 0.2 \\
		Asperity           & None & None & None & None \\
		Average slope      & -3 & 1 & -3 & 1 \\
		\hline
	\end{tabular}
	\caption{Hyper parameters for the log-amplitude gain prior $\lambda$ and the phase gain prior $\phi$. All hyperparameters are unitless. Detailed description of each parameter can be found in Section 3.4 of \citet{Arras_2021_CygA}.}
	\label{table:hyper_gains}
\end{table}

The hyperparameter setup for the log-sky prior $\psi$ is listed in Table \ref{table:hyper_logsky}. The offset mean represents the mean value of the log-sky $\psi$; thus the mean of prior sky model $\text{exp}(\psi)$ is $\text{exp}(35) \approx 10^{15}$ Jy/sr ($\approx 10^{-2}$ Jy/$\mathrm{mas^2}$). The value is allowed to vary two e-folds up and down in one standard deviation (std) of the prior. The zero mode variance mean describes the standard deviation of the offset and its standard deviation is therefore the standard deviation of the offset standard deviation. 

The next four hyperparameters are model parameters for the spatial correlation power spectrum $P_{\Psi}(\xi_{\Psi})$ in Eq. \ref{eq:sky_generative_model}. The posterior and prior power spectra of the log-sky $\psi$ are in \autoref{fig:logsky_pspec}. The average slope mean and std denote the mean and standard deviation of the slope for amplitude spectrum, which is the square root of the power spectrum.
In \autoref{fig:logsky_pspec}, the prior power spectrum samples (grey lines) follow a power law with slope mean $-6$ and standard deviation $2$. A steep prior power spectrum is chosen to suppress small scale structure in the early self-calibration stages. This prevents imprinting imaging artifacts from the noise to the final image. A relatively high standard deviation of the power spectrum is chosen to ensure flexibility of the prior model. Fluctuations and flexibility are non-trivial hyperparameters controlling the Wiener process and integrated Wiener process in the model, which determine fluctuation and flexibility of the power spectrum in a nonparametric fashion. Nonzero asperity can generate periodic patterns in the image. Thus, we used relatively small asperity parameters for our log-sky model $\psi$. Note that the power spectrum model is flexible enough to capture different correlation structures. As a result, the flexibility of the prior model can reduce biases from strong prior assumptions. 

For the self-calibration of the real data (see Section \ref{Section_resolve_real_data}), four temporal correlation kernels (amplitude gain and the phase gain for RCP and LCP mode respectively) are inferred under the assumption that antennas from homogeneous array have similar amplitude and phase gain correlation structures per polarization mode. For the self-calibration of the synthetic data (see Section \ref{Section_CLEAN_resolve_synthetic_data}), 40 individual temporal correlation kernels for the amplitude and phase gains are inferred, one per antenna and polarization mode, as the ground truth gain corruptions were generated with individual correlation structures. The mean of the amplitude gain prior model is obtained by exponentiating the offset mean of $\lambda$, which is $\text{exp}(0) = 1$ and the mean of the phase gain prior model is the offset mean of $\phi$, which is $0$ radians. Model parameters for the zero mode variance, fluctuations, and flexibility are chosen to be small to suppress extremely high gain corrections. Asperity mean and std hyper parameters are set to None since the gain solutions are not periodic. The average slope mean is $-3$ for log-amplitude and phase gain, therefore the slope mean for the power spectrum is $-6$ with the standard deviataion $2$. The broad range of the average slope parameter allows the prior model to describe different temporal correlation structures. As a result, the gain prior model is flexible enough to learn the temporal correlation structure from the data automatically. More details about prior model parameters are explained in \cite{Arras_2021_CygA}.

\section{\texttt{CLEAN} self-calibration solutions: real data} \label{Appendix_CLEAN_selfcal}\color{black}

\begin{figure}[t]
    \includegraphics[width=9cm] {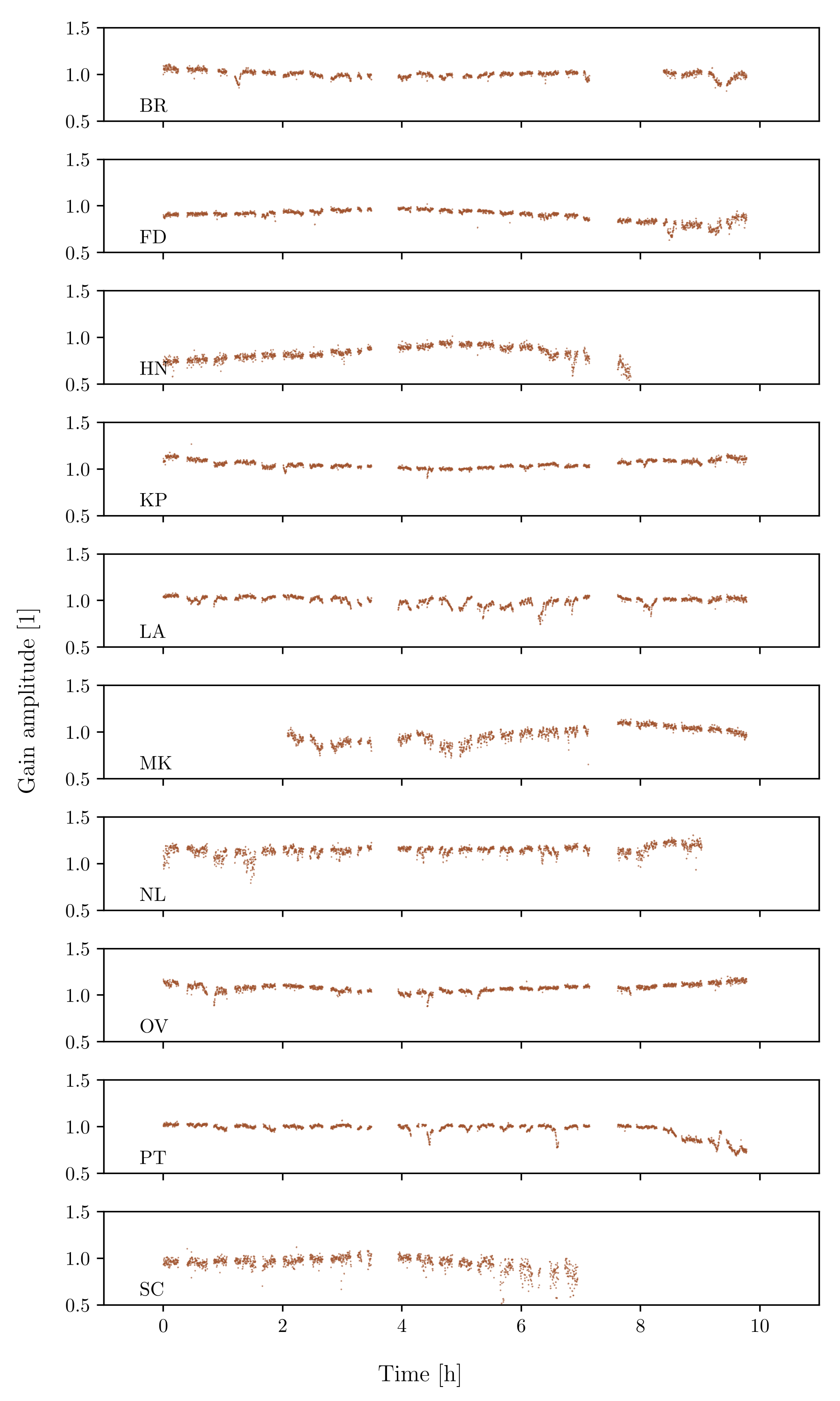}

    \caption{Real data: Amplitude gain solutions by the \texttt{CLEAN} self-calibration method.}
    \label{fig:CLEAN_selfcal_solutions_amp}
\end{figure}

\begin{figure}[t]
    \includegraphics[width=9cm] {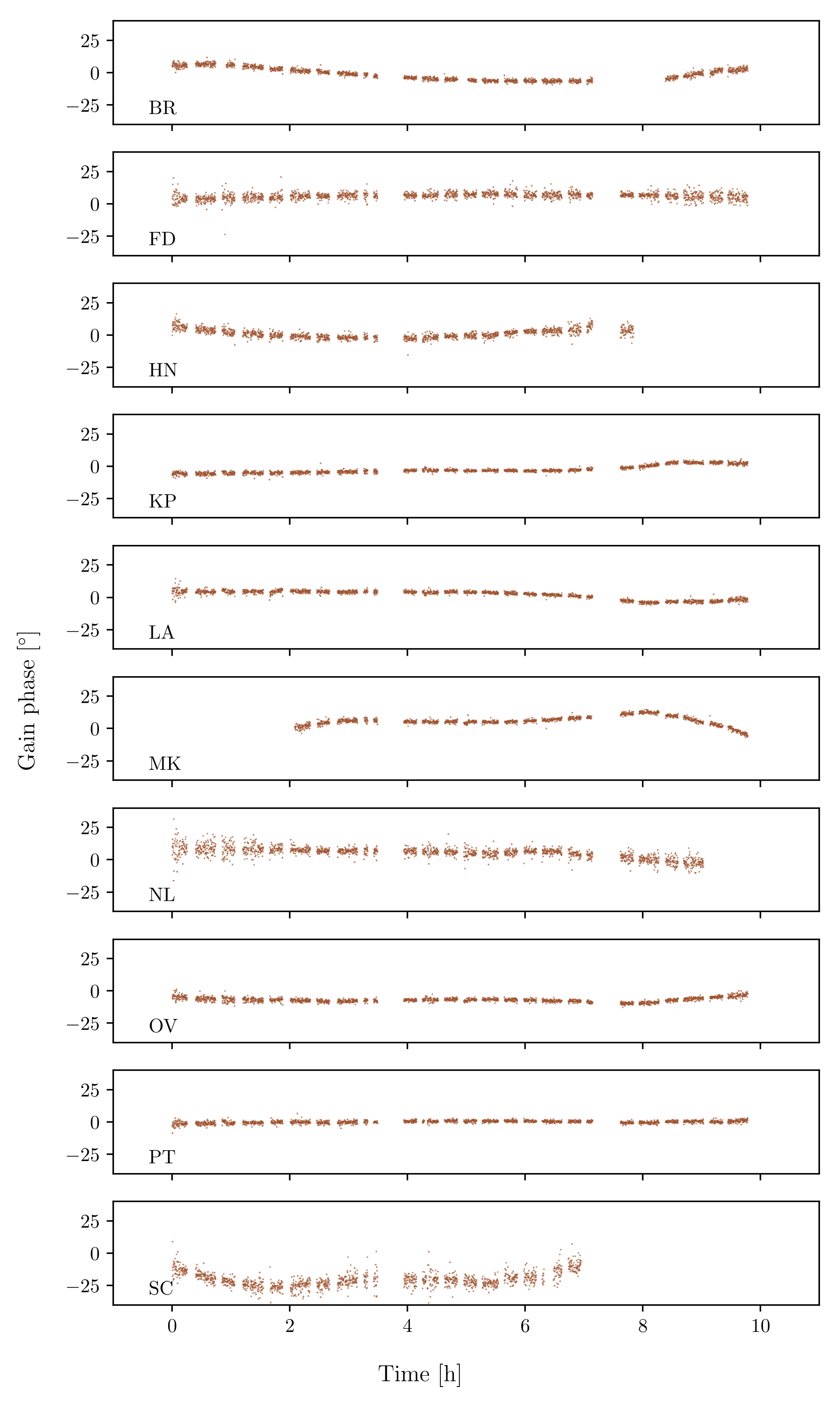}

    \caption{Real data: Phase gain solutions by the \texttt{CLEAN} self-calibration method.}
    \label{fig:CLEAN_selfcal_solutions_phase}
\end{figure}

\autoref{fig:CLEAN_selfcal_solutions_amp} and \autoref{fig:CLEAN_selfcal_solutions_phase} show the \texttt{CLEAN} self-calibration solutions for the real VLBI M87 data at 43GHz by DIFMAP software. One amplitude and one phase gain solution per each antenna are obtained because \texttt{CLEAN} image in \autoref{fig:M87_real_resolve_CLEAN} is produced by the Stokes I data averaging the RR and LL components. Since the Stokes V emission is negligible in the data, \texttt{resolve} gain solutions for RCP and LCP are very similar in \autoref{fig:amp_gains_realdata} and \autoref{fig:phase_gains_realdata}. Therefore, a high-fidelity total intensity image can be reconstructed by estimating one gain solution per each antenna in \texttt{CLEAN} self-calibration. 

We can compare self-calibration solutions from \texttt{CLEAN} and \texttt{resolve}. The gain solutions from \texttt{CLEAN} and \texttt{resolve} with scans are consistent qualitatively. As an example, the abrupt variation of the \texttt{resolve} gain amplitude for LA antenna (6h - 6.5h) in \autoref{fig:real_amp_phase_gain_LA_LCP} is due to the discrepancy of amplitude between scans. \autoref{fig:CLEAN_selfcal_solutions_amp} shows a similar behavior in the \texttt{CLEAN} gain amplitude for LA antenna (6h - 6.5h). Note that the data were flagged manually during \texttt{CLEAN} self-calibration. Outliers are often flagged during iterative \texttt{CLEAN} self-calibration and the data flagging relies on user's experience. On the other hand, in \texttt{resolve} self-calibration, the whole data after pre-calibration are used for imaging without manual flagging.

\section{\texttt{ehtim} image and self-calibration solutions: real data} \label{Appendix_ehtim_selfcal}

\begin{figure}[t]
    \includegraphics[width=\linewidth] {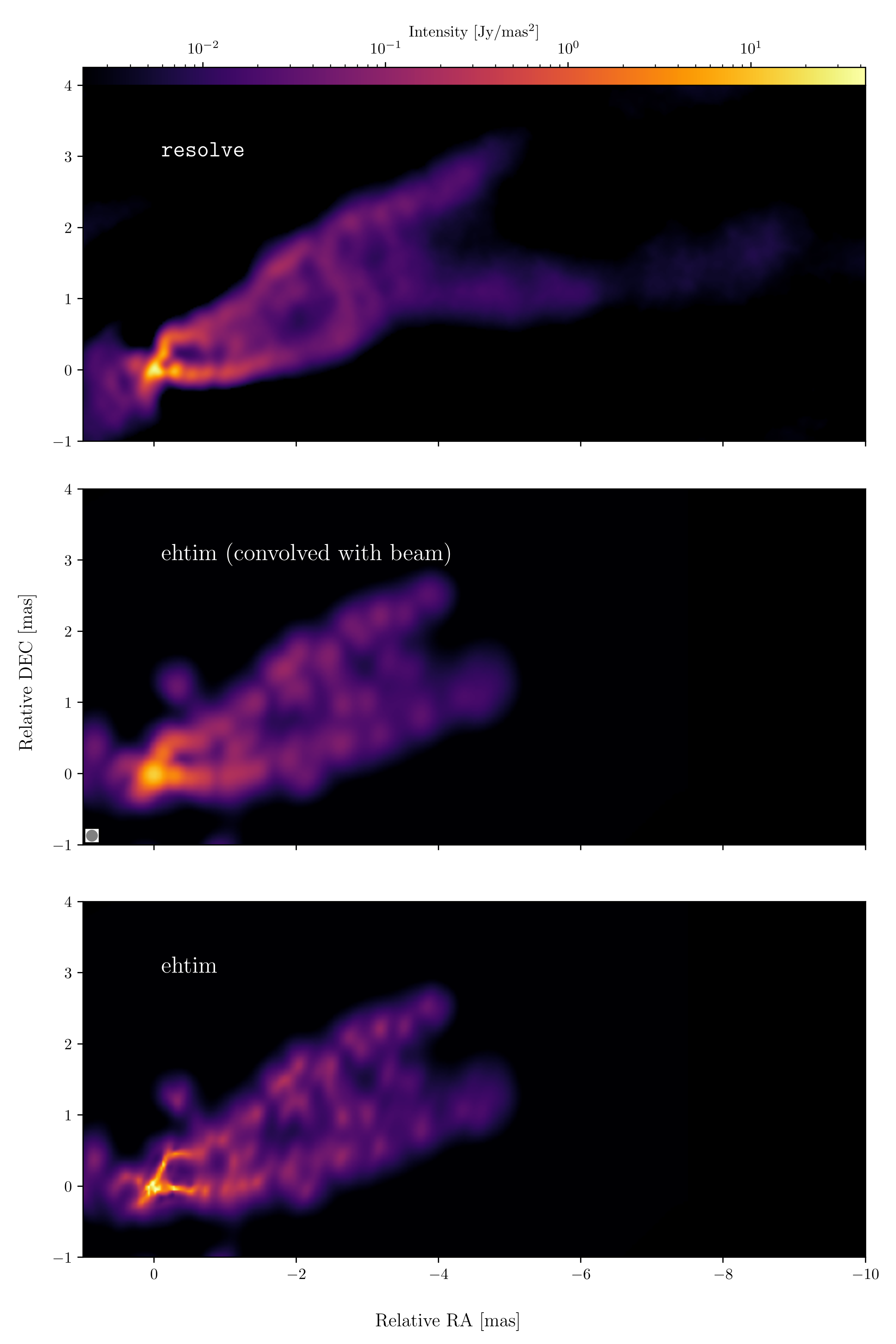}\caption{ Real data: \texttt{resolve} image (top), \texttt{ehtim} image convolved with the beam (0.167mas $\times$ 0.167mas, middle), and \texttt{ehtim} image (bottom). The beam-convolved \texttt{ehtim} image (middle) has $I_{\textrm{max}} = 14$\,Jy\,mas$^{-2}$ and \texttt{ehtim} image (bottom) has $I_{\textrm{max}} = 73$\,Jy\,mas$^{-2}$.}
    \label{fig:ehtim_sky_appendix}
\end{figure}

\begin{figure}[h]
    \includegraphics[width=9cm] {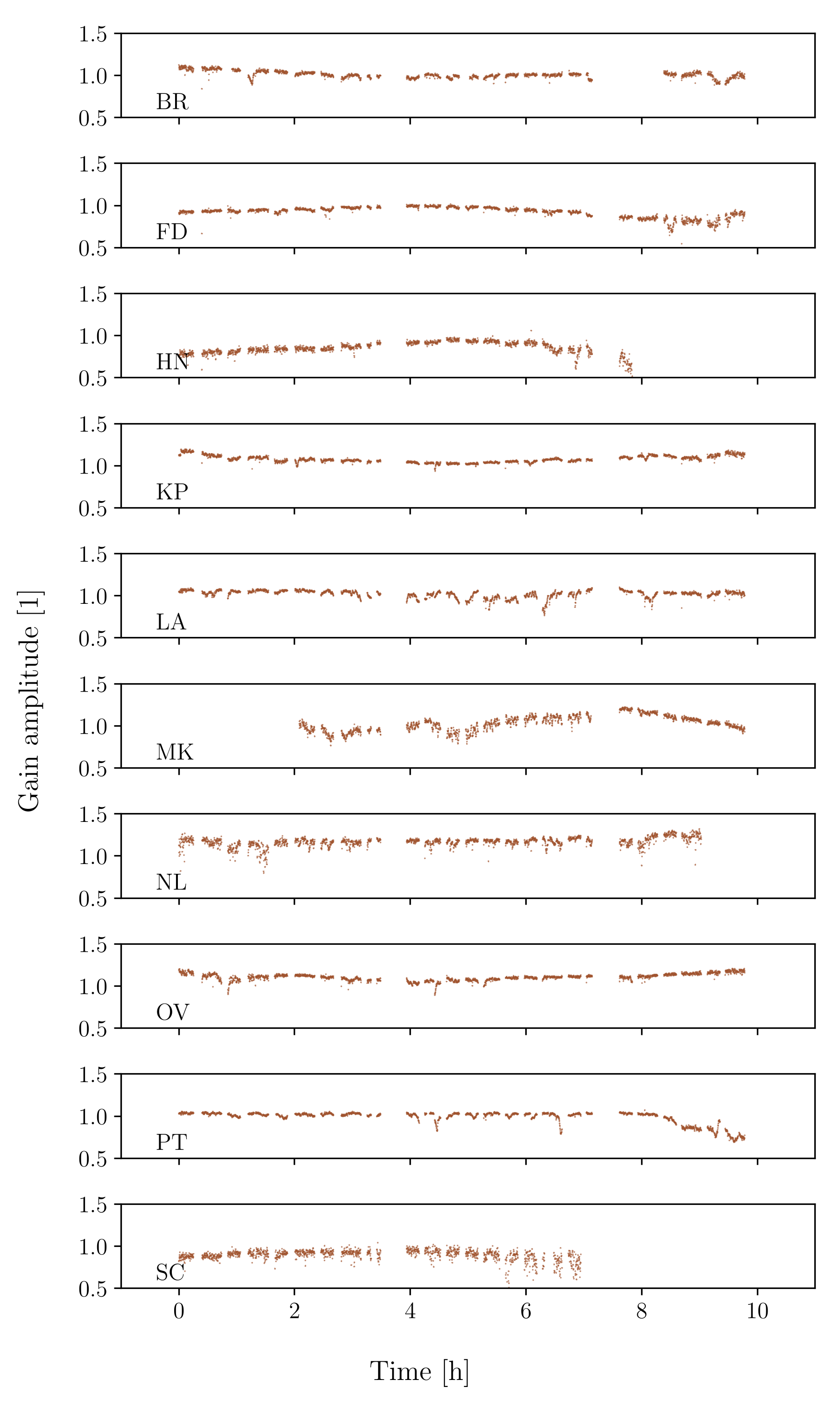}

    \caption{Real data: Amplitude gain solutions by the \texttt{ehtim} self-calibration method.}
    \label{fig:ehtim_selfcal_solutions_amp}
\end{figure}

\begin{figure}[h]
    \includegraphics[width=9cm] {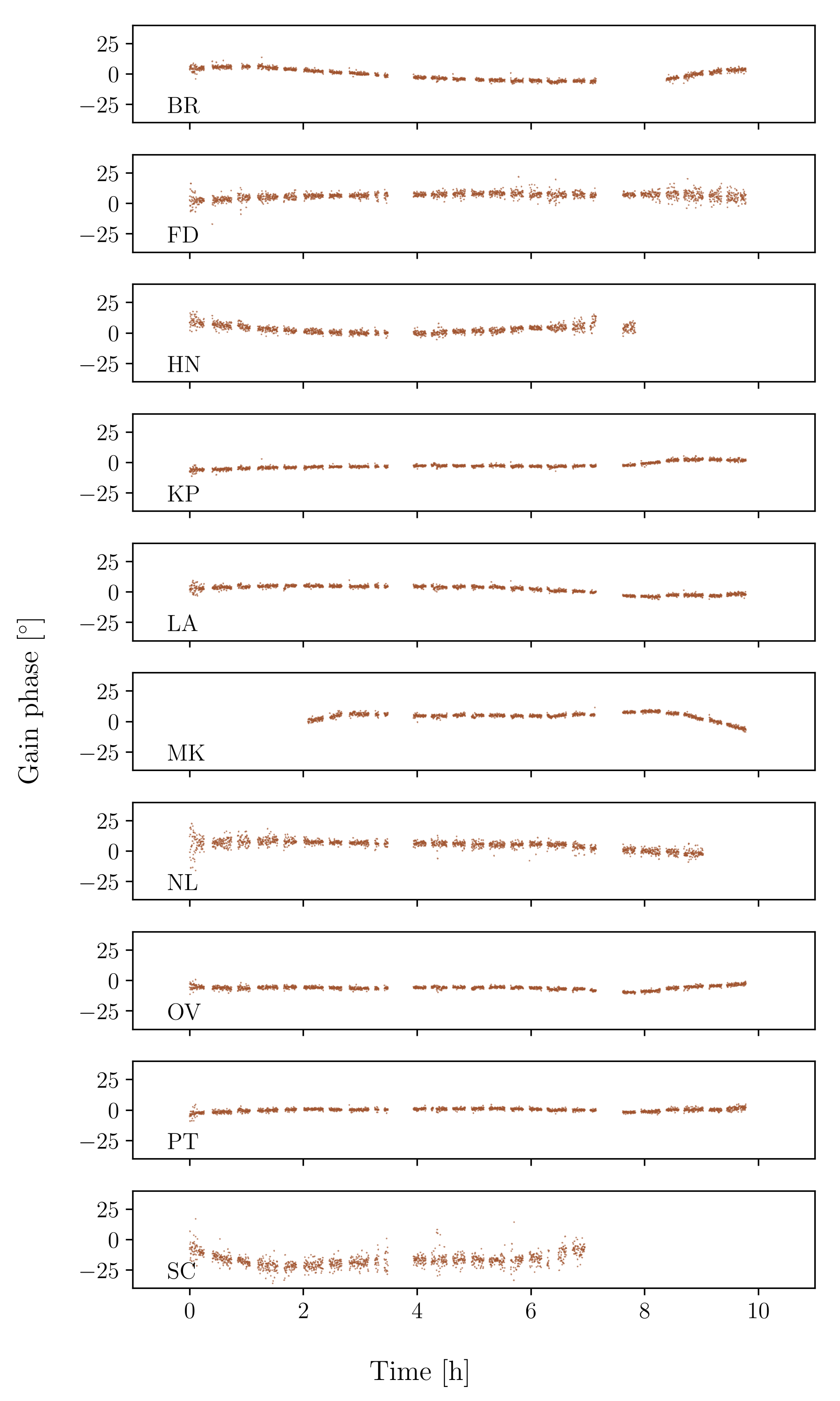}

    \caption{Real data: Phase gain solutions by the \texttt{ehtim} self-calibration method.}
    \label{fig:ehtim_selfcal_solutions_phase}
\end{figure}

\texttt{ehtim} \citep{Chael_2018_closure} is a widely used RML imaging software for VLBI data, e.g. by the Event Horizon Telescope at $230\,\mathrm{GHz}$ \citep[e.g.][]{eht_m87_paper4, eht_sgra_paper3, eht_m87_2018_paper1}, for observations with the global Millimeter VLBI Array (GMVA) at $86\,\mathrm{GHz}$ \citep[e.g.][]{Zhao2022}, observations with European VLBI network (EVN), the Very Long Baseline Array (VLBA), and the space VLBI mission RadioAstron at smaller frequencies \citep[e.g. recently][]{Kosogorov2024, Paraschos2024, Fuentes2023}, and even for related inverse problems outside of radio astronomy \citep{STIX}. To validate the \texttt{resolve} self-calibration and imaging method, we compare it to RML self-calibration and imaging performed by \texttt{ehtim}. The comparison of \texttt{resolve} and \texttt{ehtim} image is shown in \autoref{fig:ehtim_sky_appendix}. \texttt{ehtim} minimizes a weighted sum of data terms (chosen to be $\chi^2$-terms based on visibilities), and regularization terms (entropy, TV, TSV, $l_{1}$-norm, total flux constraint). To navigate this large number of parameters, usually either a parameter survey \citep[e.g.][]{eht_m87_paper4, Fuentes2023} or metaheuristics \citep{Mueller_2023_GA, Mus2024} are applied to \texttt{ehtim}. Performing a full exploration of the parameter space would exceed the scope of this comparison. Hence, for this work, we rather chose the regularization weights by manual exploration to the best of our efforts.

In \texttt{ehtim}, the self-calibration procedure is similar to \texttt{CLEAN} self-calibration. First, a model image is reconstructed by the given data. Second, residual gains are estimated by minimizing a cost function (Eq. \ref{eq:selfcal_cost_function}) based on the Fourier components of the model image. Then new data are generated by removing the estimated residual gain corruption. These procedures are continued iteratively until the desired image quality is achieved. In contrast to the \texttt{CLEAN} self-calibration, the model images computed by \texttt{ehtim} can be consistent with data since the model fits the data directly in the visibility domain.

The \texttt{ehtim} image in \autoref{fig:ehtim_sky_appendix} achieves better resolution compared to \texttt{CLEAN} image in \autoref{fig:M87_real_resolve_CLEAN}. The core regions show even sharper edge-brightening and counter jet than \texttt{resolve} image. However, the image contains sharper artifacts in the core region and the extended jet emission looks discontinuous. Therefore, we convolved \texttt{ehtim} image with the beam (0.167mas $\times$ 0.167mas) to show an image at a conservative resolution. The core, counter jet, and extended jet emission in \texttt{resolve} and beam-convolved \texttt{ehtim} images look consistent. Note that the smoothness of the image is enforced by the TV and TSV regularizers in \texttt{ehtim} reconstruction. However, since TV and TSV regularizers tend to produce rough structures (see Appendix \ref{app:RML}), these regularizers may not be optimal for recovering extended jet emission as \texttt{ehtim} was developed for the high-resolution reconstruction of compact emission. It has been demonstrated in the past that the use of extended basis functions, rather than regularizers acting in the pixel basis, may offer an advantage for these cases in RML methods \citep[see the SARA family][]{Carrillo2012, Terris2023, Wilber2023}.

\autoref{fig:ehtim_selfcal_solutions_amp} and \autoref{fig:ehtim_selfcal_solutions_phase} show the amplitude and phase gain solutions by \texttt{ehtim} self-calibration. Self-calibration by \texttt{ehtim} is performed alternating with the deconvolution. For the self-calibration of real M87 data, no temporal correlation between gain solutions is considered and small gain tolerance of (0.01, 0.05) is used to correct the gain incrementally. This means that both, gains smaller than 0.01 and larger than 0.05, are disfavored by the prior. We detect gain phase and amplitude trends that are consistent with the results obtained with \texttt{resolve} and \texttt{CLEAN}.

\section{Comparison of \texttt{resolve} and over-resolved \texttt{CLEAN} image}
\label{sec:appendix_superres}

\begin{figure}[t]
    \includegraphics[width=\linewidth] {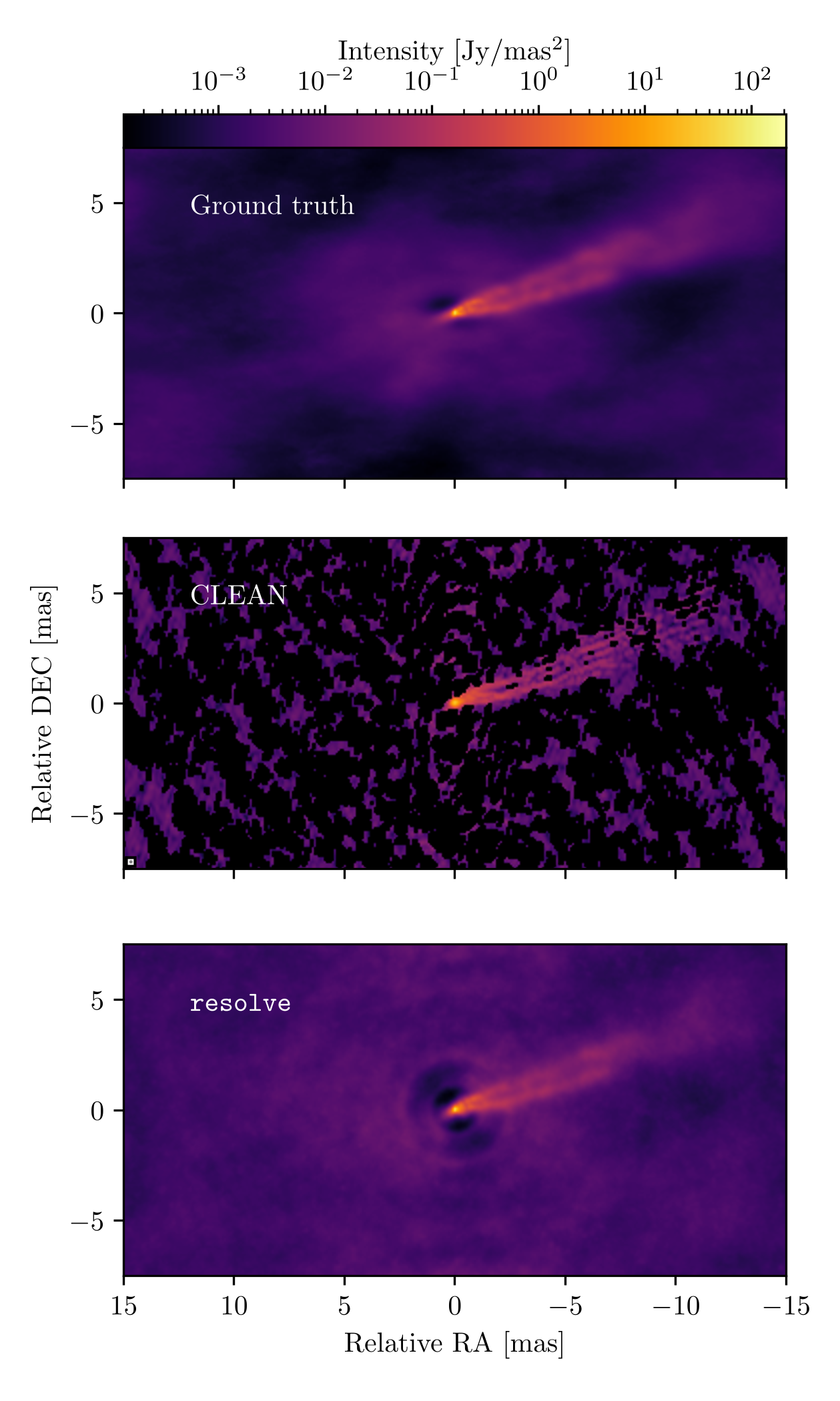}{\caption{ Synthetic data: ground truth (top) and reconstructed images obtained using \texttt{CLEAN} (middle) and \texttt{resolve} (bottom) self-calibration. The circular 0.15\,mas \texttt{CLEAN} beam is illustrated in the bottom left corner of the plot. The \texttt{CLEAN} image was masked at the lowest positive value. The unified color bar on the top of the figure shows an intensity range of the ground truth image, where maximum intensity is $I_{\textrm{max}}^{\textrm{GT}} = 209$\,Jy\,mas$^{-2}$, the minimum value is $I_{\textrm{min}}^{\textrm{GT}} = 137$\,$\mu$Jy\,mas$^{-2}$. Maximum intensity values of reconstructed images are $I_{\textrm{max}}^{\textrm{\texttt{CLEAN}}} = 34$\,Jy\,mas$^{-2}$, $I_{\textrm{max}}^{\texttt{resolve}} = 111$\,Jy\,mas$^{-2}$ correspondingly.}
    \label{fig:synthetic_sky_appendix}
    }
\end{figure}

\begin{figure*}[h]
    \centering
    \includegraphics[width=0.75\linewidth] {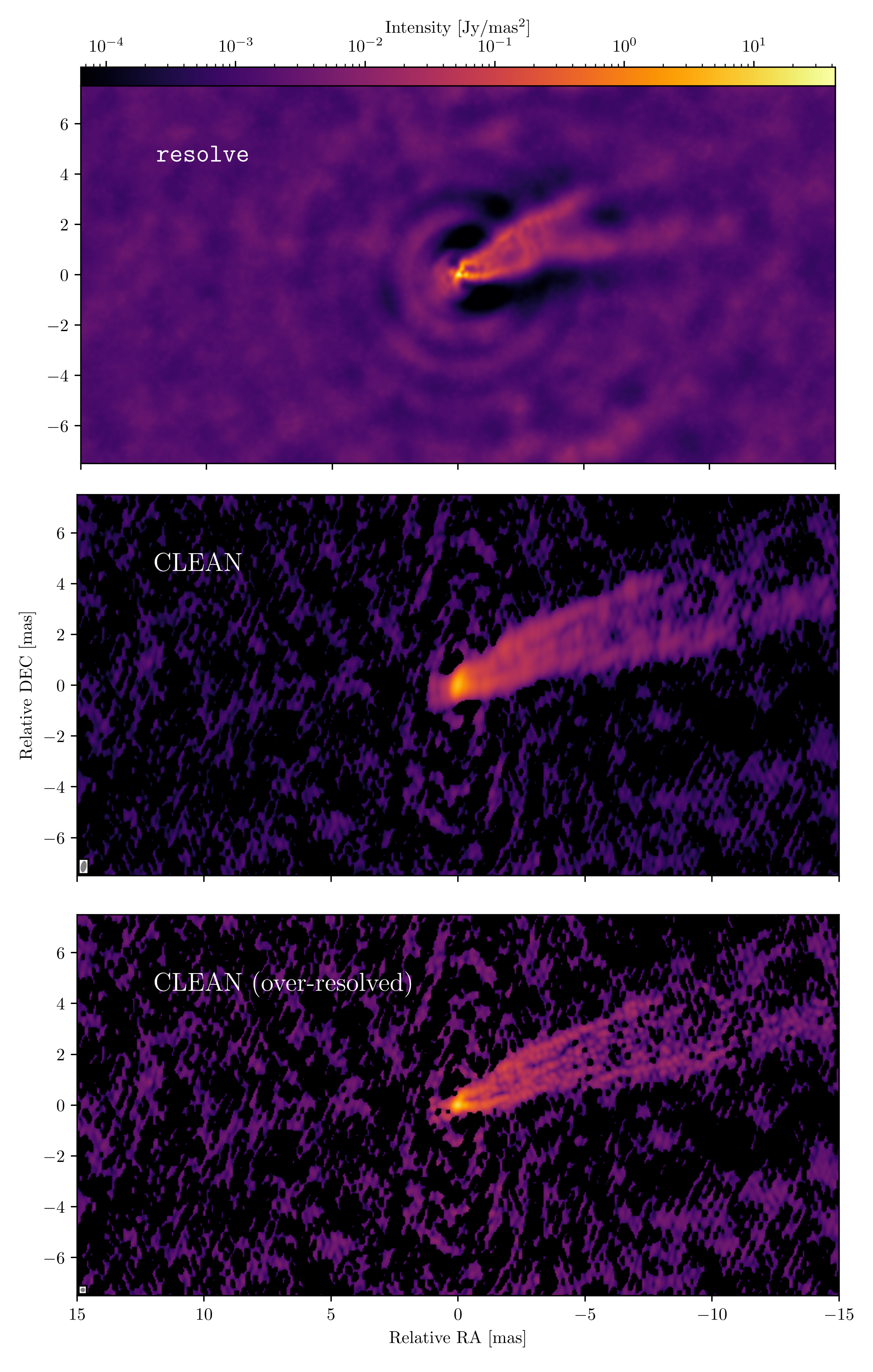}

    \caption{ M87: the posterior mean image by Bayesian self-calibration (top), the self-calibrated \texttt{CLEAN} image (middle) and the over-resolved \texttt{CLEAN} image (bottom) reconstructed from the same a-priori calibrated visibility data of VLBA observations at 43\,GHz. The unified color bar shows an intensity range from the minimum intensity up to the maximum intensity of the \texttt{resolve} image. The image obtained by the Bayesian approach has a maximum intensity $I_{\textrm{max}} = 35$\,Jy\,mas$^{-2}$ with a minimum value of $I_{\textrm{min}} =8$\,$\mu$Jy\,mas$^{-2}$. The \texttt{CLEAN} image (middle) restoring beam shown in the lower-left corner of the plot is $0.5 \times 0.2$\,mas, P.A.~$=-11^{\circ}$. The maximum intensity of the \texttt{CLEAN} reconstruction is $I_{\textrm{max}} = 6$\,Jy\,mas$^{-2}$. The over-resolved \texttt{CLEAN} image (bottom) circular restoring beam shown in the lower-left corner of the plot is 0.18\,mas. The maximum intensity of the over-resolved \texttt{CLEAN} image is $I_{\textrm{peak}} = 13$\,Jy\,mas$^{-2}$.} 
    \label{fig:M87_real_resolve_CLEAN_appendix}
\end{figure*}

\autoref{fig:synthetic_sky_appendix} and \autoref{fig:M87_real_resolve_CLEAN_appendix} show the comparison between \texttt{CLEAN} image with over-resolved beam and \texttt{resolve} image with saturated color bar for synthetic and real data respectively. In \autoref{fig:synthetic_sky_appendix}, the \texttt{CLEAN} algorithm is not able to recover small scale structures in the core of the ground truth image even with over-resolved beam. Furthermore, the extended jet in \texttt{CLEAN} image is discontinuous due to the small size of the \texttt{CLEAN} beam. In \texttt{resolve} image, the extended jet and bright core are relatively well recovered. Note that central spine in extended jet is reconstructed in \texttt{CLEAN} image. It is not pronounced in the ground truth image, can therefore be \texttt{CLEAN} artifacts \citep{Pashchenko_2023}. In \autoref{fig:M87_real_resolve_CLEAN_appendix}, the structure of core and limb-brightened region is consistent in \texttt{resolve} and \texttt{CLEAN} reconstruction with over-resolved beam. The central spine in extended jet is pronounced in the \texttt{CLEAN} image again, same as the \texttt{CLEAN} reconstruction from the synthetic data.

%
%
%

\end{document}